\documentclass{dune}
\pdfoutput=1

\input{common/preamble}

\begin{document}

%\linenumbers

\pagestyle{titlepage}

% This should be \input first thing after \begin{document}

\pagestyle{titlepage}

%\date{\scshape\today}
\date{\scshape November 30, 2022}

\title{\scshape\Large 
Snowmass Neutrino Frontier: \\
NF01 Topical Group Report\\
%\vspace{5mm}
\normalsize Three-Flavor Neutrino Oscillations
\vskip -10pt
\snowmasstitle
}

%\vspace{10mm}

\renewcommand\Authfont{\scshape\small}
\renewcommand\Affilfont{\itshape\footnotesize}

%Authors go here with affiliation indexed by [X]
\author[1]{Peter B.~Denton$^{\thanks{Topical Group Conveners}}$}
\author[*2]{Megan Friend}
\author[*3]{Mark D.~Messier}
\author[*4]{Hirohisa A.~Tanaka}

\author[5]{Sebastian B\"oser}
\author[6]{Jo\~{a}o\,A.\,B.~Coelho}
\author[7]{Mathieu~Perrin-Terrin}
\author[8]{Tom Stuttard}

\vspace{-0.5cm}
%Affiliations go here
\affil[1]{High Energy Theory Group, Physics Department, Brookhaven National Laboratory, Upton, NY 11973, USA}
\affil[2]{High Energy Accelerator Research Organization (KEK), Tsukuba, Ibaraki 305-0801, Japan}
\affil[3]{Indiana University, Bloomington, Indiana 47405, USA}
\affil[4]{SLAC National Accelerator Laboratory, Menlo Park, CA 94025, USA}

\affil[5]{University of Mainz, Germany}
\affil[6]{Universit{\'e} de Paris, CNRS, Astroparticule et Cosmologie, F-75013 Paris, France}
\affil[7]{Aix~Marseille~Univ,~CNRS/IN2P3,~CPPM,~Marseille,~France}
\affil[8]{Niels Bohr Institute, University of Copenhagen, Denmark}

\maketitle

This is the report from the Snowmass NF01 topical group and colleagues on the current status and expected future progress to understand the three-flavor neutrino oscillation picture.

\renewcommand{\familydefault}{\sfdefault}
\renewcommand{\thepage}{\roman{page}}
\setcounter{page}{0}

\pagestyle{plain} 
\clearpage
\textsf{\tableofcontents}

%\iffinal\else
%\textsf{\listoftodos}
%\clearpage
%\fi

\renewcommand{\thepage}{\arabic{page}}
\setcounter{page}{1}

\pagestyle{fancy}

% Set how header/footers look
%\newcommand{\chaptermark}[1]{%
%\markboth{Chapter \thechapter:\# 1}{}}
%\renewcommand{\chaptermark}[1]{%
%\markboth{Chapter \thechapter:\ #1}{}}
\fancyhead{}
%\fancyhead[RO,LE]{\textsf{\footnotesize \thepage}}
\fancyhead[RO]{\textsf{\footnotesize \thepage}}
%\fancyhead[LO,RE]{\textsf{\footnotesize \rightmark}}
\fancyhead[LO]{\textsf{\footnotesize \rightmark}}

\fancyfoot{}
\fancyfoot[RO]{\textsf{\footnotesize Snowmass 2021}}
\fancyfoot[LO]{\textsf{\footnotesize NF01 Topical Group Report}}
\fancypagestyle{plain}{}

\renewcommand{\headrule}{\vspace{-4mm}\color[gray]{0.5}{\rule{\headwidth}{0.5pt}}}

\clearpage

\section*{Executive Summary}
\label{sec:summary}
\addcontentsline{toc}{section}{Executive Summary}
The discovery of neutrino oscillations in 1998 and 2002 added at least seven new parameters to our model of particle physics, and oscillation experiments can probe six of them.
To date, three of those parameters are fairly well measured: the reactor mixing angle $\theta_{13}$, the solar mixing angle $\theta_{12}$, and the solar mass splitting $\Delta m^2_{21}$, although there is only one good measurement of the last parameter.
Of the remaining three oscillation parameters, we have some information on two of them: we know the absolute value of the atmospheric mass splitting $\Delta m^2_{31}$ fairly well, but we do not know its sign, and we know that the atmospheric mixing angle $\theta_{23}$ is close to maximal $\sim45^\circ$, but we do not know how close, nor on which side of maximal it is.
Finally, the sixth parameter is the complex phase $\delta$ related to charge-parity (CP) violation, which is largely unconstrained.

Determining these remaining three unknowns, the sign of $\Delta m^2_{31}$, the octant of $\theta_{23}$, and the value of the complex phase $\delta$, is of the utmost priority for particle physics.
In addition to the absolute neutrino mass scale which can be probed with cosmological data sets, they represent the only known unknown parameters in our picture of particle physics.
It is our job as physicists to determine the parameters of our model.
The values of these parameters have important implications in many other areas of particle physics and cosmology, as well as providing insights into the flavor puzzle.

To measure these parameters, a mature experimental program is underway with some experiments running now and others under construction.
In the current generation we have NOvA, T2K, and Super-Kamiokande (SK) which each have some sensitivity to the three remaining unknowns, but are unlikely to get to the required statistical thresholds.
Next generation experiments, notably DUNE and Hyper-Kamiokande (HK) are expected to get to the desired thresholds to answer all three oscillation unknowns.
Additional important oscillation results will come from JUNO, IceCube, and KM3NeT.
This broad experimental program reflects the fact that there are many inter-connected parameters in the three-flavor oscillation picture that need to be simultaneously disentangled and independently confirmed to ensure that we truly understand these parameters.

To achieve these ambitious goals, DUNE and HK will need to become the most sophisticated neutrino experiments constructed to date.
Each requires extremely powerful neutrino beams, as many measurements are statistics limited.
Each will require a very sophisticated near detector facility to measure that beam, as well as to constrain neutrino interactions and detector modeling uncertainties, which are notoriously difficult in the energy ranges needed for oscillations.
To augment the near detectors, additional measurements and theory work are crucial to understand the interactions properly, see NF06 \cite{Balantekin:2022jrq}.
Finally, large highly sophisticated far detectors are required to be able to reconstruct the events in a large enough volume to accumulate enough statistics.
DUNE will use liquid argon time-projection chamber (LArTPC) technology most recently demonstrated with MicroBooNE.
LArTPCs provide unparalleled event reconstruction capabilities and can be scaled to large enough size to accumulate the necessary statistics.
HK will expand upon the success of SK's large water Cherenkov tank and build a new larger tank using improved photosensor technology.

It is fully expected that with the combination of experiments described above, a clear picture of three-flavor neutrino oscillations should emerge, or, if there is new physics in neutrino oscillations (see NF02 \cite{Karagiorgi:2022fgf} and NF03 \cite{Coloma:2022dng}), that should fall into stark contrast in coming years.

\clearpage

\FloatBarrier
\section{Introduction and Current Three-Flavor Status}
\label{sec:introduction}

\subsection{Neutrino Oscillations in Particle Physics}

The Standard Model has been highly successful, but we know it cannot be a 
complete fundamental physics theory. The Standard Model does not provide a 
source of dark matter and dark energy whose imprints on the universe we see 
in the night sky through our telescopes. The Standard Model does not explain
the vast difference in strength between the electric, weak, and strong 
forces and gravity.
The Standard Model does not explain the hierarchy of masses of quarks and leptons.
The Standard Model does not explain why charge-parity (CP) is sometimes violated and sometimes conserved.
And the Standard Model does not explain the mechanism behind neutrino masses, nor why they are hierarchically different from the other fermions. 
Of these four, the effects of neutrino masses are the only sign of physics 
outside the Standard Model which, to date, we can manipulate in our laboratories.

While there are various simple ways to introduce neutrino masses to the Standard Model, 
these approaches leave many questions unanswered. They do not account for the 
unusual smallness of the neutrino masses relative to their charged partners; 
they do not explain why neutrino mixing is large, whereas mixing in other 
sectors of the Standard Model is small. They don't predict patterns or 
symmetries in the masses or answer a fundamental question about the 
differences between neutrinos, antineutrinos, and lepton charge conservation.

Since the discovery of neutrino oscillations in 1998, a global program has 
been developed to explore neutrino masses and their mixing using a wide variety 
of natural and artificial sources. This program has narrowed the allowed ranges 
of neutrino mass splittings and shown that all neutrino mixing is relatively, 
perhaps surprisingly, large. However, many questions remain. Do neutrinos 
follow a ``normal hierarchy'' that would associate most of the electron 
flavor with the lightest states or an ``inverted hierarchy'' where the 
electron flavor is mainly in the heavier states? Do neutrino masses and 
mixing possess new symmetries? Do neutrino oscillations violate charge-parity?
If, as many think, the lightness of neutrino masses is associated with 
physics approaching the scale of grand unification, will their 
oscillations contain echos of new physics at these energy scales?

The answers to these questions will demand levels of precision not yet 
seen in neutrino physics. They will place new demands on the experimental 
program: larger, more precise detectors, a more accurate understanding 
of the mechanisms which produce neutrinos, and the physics behind the 
interactions which experiments use to detect them.

This report documents the current and future program to measure neutrino 
oscillations to precisely elucidate the now standard 
Pontecorvo–Maki–Nakagawa–Sakata (PMNS) framework of 
neutrino masses and mixings. Precise measurements of the PMNS framework 
will answer questions about new symmetries in neutrino mixing, enable 
tests of predictions of neutrino mass models, and resolve the neutrino mass 
hierarchy. These measurements will establish a baseline understanding 
of neutrino oscillations upon which theorists and experimenters 
will build searches for additional physics beyond the 
Standard Model associated with neutrinos.

\subsection{Current Knowns and Known Unknowns in Neutrino Oscillations}
Given many oscillation experiments over the last several decades, a clear picture of the overall framework of three-flavor oscillations has emerged.
The details are discussed in more detail in section \ref{ssec:role of each} below, but generally we have two mass-squared differences: $\Delta m^2_{21}\sim+7.5\times10^{-5}$ eV$^2$ and $\Delta m^2_{31}\sim\pm2.5\times10^{-3}$ eV$^2$.
We know two of the mixing angles fairly well: $\theta_{13}\sim9^\circ$ and $\theta_{12}\sim34^\circ$.
The third mixing angle is close to maximal, but somewhat uncertain, $\theta_{23}\sim45^\circ$, and the complex phase $\delta$ is largely unconstrained.

The three primary goals of the coming years of neutrino oscillations are to 1) determine the sign of $\Delta m^2_{31}$, known as the atmospheric mass ordering question, 2) determine whether $\theta_{23}$ is more than or less than $45^\circ$ and how close to maximal it is, known as the octant question, and 3) measure $\delta$ and determine if $\sin\delta=0$, and thus CP conservation, can be excluded or not.

While some of these parameters can be assessed at an individual experiment (with input from orthogonal experiments), some parameters will first be measured as a result of global analyses of all relevant neutrino data, known as global fits.
There are three primary global fit groups at the moment, all based in Europe \cite{Capozzi:2021fjo,deSalas:2020pgw,Esteban:2020cvm}.
While they generally agree on some parameters like $\theta_{13}$, there are some differences among them for $\theta_{12}$, $\Delta m^2_{21}$ and $\theta_{23}$.

\subsection{Neutrino oscillations and the Previous P5}

The study of neutrino oscillations featured prominently
in the previous P5 report~\cite{p5-2014} which identified
``Pursue the physics associated with neutrino mass'' as
one of five key science drivers for U.S. high energy
physics. P5 identified the three key questions for neutrino
physics as 
``What is origin of neutrino mass?'',
``How are the neutrino masses ordered?'', and
``Do neutrinos and antineutrinos oscillate differently?''.

In the spirit of its Recommendation 1, 
``Pursue the most important 
opportunities wherever they are, and host unique, 
world-class facilities that engage the global 
scientific community'',
P5 recommended an ambitious neutrino oscillation
program to be based at the proton source at Fermilab
(Recommendations 12 and 13).
On this point, P5 was very specific, recommending
the formation of an international collaboration
to build a neutrino beam capable of delivering
1.2~MW with the possibility of upgrades to multi-megawatt
power. This beam is to target an underground 
liquid argon detector of 40~kt fiducial mass located at a 
baseline of 1350~km in Lead, SD.
The program, known as the Deep Underground Neutrino Experiment (DUNE) is expected to begin data taking in 2029 and
deliver 3$\sigma$ sensitivity for 75\% of $\delta_{\rm CP}$ 
values by 2035.
Recommendation 13 concluded that this experiment
``is the highest priority large project in its timeframe.''.

To support this program, P5 recommended that Fermilab
proceed with construction of PIP-II (Recommendation 14)
to achieve multi-megawatt beam powers and 
set goals of reaching an
exposure of 600 kt $\cdot$ MW $\cdot$ yr, and precise
control of systematic uncertainties:
$\pm 1$\% signal uncertainty, and
$\pm 5$\% background uncertainty for
electron neutrino appearance in neutrino and antineutrino
beams. This control of systematic uncertainties 
is unprecedented in
long baseline physics and places ambitious requirements
on the experiment near detector suite and may 
also necessitate
ancillary measurements of neutrino cross-sections
and the production of hadrons which produce 
neutrino fluxes.
These ancillary measurements align with P5's 
Recommendation 4; 
``Maintain a program of projects of all
scales, from the largest international projects to mid- and
small-scale projects.''

This P5 vision for building a U.S. based long baseline
program built on PIP-II and LBNF came with some 
choices. Recommendation 15 of the 2014 P5 report 
advised against new experimental projects on the NuMI beamline. The committee also recommended termination of 
U.S. participation in the MICE program of neutrino factory 
R\&D and recommended against advancing a 
proposal (NuSTORM) to construct a muon storage 
ring to conduct short baseline measurements 
and exercise ideas for a future neutrino factory.
The committee endorsed 
continued development of the idea to use 
high intensity cyclotrons to measure CP violation
in the neutrino sector (IsoDAR and Daedalus)
and encouraged further exploration of the potential
to measure neutrino oscillations at the South Pole 
using the IceCube neutrino telescope.

\FloatBarrier
\section{Three-Flavor Oscillation Theory}
\label{sec:theory}
In this section we briefly review the standard paradigm of oscillation theory in vacuum and in matter and the role of each of the six oscillation parameters.
Notably, there are three major current unknowns that the community is attempting to determine while simultaneously making precision measurements of all the oscillation parameters.
Additional discussion of the role of theory supporting neutrino oscillations can be found in \ref{sec:osc theory support}.

\subsection{Neutrino Oscillation Probabilities}
The discovery of neutrino oscillations, coming from Super-Kamiokande \cite{Super-Kamiokande:1998kpq} and SNO \cite{SNO:2002tuh} data confirmed the existence of neutrino oscillations at high $>5\sigma$ significance independent of theoretical flux predictions.
This immediately added at least seven new parameters to our model of particle physics: three masses and four parameters related to the mixing between the weak interaction basis and the mass basis.
Moreover, since there is not a single obvious minimal mechanism for neutrino mass generation, there are likely additional parameters and particles waiting to be discovered at scales other than the neutrino scale $\sim0.01$ eV.
Nonetheless the neutrino oscillation phenomenon is sensitive to six of the seven parameters and measuring these parameters has been the focus of an intense global effort over the last several decades and is the focus of this report.

Neutrinos are produced in flavor states which are linear combinations of the mass states.
Since neutrinos propagate in the mass states (i.e.~the mass states are the eigenstates of the Hamiltonian), quantum mechanical interference happens during propagation.
This interference is related to the difference in the momenta or energy of the states and is proportional to $m_i^2-m_j^2\equiv\Delta m^2_{ij}$.
Given the known values of the $\Delta m^2$'s and the conveniently available neutrino energies, it is somewhat a coincidence that it is feasible to see neutrino oscillations on human scales.
Atmospheric neutrinos experience their oscillations across a range of energies easily accessible for a range of distances smaller than the diameter of the Earth.
Reactor neutrinos experience their oscillations over baselines of $\sim1$ km or $\sim30$ km which make for relatively easy placement of detectors.

The probability that a neutrino that begins as flavor $\alpha$ is detected in a charged-current (CC) interaction as flavor $\beta$ is
\begin{equation}
P_{\alpha\beta}=\delta_{\alpha\beta}-4\sum_{i>j}\Re(U_{\alpha i}^*U_{\beta i}U_{\alpha j}U_{\beta j}^*)\sin^2\left(\frac{\Delta m^2_{ij}L}{4E}\right)\pm8J\prod_{i>j}\sin\left(\frac{\Delta m^2_{ij}L}{4E}\right)\,,
\end{equation}
where $L$ is the distance traveled, $E$ is the neutrino energy, $J\equiv\Im(U_{\alpha i}^*U_{\beta i}U_{\alpha j}U_{\beta j}^*)$ is the Jarlskog coefficient governing CP violation \cite{Jarlskog:1985ht}, the upper (lower) sign is for (anti-)neutrinos for $P_{\mu e}$ and is cyclic in flavor, and $U$ is the PMNS matrix \cite{Pontecorvo:1957cp,Maki:1962mu} given by
\begin{equation}
|\nu_i\rangle=U_{\alpha i}|\nu_\alpha\rangle\,.
\end{equation}
There are many valid parameterizations of the mixing matrix \cite{Denton:2020igp} and the one that is phenomenologically useful and standard in the field is
\begin{align}
U&=
\begin{pmatrix}
1\\&c_{23}&s_{23}\\&-s_{23}&c_{23}
\end{pmatrix}
\begin{pmatrix}
c_{13}&&s_{13}e^{-i\delta}\\&1\\-s_{13}e^{i\delta}&&c_{13}
\end{pmatrix}
\begin{pmatrix}
c_{12}&s_{12}\\-s_{12}&c_{12}\\&&1
\end{pmatrix}\\&=
\begin{pmatrix}
c_{12}c_{13}&s_{12}c_{13}&s_{13}e^{-i\delta}\\
-s_{12}c_{23}-c_{12}s_{13}s_{23}e^{i\delta}&c_{12}c_{23}-s_{12}s_{13}s_{23}e^{i\delta}&c_{13}s_{23}\\
s_{12}s_{23}-c_{12}s_{13}c_{23}e^{i\delta}&-c_{12}s_{23}-s_{12}s_{13}c_{23}e^{i\delta}&c_{13}c_{23}
\end{pmatrix}
\end{align}
where $c_{ij}$, $s_{ij}$ are $\cos\theta_{ij}$, $\sin\theta_{ij}$ respectively.
If neutrinos have a Majorana mass term then there are two additional complex phases, but they are not measurable in neutrino oscillation experiments as the effect is suppressed by $\mathcal O((m/E)^2)\lesssim10^{-14}$ or smaller for all neutrino oscillation experiments.

\subsection{Matter Effect}
The above discussion of neutrino oscillations is valid only in the vacuum or sufficiently low densities such as the Earth's atmosphere.
In other environments, the presence of electrons modifies the propagation of the states and, notably, does so in a basis that is not the usual propagation basis which is the mass basis \cite{Wolfenstein:1977ue}.
This modifies the oscillation probabilities in a significant way that is relevant in a number of environments.
In fact, for neutrinos produced in the Sun, it is the dominant effect.
It is because of this fact that the solar mass ordering was determined: that is, that $\Delta m^2_{21}$ is positive\footnote{It is useful to define the numbering of the mass states by $|U_{e1}|^2>|U_{e2}|^2>|U_{e3}|^2$ since those elements of the mixing matrix are all fairly well measured. If one adopts a different definition then one could instead reframe the solar mass ordering measurement as a measurement that $\theta_{12}<45^\circ$ in the usual parameterization of the mixing matrix.}.

The same matter effect can be used to determine the atmospheric mass ordering which is the strategy DUNE will leverage.
In the normal mass ordering DUNE will see a relatively large number of $\nu_e$'s and a relatively small number of $\bar\nu_e$'s and vice-versa in the inverted ordering.
While writing the neutrino oscillation probabilities for the different channels in vacuum as a function of the six underlying oscillation parameters is relatively straightforward, doing so in the presence of matter is significantly more complicated, even if the matter density is essentially constant \cite{cardano:1545aa,Barger:1980tf,Zaglauer:1988gz,Kimura:2002wd,Denton:2019ovn}.
As such numerous approximations have been made to better understand the role of the matter effect on long-baseline neutrino oscillations \cite{Barenboim:2019pfp}.

In the Sun the matter effect is fairly large relative to the mass splittings leading to an approximate alignment between the $\nu_e$ state and the second eigenstate of the propagation basis for the higher energy $^8$B neutrinos \cite{Mikheyev:1985zog}.
Then, since the density profile in the Sun changes sufficiently slowly towards the surface \cite{Parke:1986jy}, the neutrinos are emitted in a dominantly $\nu_2$ state with subleading $\nu_1$ and $\nu_3$ contributions.
En route to the Earth, the neutrinos decohere \cite{deHolanda:2003tx} and the probability for detection is $P_{e\beta}=\sum_i|\hat U_{ei}|^2|U_{\beta i}|^2$ where $\hat U$ is the matrix that diagonalizes the Hamiltonian at densities of the production region.
In addition, solar neutrinos detected at night pass through the Earth and the decohered states begin oscillations again leading to the day-night effect for which SK has reported preliminary evidence for \cite{Super-Kamiokande:2016yck} and DUNE and HK may be able to measure well \cite{Capozzi:2018dat,DUNE:2020ypp,Abe:2018uyc}.

The matter effect has only been significantly detected in the Sun by a comparison of solar neutrino data and reactor neutrino data.
Expected improvements on the matter effect in the Earth will come from nighttime solar neutrinos at DUNE and HK, long-baseline accelerator neutrinos at DUNE \cite{Kelly:2018kmb}, and low-energy atmospheric neutrinos at DUNE \cite{Kelly:2021jfs,Denton:2021rgt}.

\subsection{Role of Each Oscillation Parameter}
\label{ssec:role of each}
\begin{figure}
\centering
\includegraphics[width=\textwidth]{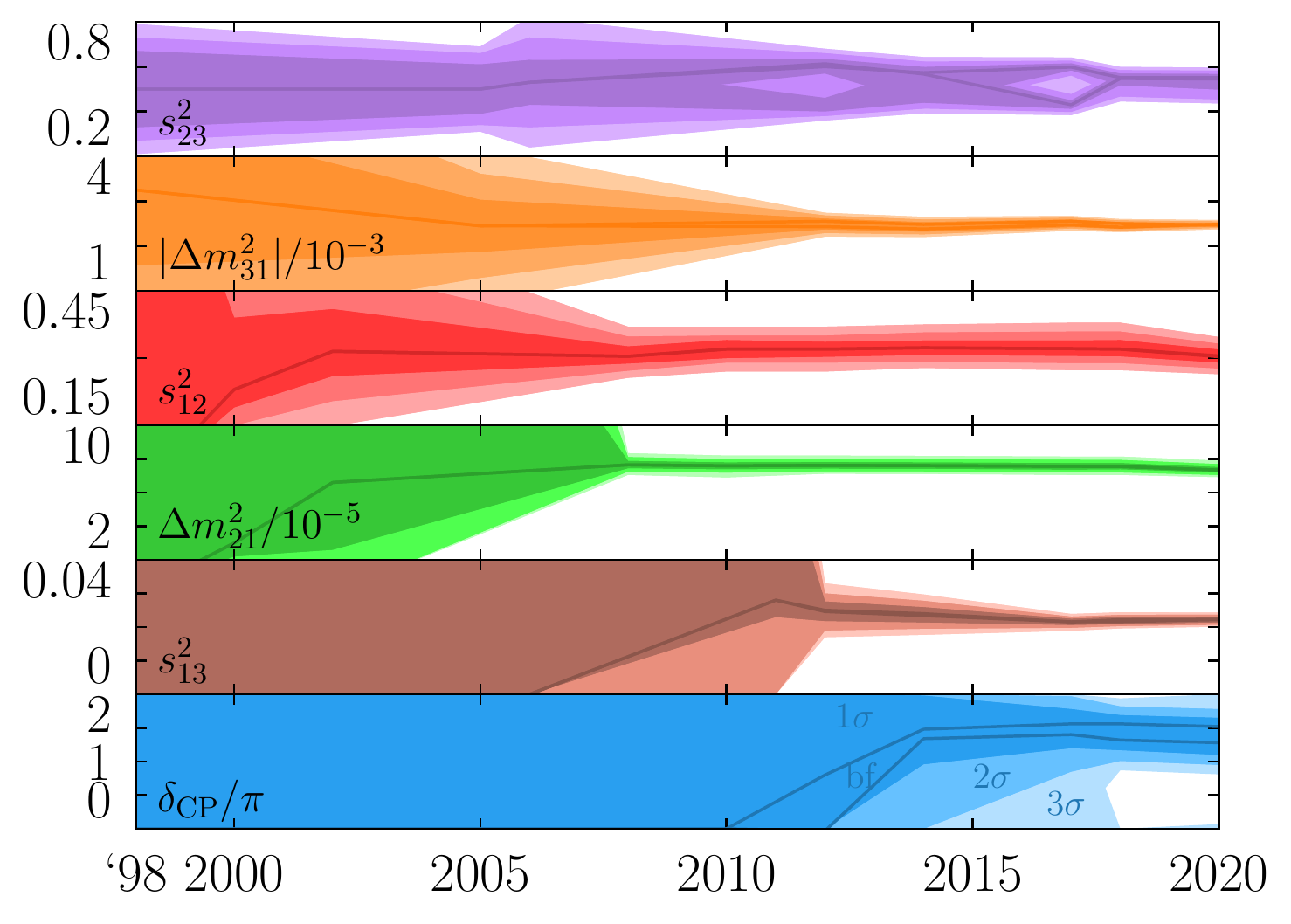}
\caption{The six oscillation parameters listed in the order they were first measured, and the evolution of our understanding of them.
Data comes from \cite{Super-Kamiokande:1998kpq,Gonzalez-Garcia:2000opv,Maltoni:2002ni,Super-Kamiokande:2005mbp,Super-Kamiokande:2006jvq,Schwetz:2008er,Gonzalez-Garcia:2010zke,T2K:2011ypd,Forero:2012faj,Forero:2014bxa,deSalas:2017kay}.}
\label{fig:history of parameters}
\end{figure}

Due to the rich phenomenology of neutrino oscillation physics, the human and planet sized oscillation lengths, and wide range in available neutrino energies, the interplay of the six oscillation parameters is quite complicated.
The history of their measurements is shown in Fig.~\ref{fig:history of parameters}.
The six oscillation parameters affect oscillations in particular ways discussed here.

\paragraph{$\boldsymbol{\Delta m^2_{31}}$}
Neutrino oscillation experiments are only sensitive to the difference of mass squareds since neutrino oscillations occur from the relative accumulated phase.
Given three neutrinos, there are thus two degrees-of-freedom often quantified as $\Delta m^2_{31}$ and $\Delta m^2_{21}$; $\Delta m^2_{32}$ then follows in a straightforward fashion as $\Delta m^2_{32}=\Delta m^2_{31}-\Delta m^2_{21}$.
$\Delta m^2_{31}$ is known as the atmospheric mass squared difference and is known to be $\sim\pm2.5\times10^{-3}$ eV$^2$.
An oscillation maximum or minimum happens when $\Delta m^2L/(4E)\simeq n\pi/2$ where $n$ is some integer; there are additional corrections due to the other $\Delta m^2$'s as well as the matter effect.
These $\Delta m^2$'s are sometimes referred to as frequencies since they dictate when the probability for a neutrino to interact with a certain charged lepton is a maximum or a minimum.
Most experiments are probing the first oscillation maximum or minimum.
This value for $\Delta m^2_{31}$ corresponds to oscillations at a baseline of 500 m for 1 MeV neutrinos (e.g.~reactor neutrinos) or 500 km for 1 GeV neutrinos (e.g.~accelerator or atmospheric neutrinos).
There is an impressive level of agreement on $|\Delta m^2_{31}|$ among a total of seven experiments, reactor, accelerator, and atmospheric, spanning about four orders of magnitude in energy and baseline \cite{DayaBay:2018yms,RENO:2018dro,MINOS:2013utc,T2K:2021xwb,NOvA:2021nfi,Super-Kamiokande:2019gzr,IceCube:2017lak,IceCube:2019dqi}.
The sign of $\Delta m^2_{31}$ is undetermined which is one of the three major known unknowns in neutrino oscillations; some hints currently suggest that it is positive, although the significance is at the $\sim1.5-3\sigma$ level from global analyses \cite{Capozzi:2021fjo,deSalas:2020pgw,Esteban:2020cvm}.
Determining the sign of $\Delta m^2_{31}$ has important implications for neutrinoless double beta decay, cosmological measurements of the sum of the neutrino masses, kinematic end point mass measurements, supernova neutrinos, and detections of the cosmic neutrino background.

\paragraph{$\boldsymbol{\Delta m^2_{21}}$}
The other $\Delta m^2$ is known as the solar $\Delta m^2$ and is $\Delta m^2_{21}\sim+7.5\times10^{-5}$ eV$^2$.
This corresponds to oscillations at 17 km for 1 MeV neutrinos.
The absolute value of $\Delta m^2_{21}$ was determined by KamLAND \cite{KamLAND:2013rgu} and will be further measured by JUNO at high precision \cite{JUNO:2015sjr}.
$\Delta m^2_{21}$ can also be determined with solar neutrinos, although with generally worse resolution.
Nonetheless, the matter effect in the Sun combined with the fact that the disappearance probability decreases with energy from $\sim0.55$ at $E\lesssim1$ MeV for pp neutrinos \cite{Kaether:2010ag,SAGE:2009eeu,BOREXINO:2014pcl} to $\sim0.3$ at $E\gtrsim5$ MeV for $^8$B neutrinos \cite{Cleveland:1998nv,SNO:2011hxd,BOREXINO:2014pcl,Super-Kamiokande:2016yck} indicates that $\Delta m^2_{21}$ is positive.
In addition, the Earth's matter effect for nighttime solar neutrinos provides some information about $\Delta m^2_{21}$.
Extracting the complex phase $\delta$ in long-baseline accelerator neutrinos depends on the precision of this parameter.

\paragraph{$\boldsymbol{\theta_{23}}$}
The atmospheric mixing angle is $\theta_{23}$ and is known to be close to maximal: $\sim45^\circ$.
Thus the $\nu_\mu$ disappearance probability at the first oscillation minimum is close to zero.
This parameter is currently measured dominantly from $\nu_\mu$ disappearance measurements at MINOS \cite{MINOS:2013utc}, T2K \cite{T2K:2021xwb}, NOvA \cite{NOvA:2021nfi}, IceCube \cite{IceCube:2017lak,IceCube:2019dqi}, and SK \cite{Super-Kamiokande:2019gzr}.
These measurements constrain $\sin^2(2\theta_{23})$ which seems to be close to 1, and thus can only tell how close to 45$^\circ$ $\theta_{23}$ is, but not whether it is preferred to be below 45$^\circ$ or above; this is known as the octant problem and is one of the three big known unknowns in neutrino oscillations.
The octant question tells us whether the mass state that is least $\nu_e$ (typically defined as $\nu_3$) is more $\nu_\mu$ (upper octant) or more $\nu_\tau$ (lower octant).
Determining the octant likely requires an appearance measurement and is a goal of current accelerator experiments NOvA and T2K and the upcoming experiments DUNE and HK.
Existing data is inconclusive on the octant question.

\paragraph{$\boldsymbol{\theta_{12}}$}
The solar mixing angle is $\theta_{12}$ and is $\sim34^\circ$.
This indicates that in long-baseline reactor neutrinos at the $\Delta m^2_{21}$ oscillation minimum that the $\nu_e\to\nu_e$ probability drops to $\sim15\%$.
This parameter has been fairly well measured by KamLAND \cite{KamLAND:2013rgu} and will be well measured by JUNO \cite{JUNO:2015sjr}, both with reactor neutrinos.
The solar mixing angle can also be measured with solar neutrinos as well as nighttime solar neutrinos with comparable precision as the reactor experiments.

\paragraph{$\boldsymbol{\theta_{13}}$}
The third and final mixing angle measured, $\theta_{13}$ is known as the reactor mixing angle.
Based on theoretical predictions, it was anticipated by many to be quite small or zero.
$\theta_{13}$ plays a role in both $\Delta m^2_{21}$ and $\Delta m^2_{31}$ measurements.
For example, in solar neutrinos there is a contribution from $\theta_{13}$ since some of the $\nu_e$'s escape the Sun as $\nu_3$ for all solar neutrino energies and this fraction is $s_{13}^2$.
It can also be measured in long-baseline accelerator experiments and it plays a role in CP violation, as do all of the other parameters, see below.
Finally, the best way to probe $\theta_{13}$ is with reactor neutrinos at a baseline of $\sim1$ km where Daya Bay \cite{DayaBay:2018yms}, RENO \cite{RENO:2018dro}, and Double Chooz \cite{DoubleChooz:2019qbj} all agree in their measurements of $\theta_{13}\sim8.5^\circ$ with excellent precision.

\paragraph{$\boldsymbol{\delta}$}
There is one guaranteed complex phase in the mixing matrix known as $\delta$.
Additional phases may be physical if lepton number is violated e.g.~by the presence of a Majorana mass term, although the effect of these phases on neutrino oscillations is completely negligible.
In addition, $\delta$ is only physical if all three neutrino masses are different and all three mixing angles are non-zero; the discovery of non-zero $\theta_{13}$ was the final piece to open the door to a physical $\delta$.
Since this phase is complex it governs the difference between neutrinos and antineutrinos and thus dictates the amount of CP violation in the lepton mixing matrix.
The simplest measurement of $\delta$ results from comparing appearance measurements of neutrinos and antineutrinos which is sensitive to $\sin\delta$, although this channel provides no information about the sign of $\cos\delta$; this is the scenario for T2K and HK to a good approximation.
Meanwhile, $\cos\delta$ can be determined either in the presence of matter (this is DUNE's approach) or by an extremely careful analysis of both $\nu_e$ and $\nu_\mu$ (or $\nu_\tau$) disappearance data which is likely not feasible in upcoming experiments.
Currently T2K shows a mild preference for $\delta\sim270^\circ$ \cite{T2K:2021xwb} while NOvA, which has comparable sensitivity, is compatible with most values of $\delta$ \cite{NOvA:2021nfi}.
Determining CP violation in the lepton sector has important implications for understanding CP violation in the larger context where we see that the weak interaction violates CP but the strong interaction doesn't and the quark mass matrix violates CP but only a little bit.
The Jarlskog invariant, which usefully quantifies the ``amount'' of CP violation \cite{Jarlskog:1985ht}, for the quark matrix is $J_{\rm CKM}=+3\times10^{-4}J_{\max}$ while for leptons it could be much larger, $|J_{\rm PMNS}|<0.34J_{\max}$ where $J_{\max}\equiv\frac1{6\sqrt3}\approx0.096$.
Understanding this mystery of CP violation is a top priority in particle physics.

The three unknowns presented here: the atmospheric mass ordering, the octant of $\theta_{23}$, and the value of $\delta$ represent three of the four known unknowns in particle physics with the absolute neutrino mass scale being the fourth parameter.
We show the preferred values of the parameters in the lepton mixing matrix and the quark mixing matrix in table \ref{tab:pmnsckm}.

\begin{table}
\centering
\caption{The preferred values of the parameters governing the lepton and quark mixing matrices in the usual parameterization.
The quark parameters are quite well measured while $\theta_{23}$ for the leptons has modest uncertainties and $\delta$ for the leptons is largely undetermined.
Note that the masses for all of the quarks are extremely well known, the masses for neutrinos are largely unconstrained.}
\begin{tabular}{c|cccc}
 & $\theta_{23}$ & $\theta_{13}$ & $\theta_{12}$ & $\delta$ \\\hline
Leptons & $\sim45^\circ$ & $8.5^\circ$ & $34^\circ$ & ?\\
Quarks & $2.4^\circ$ & $0.20^\circ$ & $13^\circ$ & $69^\circ$
\end{tabular}
\label{tab:pmnsckm}
\end{table}

\subsection{Flavor Model Predictions and Desired Precision}
Understanding the flavor structure of particle physics: why the fermion masses are distributed as they are and why their mixing with the weak interaction has the structure it has, is one of the biggest open questions in understanding our model of particle physics.
Neutrino oscillation measurements of the remaining mixing unknowns, the atmospheric mass ordering, the octant of $\theta_{23}$ and the value of $\delta$, will provide a wealth of information to enhance our knowledge of the three family structure of particle physics.
Before these measurements are made it is important to consider the model predictions which can, in turn, motivate a desired level of precision for these parameters.

\begin{figure}
\centering
\includegraphics[width=3in]{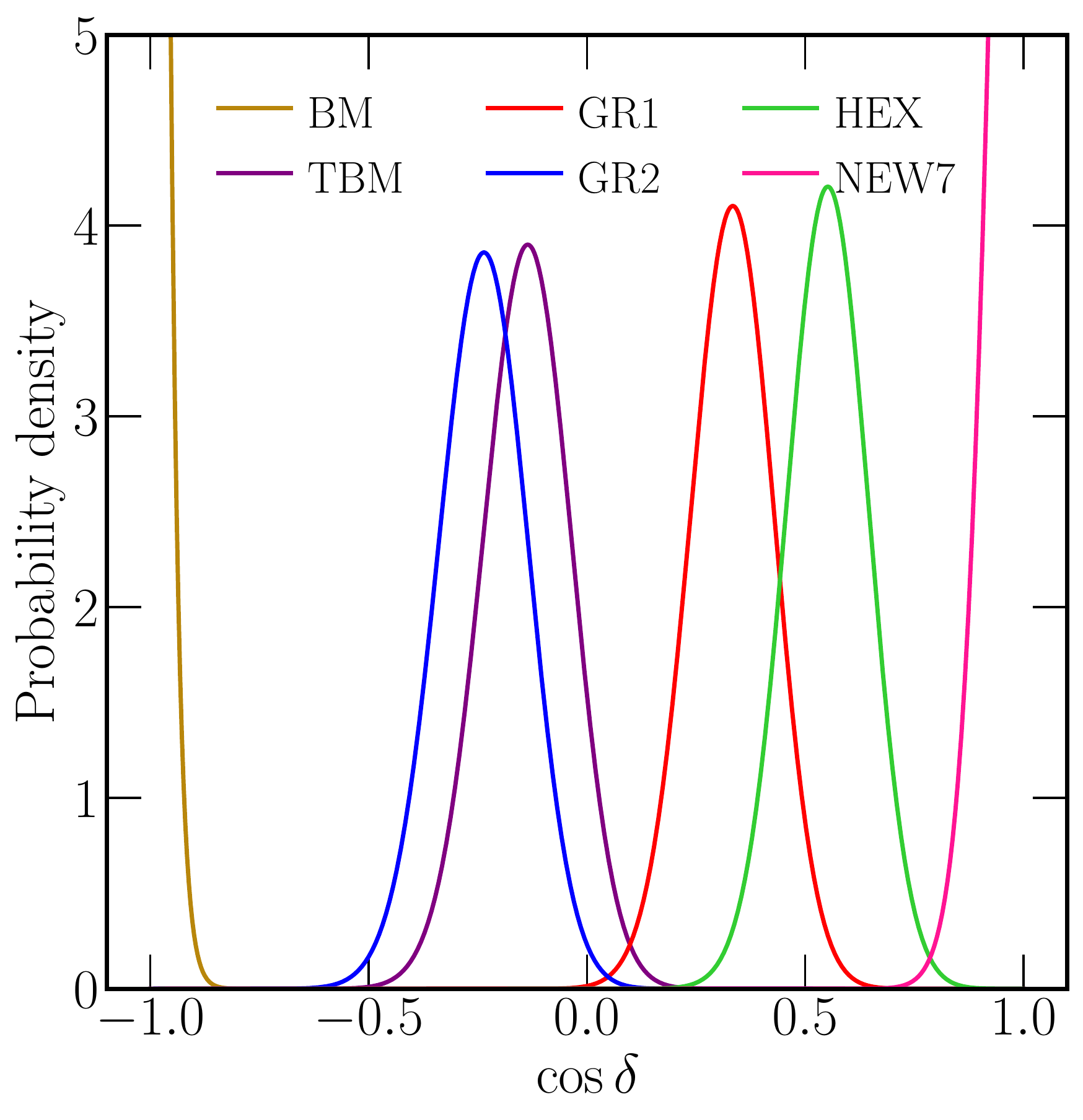}
\caption{Predicted ranges of $\cos\delta$ given existing known oscillation parameters under several example model predictions.
Figure from \cite{Everett:2019idp}.}
\label{fig:cosdelta preds}
\end{figure}

Different flavor structures make different predictions, notably in terms of the complex phase.
Many of these predictions are functions of $\cos\delta$.
For example, various models based on discrete symmetries are of the form $U_{\rm PMNS}=U_e^\dagger U_\nu$ where $U_e$ has a (12) rotation inspired by the Cabibbo angle and $U_\nu$ contains two discrete rotations that are often fixed such that the $\sin^2\theta_{ij}$ form a ratio of small integers.
Given the existing oscillation constraints one can then extract a range of expected values for e.g.~$\cos\delta$ shown in Fig.~\ref{fig:cosdelta preds}.
In this we see that, for example, some models such as Golden Ratio (GR) 1 and 2 predict quite different values of $\cos\delta$ but similar values of $|\cos\delta|$ emphasizing the need to measure the sign of $\cos\delta$.
Being able to differentiate among these models should be a goal for the upcoming measurements of $\delta$.
In addition, these models predict specific correlations of the values of the other mixing parameters, $\theta_{12}$, $\theta_{13}$, and $\theta_{23}$ that require more precise measurements to discriminate among the models.
Thus improving the precision of all mixing parameters, not just $\delta$ and $\theta_{23}$, is very important to add clarity to the flavor puzzle, see e.g.~\cite{Gehrlein:2022nss} for a recent overview.

\section{Three-Flavor Neutrino Oscillation Facilities}
\subsection{JUNO}

The JUNO experiment is the successor to the very successful 
program of experiments which used nuclear reactors as a source for 
studies of $\bar{\nu}_e \rightarrow \bar{\nu}_e$ oscillations.
Nuclear reactors are an abundant source 
of $\bar{\nu}_e$ with energies below 8~MeV.
When measured at distances of 1 -- 2~km,
reactor neutrino spectra have provided precise measurements of 
$\sin^2 2 \theta_{13}$ and 
$\Delta m^2_{31}$ and the near detectors used in these experiments
have made high statistics measurements of reactor neutrino fluxes~\cite{DayaBay:2021dqj,RENO:2020dxd,DoubleChooz:2020vtr}.

\begin{figure}[!htb]
\begin{centering}
\includegraphics[width=6in]{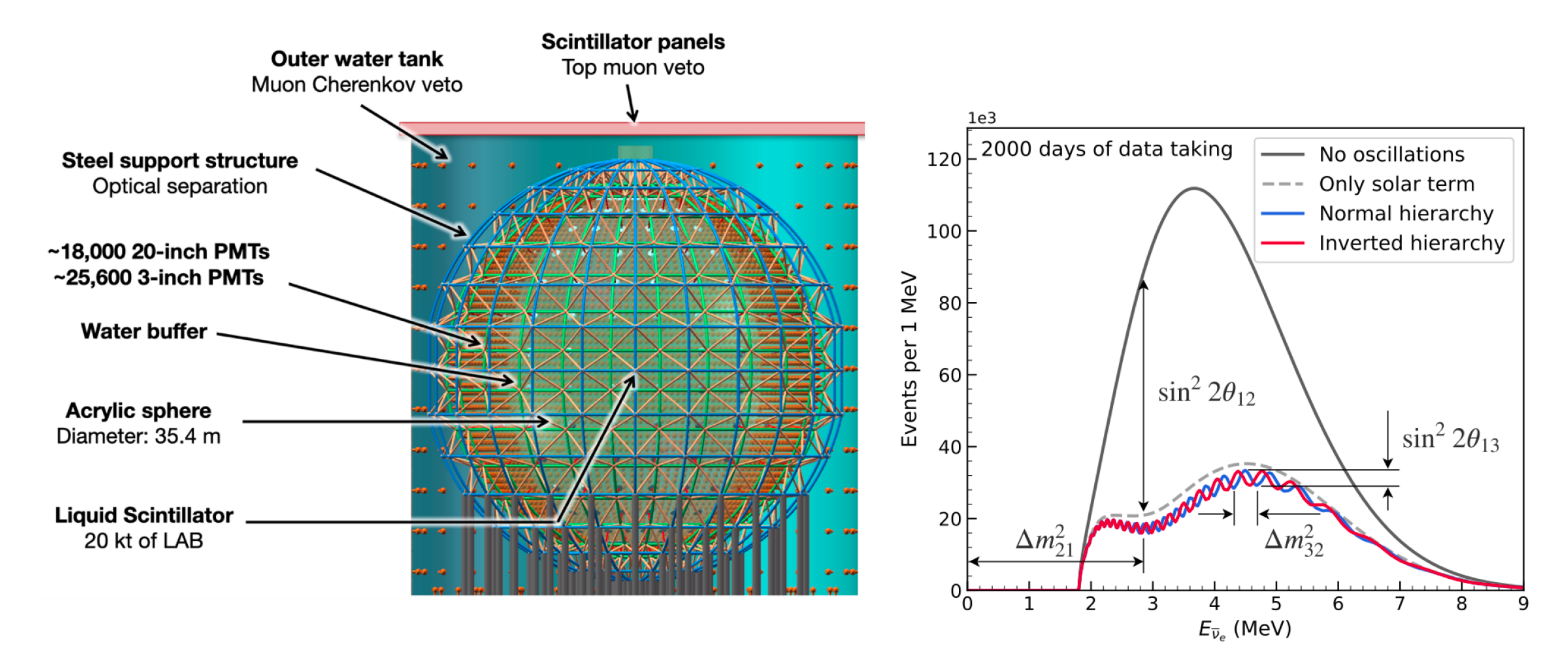}
\caption{
Left: A schematic of the JUNO detector.
Right: Predicted $\bar{\nu}_e$ spectra for the JUNO experiment
showing the dependence on neutrino oscillations parameters.
\label{fig:juno_fig}
}
\end{centering}
\end{figure}
JUNO (Jiangmen Underground Neutrino Observatory)
is constructing a 35~m diameter 
detector containing 20~kt of 
liquid scintillator 
700~m underground in Kaiping, South China at a distance of 53~km 
from the reactor cores of the Yangjiang and Taishan nuclear power plants.
This $\simeq 25 \times$ increase in baseline over the Daya Bay experiment
will give JUNO unique sensitivity to the solar oscillation parameters
$\Delta m^2_{21}$, and $\sin^2 2 \theta_{12}$ among the next generation 
experiments. JUNO projects that measurements of 
$\sin^2 2 \theta_{12}$,
$\Delta m^2_{21}$, and 
$\Delta m^2_{32}$
with reach $\simeq 1\%$ precision in six years of data taking.

JUNO will instrument its 
volume with 43,000 photomultipliers to reach 75\% photo-cathode coverage
yielding an expected energy resolution of $3\%$.
This will allow JUNO to provide information on the atmospheric mass ordering at up to the $3~\sigma$ level by observing the phase of the two ``fast'' oscillations from $\Delta m^2_{31}$ and $\Delta m^2_{32}$ in the recoil electron energy spectrum, see Fig.~\ref{fig:juno_fig}.

Construction of JUNO began in 2015 and construction of the detector
started in 2022. JUNO expects to complete the detector in 2023.

\subsection{Fermilab/SURF Program}

\subsubsection{NOvA}
The NOvA experiment uses the Neutrinos from the Main Injector (NuMI) beam,
originally constructed for the MINOS experiment~\cite{MINOS:2020llm}
to measure
$\nu_\mu \rightarrow \nu_\mu$,
$\nu_\mu \rightarrow \nu_e$,
$\bar{\nu}_\mu \rightarrow \bar{\nu}_\mu$,
$\bar{\nu}_\mu \rightarrow \bar{\nu}_e$,
in a relatively narrow-band beam peaked at 2~GeV.
The neutrino beam passes through a near detector located 1~km
from the proton target and then through a second 14,000~ton
detector located 810~km away in Northern Minnesota.
The detectors employ liquid scintillator segmented into
4~cm $\times$ 6~cm $\times$ 15~m cells
by a reflective PVC structure. 
NOvA's segmentation provides for 3.5\% muon energy resolution 
coupled with a 25\% hadronic energy resolution giving 
an overall neutrino energy resolution of 8--10\% for 
$\nu_\mu$-charged current events crucial for
measurements of $\nu_\mu \rightarrow \nu_\mu$
oscillations which determine the parameters
$| \Delta m^2_{32}|$ and $\sin^2 2 \theta_{23}$.
The use of low $Z$ materials
allows NOvA to efficiently separate electrons from photons
which is crucial for measurements of electron neutrino 
appearance which are used
for the measurement of the $\theta_{23}$ octant, atmospheric mass 
ordering, and $\delta_{\rm CP}$.
Due to the long baseline,
$L = 810$~km, the choice of mass ordering modifies the 
$\nu_\mu \rightarrow \nu_e$ and
$\bar{\nu}_\mu \rightarrow \bar{\nu}_e$ 
oscillation probabilities
by $\pm 20$\%. Of planned long-baseline accelerator 
experiments, NOvA's baseline
will remain the longest until the DUNE 
experiment begins operations.

\begin{figure}[!htb]
\begin{centering}
\includegraphics[width=5in]{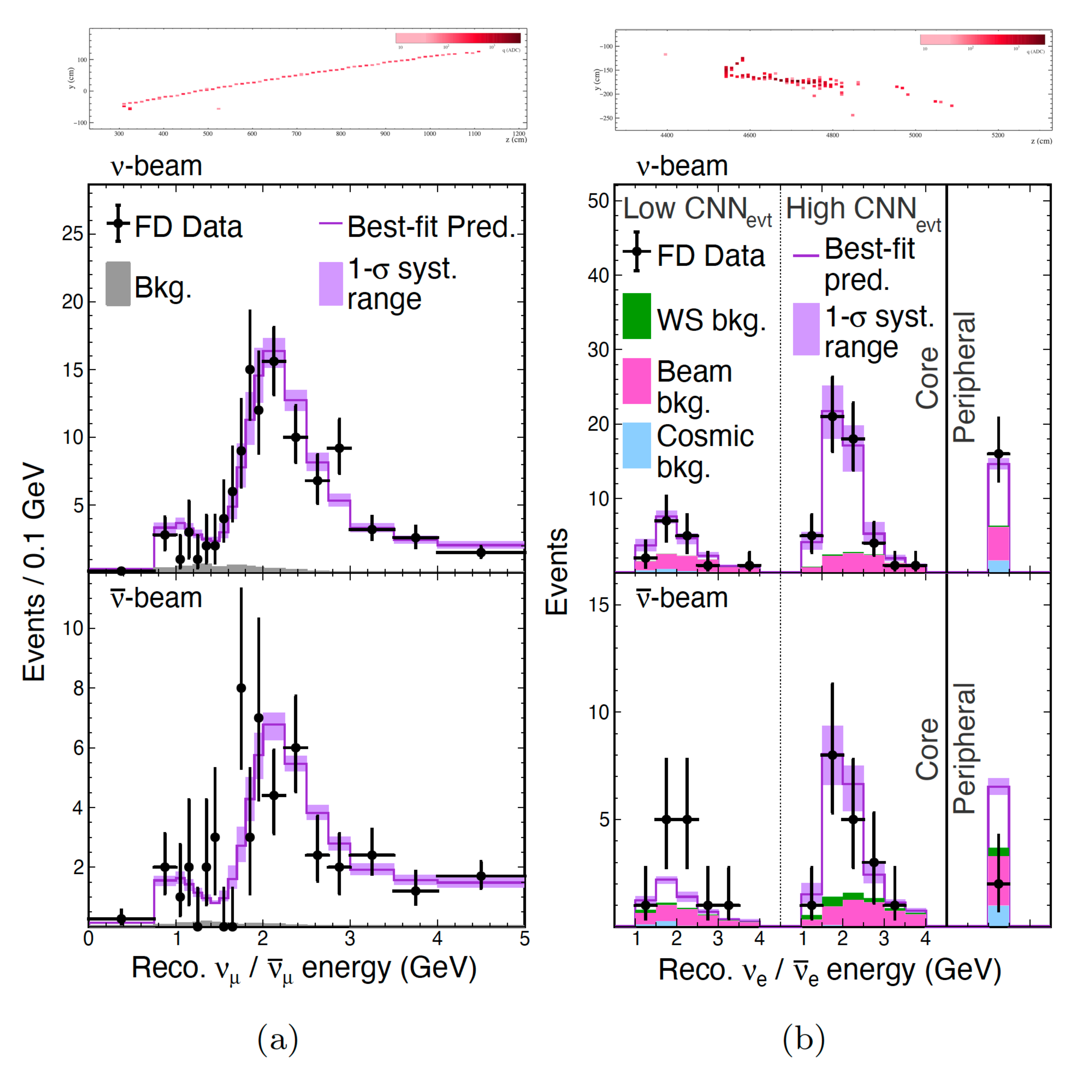}
\caption{
The $\nu_\mu$ and $\bar{\nu}_\mu$ neutrino
spectra (left panels) and
$\nu_\mu \rightarrow \nu_e$ and 
$\bar{\nu}_\mu \rightarrow \bar{\nu}_e$ (right)
spectra recorded by the NOvA far detector.
~\cite{NOvA:2021nfi}.
\label{fig:nova_spectra}
}
\end{centering}
\end{figure}

\begin{figure}[!htb]
\begin{centering}
\includegraphics[width=7in]{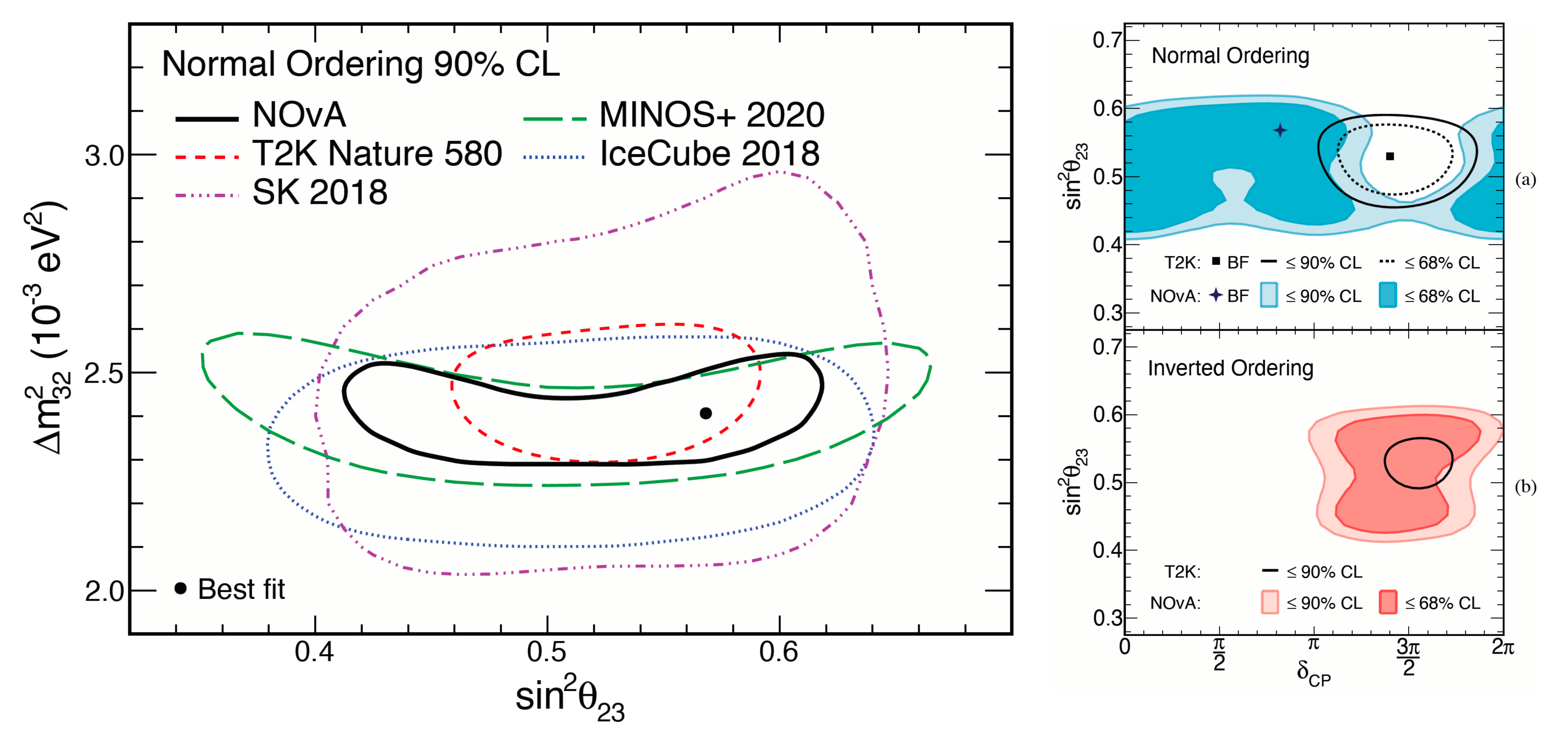}
\caption{
Left: The best fit parameters for
$\Delta m^2_{32}$ and $\sin^2 \theta_{23}$
from the NOvA experiment and others.
Right: The best-fit values of $\sin^2 \theta_{23}$ and $\delta_{\rm CP}$ to NOvA data for the normal neutrino
mass ordering (a), and the inverted neutrino mass
ordering (b). Also shown are recent best-fits to 
T2K data~\cite{T2K:2020nqo}.
\label{fig:nova_results}
}
\end{centering}
\end{figure}

NOvA has been taking data since 2014 releasing new data on standard three-flavor
oscillations each year since its first results in 2016. 
During that time the accelerator power to NuMI has 
steadily increased throughout NOvA's run.
Typical running today averages 650~kW to 730~kW depending
on other beam consumers at Fermilab. An hourly peak power of
893~kW has been demonstrated
exceeding the design power of 700~kW.
In total, NOvA has recorded
$13.6 \times 10^{20}$ protons-on-target (POT)
in neutrino beam and 
$12.5 \times 10^{20}$ POT in anti-neutrino beam,
yielding
214 $\nu_\mu \rightarrow \nu_\mu$ events,
59 $\nu_\mu \rightarrow \nu_e$ events,
103 $\bar{\nu}_\mu \rightarrow \bar{\nu}_\mu$, and,
19 $\bar{\nu}_\mu \rightarrow \bar{\nu}_e$ events in the far detector.
The spectra of these events (Fig.~\ref{fig:nova_spectra}) give
$\Delta m^2_{32} = (2.41 \pm 0.07) \times 10^{-3}~{\rm eV}^2$,
$\sin^2 \theta_{23} = 0.57 ^{+0.03}_{-0.04}$,
and
$\delta_{\rm CP} = 0.82^{+0.27}_{-0.87}$.
NOvA has not observed a large asymmetry in the rates of
$\nu_\mu \rightarrow \nu_e$ vs. 
$\bar{\nu}_\mu \rightarrow \bar{\nu}_e$
oscillations and hence does not favor large CP violation and has
excluded parameter combinations in the inverted
mass ordering which would produce a large 
$\nu_\mu \rightarrow \nu_e$ vs.~$\bar{\nu}_\mu \rightarrow \bar{\nu}_e$
asymmetry (Fig.~\ref{fig:nova_results}).
NOvA's observation is different from T2K which has seen
a large asymmetry in these rates and the two experiments are
in slight tension.
NOvA's modest asymmetry also has not produced a clear resolution
of the atmospheric mass ordering; NOvA data favors the normal
ordering at 1~$\sigma$.

NOvA's statistical uncertainties are roughly 2$\times$ larger than its
systematic uncertainties and the experiment will remain
statistics limited until the projected end of its run in 2026 when
a shutdown of NuMI becomes necessary for PIP-II construction.
The NuMI beam intensity is projected to increase
to 900~kW by 2024 allowing NOvA to double its current data set prior to the planned end of its run.
This final data set would have
3~$\sigma$ sensitivity to resolve the neutrino mass ordering
for $\simeq$50\% of $\delta_{\rm CP}$ values and up to
$>4~\sigma$ sensitivity for the most favorable parameters. The
experiment projects 2~$\sigma$ sensitivity to CP violation
for $\simeq$20\% of $\delta_{\rm CP}$ values.

\subsubsection{DUNE}
The DUNE experiment is the next step in the evolution of the Fermilab
program. The DUNE collaboration was formed to realize the 2014 P5 
vision of a best-in-class long-baseline experiment based at Fermilab.

For the experiment, a new neutrino beamline is under construction
directed from Fermilab to the Sanford Underground Research Facility (SURF) in
South Dakota at a distance of 1285~km. This beamline will use protons from an upgraded Main Injector fed from the PIP-II accelerator. 
DUNE will use these protons 
to make an intense and 
$\simeq 90\%$ pure beam of muon neutrinos and antineutrinos
ranging in energy from 0.5~GeV up to 10~GeV
as part of the near site of its Long-Baseline Neutrino Facility (LBNF)
which will house both the neutrino beam and near detector
for the experiment.
The combination of a very long baseline and wide-spectrum 
beam are unique features of the DUNE beamline.
Construction of the PIP-II source is underway, as
is the construction of LBNF.
\begin{figure}[htbp]
\centering
\includegraphics[width=0.8\textwidth]{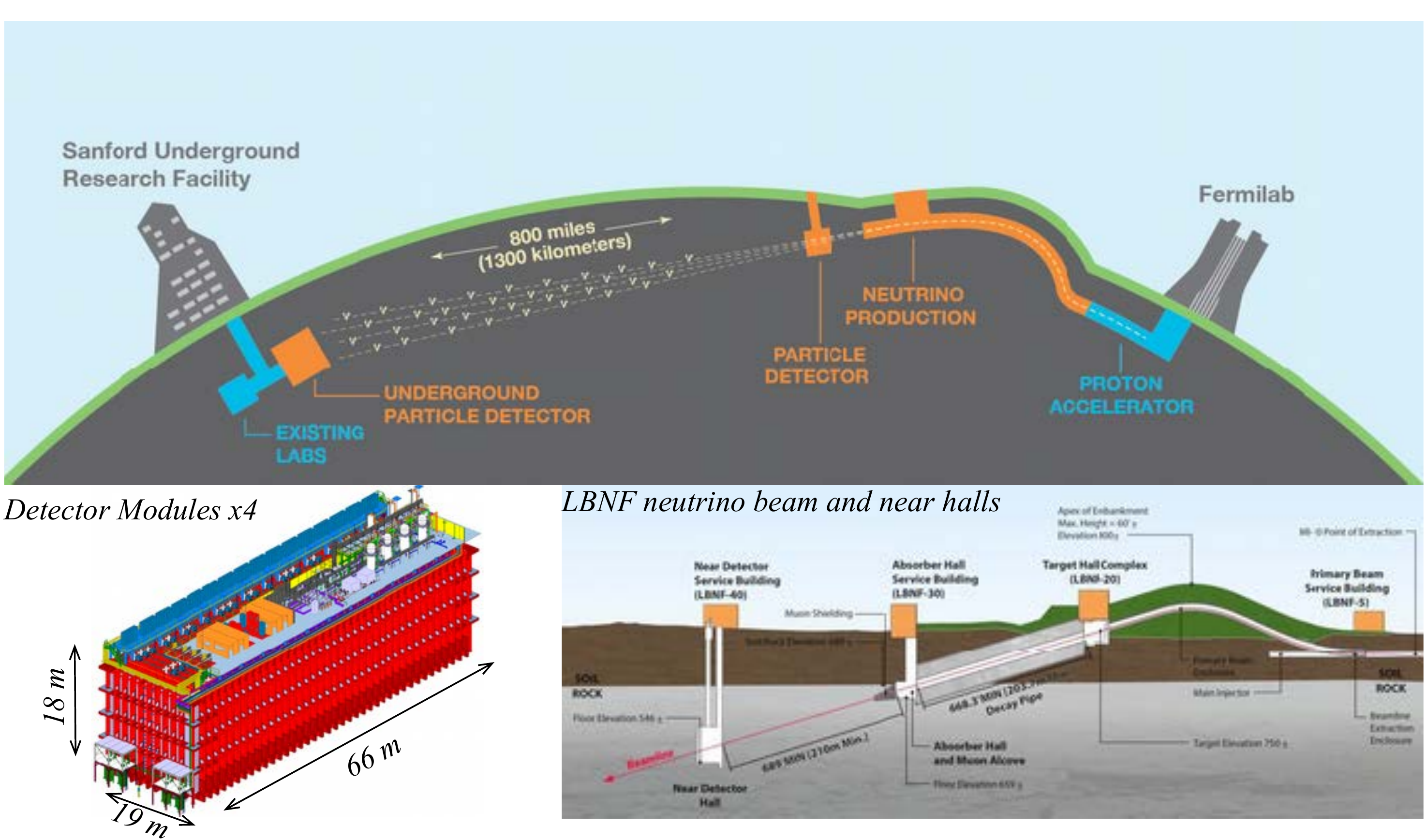}
\caption{
Overview of the DUNE experiment. The Long-Baseline Neutrino Facility will house the beamline
and near detector caverns at Fermilab. At the far site, $L = 1285$~km away,
DUNE will
construct four liquid argon time projection chambers.
\label{fig:dune-overview}
}
\end{figure}

To receive this beam, DUNE intends to construct its far detector
in four 18~m $\times$ 19~m $\times$ 66~m modules to be 
constructed and deployed in stages.
Excavation of the caverns to house these modules
is underway. Design of two of these modules
are in advanced stages, and construction is underway.
These modules will instrument a fiducial mass of 10 or
more kilotons of liquid argon using a time projection chamber
capable of $\simeq 5$~mm ($\simeq X_0/30$) spatial resolution
and waveguides to collect scintillation light produced by 
the liquid argon.
This large size and high resolution will allow for detailed
reconstriction of the interactions of neutrinos with $\mathcal{O}(1$~GeV$)$
of energy and a very efficient tag of the incident neutrino flavor
through separation of muons from pions and electrons from photons.

The DUNE TPCs build on considerable experience with liquid argon
gained through the ICARUS~\cite{Farnese:2019xgw} and 
MicroBooNE~\cite{Fleming:2012gvl} experiments.
The first of these modules will use a so-called single-phase technology. The ionization electrons will be drifted horizontally to three
layers of wires to read their induction and collection signals.
For the second module, the design will exploit the long drift lengths
demonstrated and drift electrons vertically.
This dual-phase, vertical drift technology offers several potential 
simplifications to the construction of the detector and an increase
in its fiducial fraction. Plans for the third and fourth
modules are still taking shape~\cite{MOOD}.

\begin{figure}[htpb]
\centering
\includegraphics[width=0.90\textwidth]{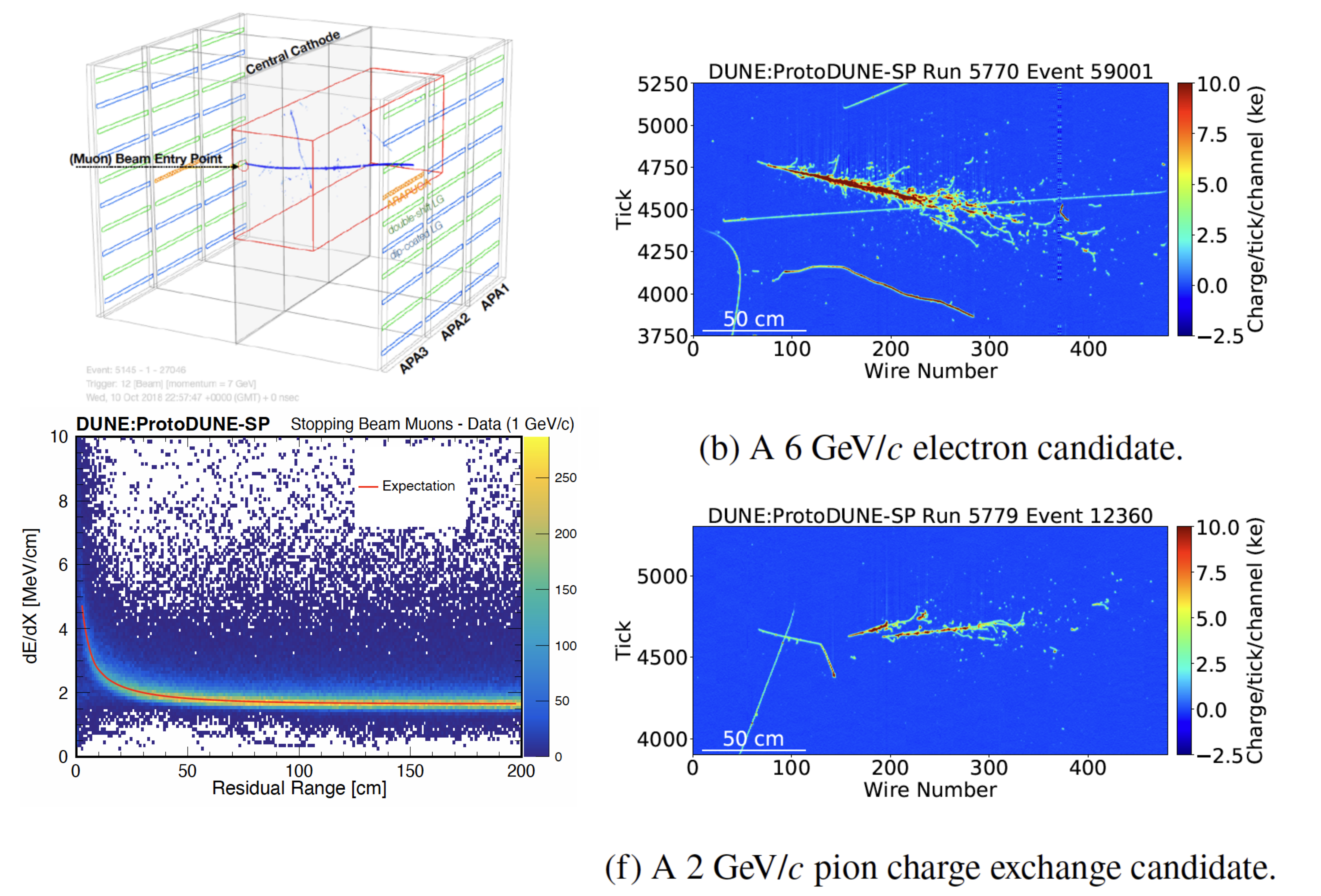}
\caption{
A sample of data from the ProtoDUNE run of the horizontal,
single-phase, prototype detector.
Top left: A 3D reconstruction of a beam muon passing through
the detector. Bottom left: The measured muon $dE/dx$ profile
for muons stopping in the detector.
Top right: A 6~GeV electron shower.
Bottom right: A candidate conversion of a charged to neutral pion.
\label{fig:protodune}
}
\end{figure}

The collaboration has tested the horizontal drift technology 
in a large-scale prototype built and operated in a test
beam at CERN~\cite{DUNE:2020cqd} and a run using the vertical drift 
technology is in progress. Figure~\ref{fig:protodune} shows some 
results published from the horizontal prototype run.

To fully exploit the potential of the detectors and beam, DUNE will
require several advances in the understanding of its neutrino
flux, the properties of neutrino interactions,
and the efficiency of the detector performance. To meet these goals, 
DUNE will construct a suite of near detectors~\cite{DUNE:2021tad}.
The downstream detector, System for on-Axis Neutrino
Detection, (SAND) will use a straw tube tracker placed inside
a magnet and calorimeter reused from the KLOE experiment.
The goal is to measure the momentum spectrum and charge-sign
of muons produced by muon neutrino interactions to ensure
that the neutrino flux is stable and well understood in the on-axis location.
Upstream, a LArTPC system consisting of an array of modular LArTPCs optimized for the high multiplicity environment (ND-LAr) and a  magnetized iron scintllator muon spectrometer (TMS) downstream to analyze muons which exit the back of ND-LAr, will enable the experiment to measure and control uncertainties in the flux, cross section, and detector response in order to predict the expected far detector spectrum \cite{DUNE:2021tad}. The ability to perform this extrapolation precisely is essential for extracting the oscillating parameters, which also depends on obtaining measurements on a near detector which is functionally the same in fundamental aspects as the far detector.

While the SAND detector remains in place, the ND-LAr+TMS system
will be installed on rails allowing them to move
to different off-axis positions in the DUNE beam. As the neutrino
spectrum peaks at a different and lower energies at
each position, the ensemble forms a complete set of neutrino
spectra which allow for the synthesis of oscillated spectra
as the far detector. This data-driven approach,
known as DUNE-PRISM~\cite{DUNE:2021tad} to the near to far 
extrapolation is an essential piece of the DUNE 
program to mitigate systematic
uncertainties. 
To reach the P5 goals
of controlling signal and background uncertainties to 1\% and 5\%
respectively, the collaboration plans to replace this spectrometer
with gaseous argon TPC. This low-density TPC will allow for much more
detailed reconstruction of neutrino event topologies in argon 
enabling systematic uncertainties to be controlled to a precision 
matched to the ultimate exposure of the DUNE far detector~\cite{DUNE:2022yni}.

\begin{figure}[htpb]
\centering
\includegraphics[width=0.5\textwidth]{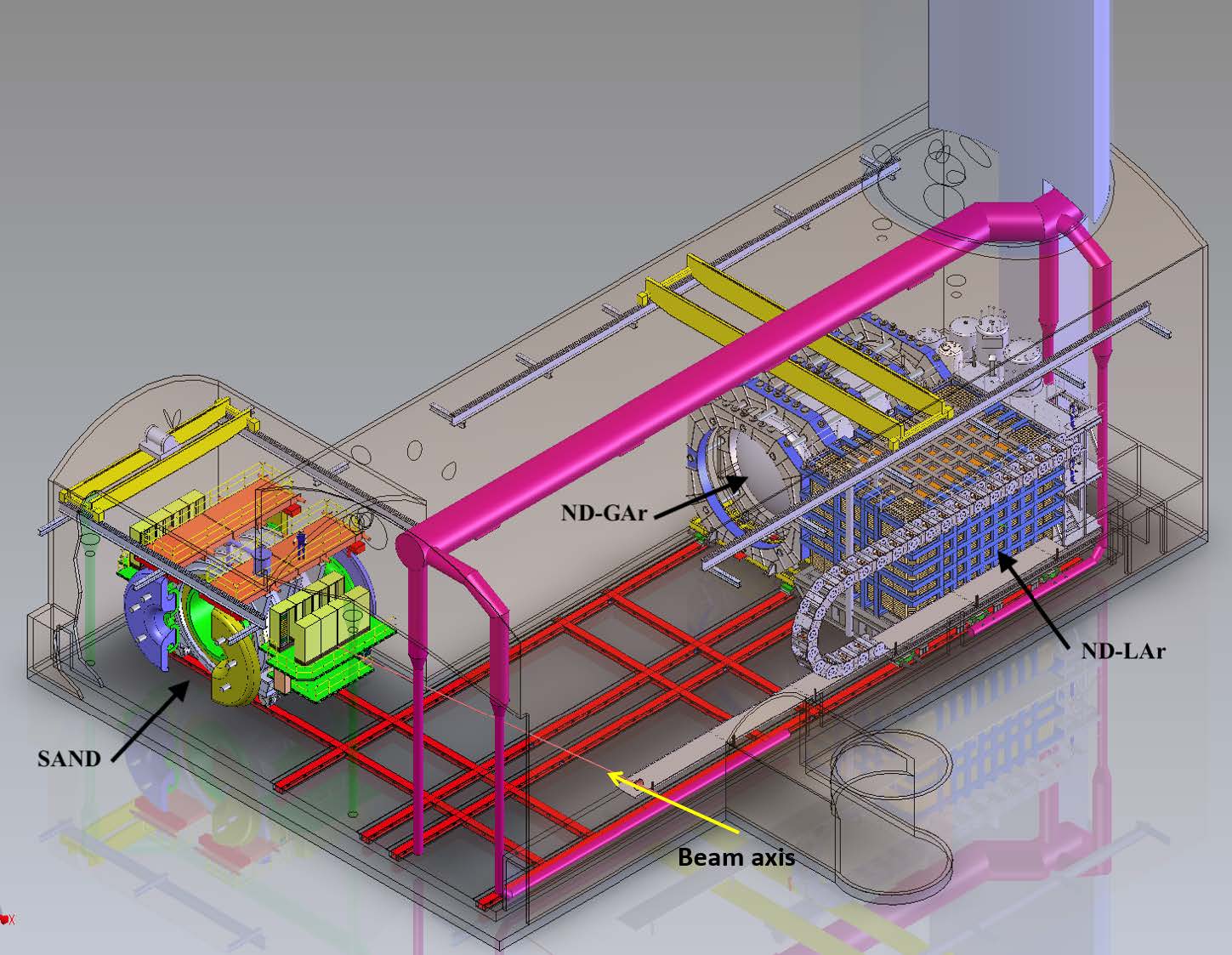}
\caption{
The layout of the DUNE near detector is shown here, with the
front two components in their most off-axis configuration.
Note, the ``day one'' DUNE near detector replaces
the ND-GAr detector with a temporary muon spectrometer.
\cite{DUNE:2021tad}
}
\end{figure}

\begin{figure}[htbp]
\centering
\includegraphics[width=0.90\textwidth]{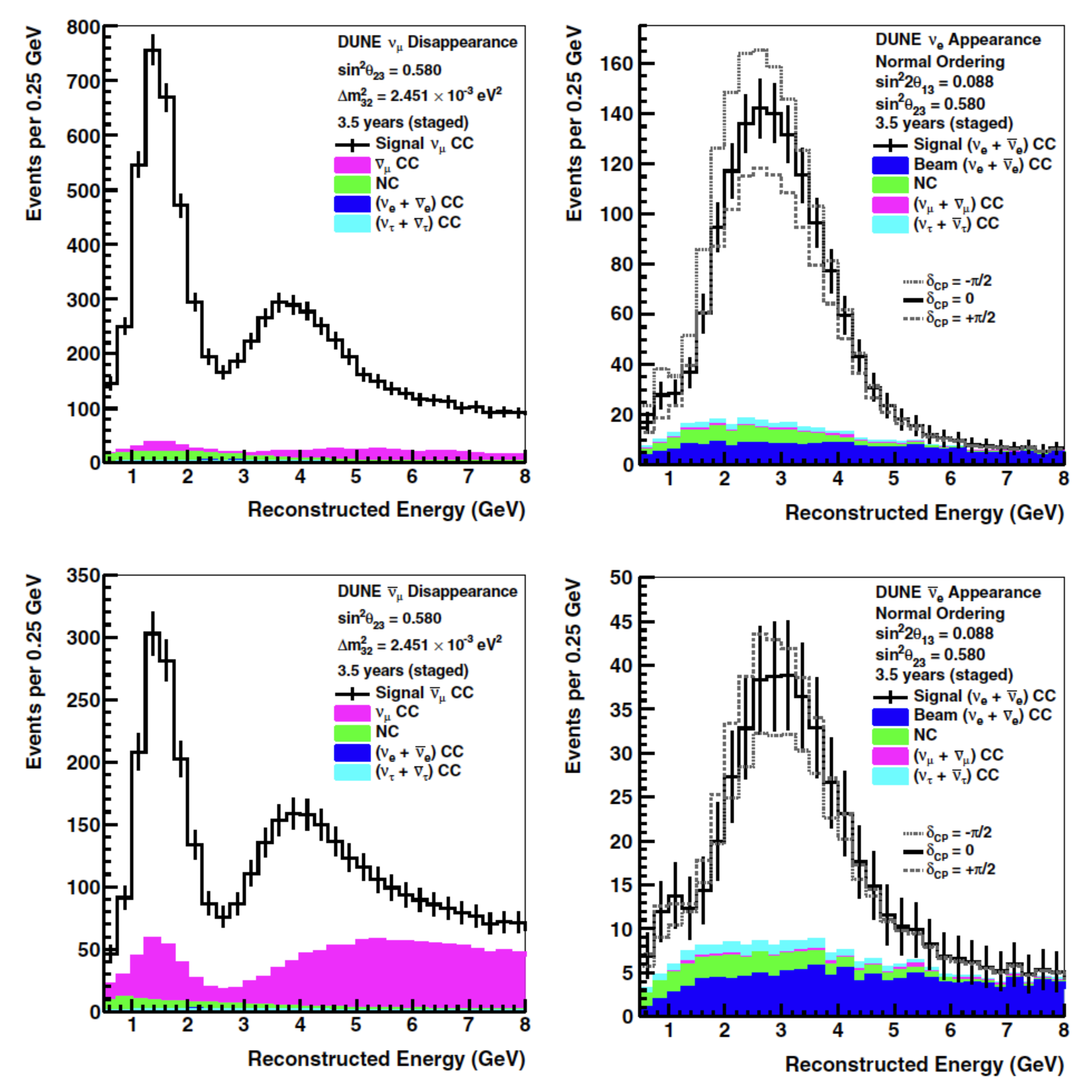}
\caption{
Left: 
The oscillated muon neutrino (top)
and 
muon antineutrino (bottom)
spectra predicted at the DUNE far detector
Right: The electron neutrino (top) 
and anti-electron (bottom)
appearance spectra at the DUNE far detector.
The calculations assume currently best-fit PMNS parameters.
The spectra are normalized to a combined 336 kt-MW-year
exposure divided equally between neutrino and antineutrino beam running.
\cite{DUNE:2020jqi}
\label{fig:dune-spectra}
}
\end{figure}
\begin{figure}[htbp]
\centering
\includegraphics[width=0.90\textwidth]{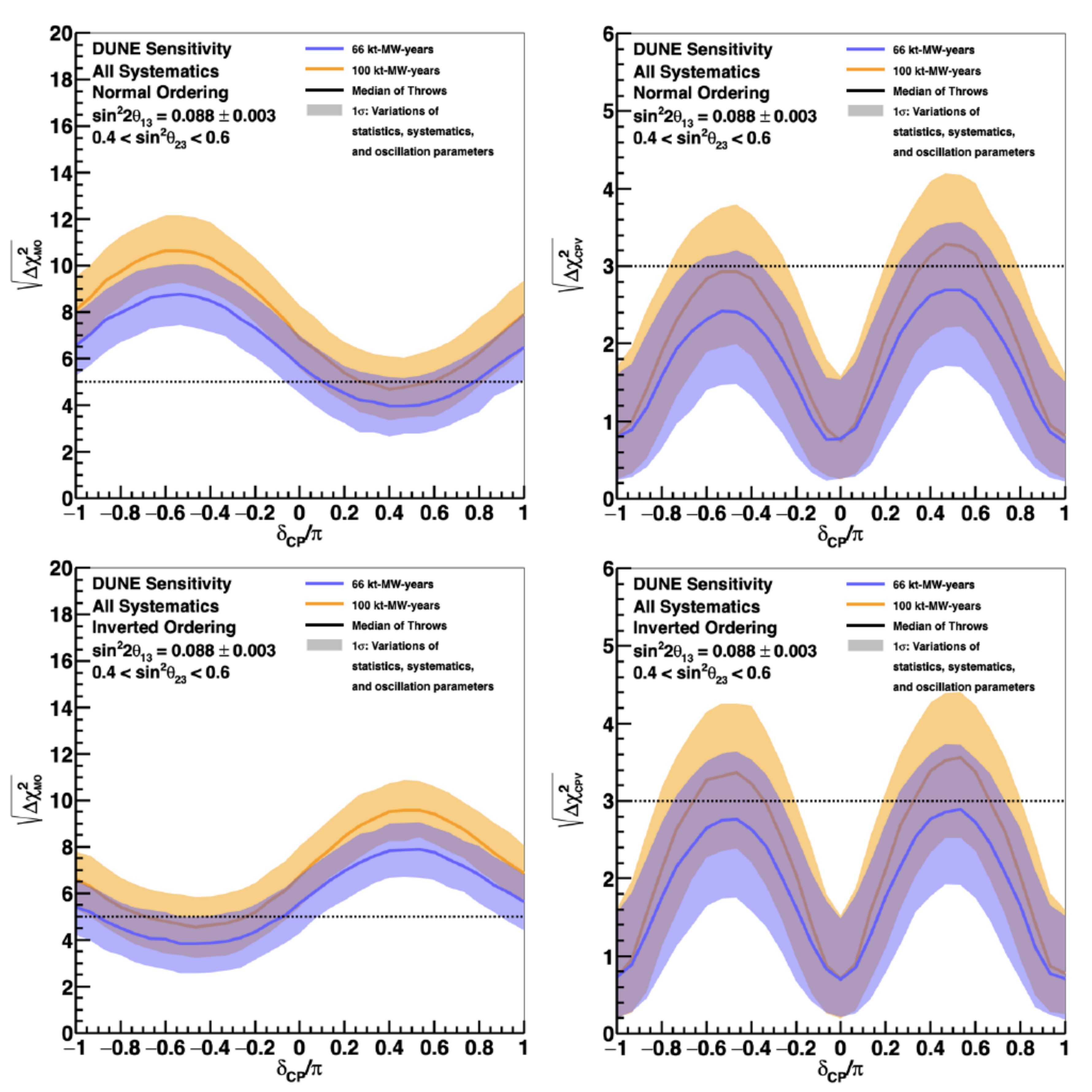}
\caption{
The early DUNE sensitivity to the neutrino mass ordering (left panels)
and CP violation (right panels) calculated assuming the neutrino
mass ordering is normal (top panels) and inverted (bottom panels).
The two curves show the median sensitivity for 
66 and 100 kt-MW-year exposures;
the bands indicate the $1\sigma$ range of these curves based on
variations in oscillation parameter assumptions, statistics, and systematics.
The dotted lines indicate where the experiment reaches 
$5\sigma$ resolution of the neutrino
mass ordering and $3~\sigma$ observation of CP
violation.
\label{fig:dune-early-sensitivity}
}
\end{figure}
\begin{figure}[htbp]
\centering
\includegraphics[width=0.8\textwidth]{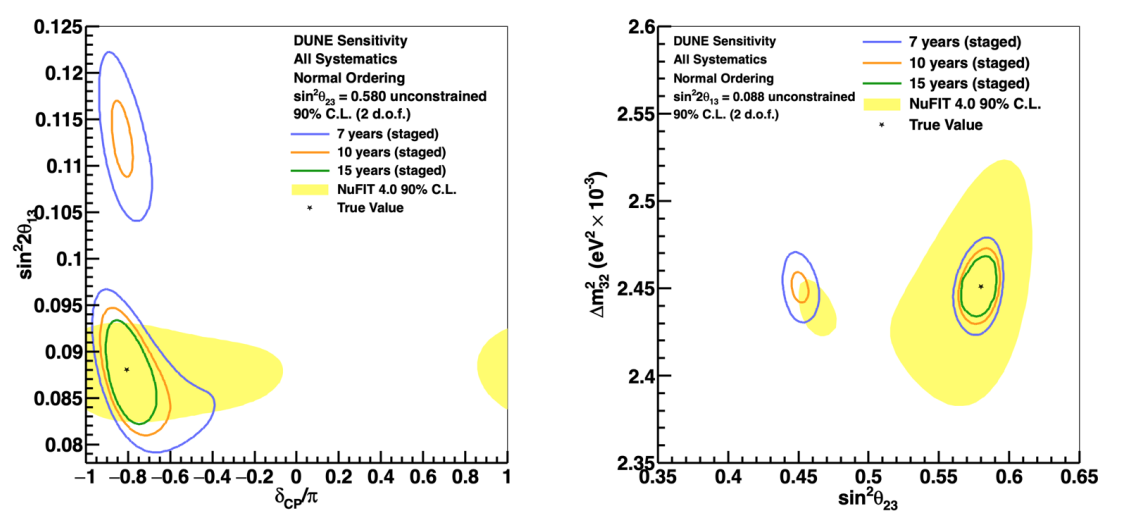}
\caption{
90\% confidence intervals for 
$\sin^2 2\theta_{13}$ -- 
$\delta_{\rm CP}$ (left),
and
$\sin^2 \theta_{23}$ --
$\Delta m^2_{32}$ (right)
after seven, ten, and fifteen years of staged DUNE operations. Yellow
regions indicate recent global fits from the NuFIT 4.0 
group~\cite{Esteban:2018azc,nuft40}.
\label{fig:dune-measurements}
}
\end{figure}

DUNE will make precise measurements of the effects of oscillations
on the muon neutrino and antineutrino spectra at its far detector
and the resulting appearance of electron neutrinos and antineutrinos 
(Fig.~\ref{fig:dune-spectra}).
These measurements will enable
precise determinations of the probabilies of
$\nu_\mu \rightarrow \nu_\mu$,
$\bar{\nu}_\mu \rightarrow \bar{\nu}_\mu$,
$\nu_\mu \rightarrow \nu_e$, and,
$\bar{\nu}_\mu \rightarrow \bar{\nu}_e$ oscillations
and the PMNS parameters
$\Delta m^2_{32}$,
$\theta_{23}$,
$\theta_{13}$,
and $\delta_{\rm CP}$.
DUNE's long-baseline provides excellent sensitivity to the matter
effect on neutrino oscillations and hence strong resolving power
for the neutrino mass ordering.
Figure~\ref{fig:dune-early-sensitivity}
shows the early sensitivity of the DUNE experiment
to the mass-ordering and CP violation.
This early data ($\simeq$100 kt-MW-yr exposure) 
can provide $5~\sigma$ resolution of
the atmospheric mass ordering and $3~\sigma$ sensitivity to
CP violation if it is maximal. 
Figure~\ref{fig:dune-measurements} shows
the ultimate precision of the experiment.
The high precision of these standard three-flavor measurements in a wide-band beam
will provide a platform for searches of additional physics
beyond the standard model that might impact neutrino oscillations.

In addition to the long-baseline neutrino physics program
DUNE will collect 40,000 atmospheric neutrino 
interactions~\cite{DUNE:2015lol} of which 800 will 
be $\nu_\tau$ charged-current interactions.
This sample, measured with the unprecedented detail afforded by
the liquid argon TPC will present several opportunities
for a new level of precision in the study of 
$\nu_\mu \rightarrow \nu_\tau$ oscillations~\cite{dune-nutau-loi}.

Finally, DUNE will also measure solar neutrinos providing considerably improved solar measurement of the solar oscillation parameters $\theta_{12}$ and $\Delta m^2_{21}$ \cite{Capozzi:2018dat,Caratelli:2022llt}, although they will not be competitive with the precision of JUNO.

\subsection{J-PARC/Kamioka Program}
The J-PARC/Kamioka program consists of a neutrino beam produced at the Japan Proton Accelerator Research Complex (J-PARC) in Tokai village, Ibaraki prefecture, Japan.  A suite of near detectors which measure the neutrino beam flux and neutrino interactions is also located in Tokai.  A large-scale water Cerenkov detector operated as part of the Kamioka Observatory, is installed 295~km away from the beam neutrino source, near Kamioka town, Gifu Prefecture, Japan, and acts as a stand-alone neutrino (and proton decay) detector, as well as a far detector for the J-PARC neutrino beam.

\subsubsection{Super-Kamiokande (SK)}
\label{sec:sk}
The Super-Kamiokande (SK) detector is a massive water Cherenkov detector installed 1,000 m underground in the Mozumi Mine in western Japan \cite{Fukuda:2002uc}.  The detector contains 50~kt of ultra-pure water instrumented by \(>\)10,000 50~cm inward facing Photo Multiplier Tubes (PMTs).  An optically separated, highly instrumented outer detector is used to veto particles with initial interaction points outside of the tank volume.  The SK detector has excellent PID performance between showering (electron-like) and non-showering (muon-like) particles at high energies, with a \(\sim\)1\% mis-PID rate at 1~GeV.  SK has been operating with continual improvements since 1996.

SK was initially conceived as a massive, highly instrumented, water Cherenkov detector, designed to resolve the atmospheric neutrino anomaly, resolve the solar neutrino puzzle, and search for nucleon decay. 
%SK is, and likely will be for years to come, the leading operating experiment sensitive to a burst of neutrinos from a supernova in the Milky Way.
SK also serves as the T2K far detector, as described in Sec.~\ref{sec:t2k}.  The discussion in this SK text will focus on the SK sensitivity to three-flavor-oscillations via atmospheric neutrino oscillations as T2K is a separate collaboration.

SK recently entered a new phase of operation following upgrade work in 2018. The detector was refurbished and the water purification system was upgraded to allow operation with dissolved gadolinium sulfate, Gd\(_2\)(SO\(_4\))\(_3\), 0.2\% by mass. First operation of this new phase of running, called SK-Gd, commenced in 2020 with 0.02\% loading.  Additional loading is planned for the near future.  Neutron capture on gadolinium will improve differentiation between neutrinos and antineutrinos, as well as improving energy reconstruction using neutron counting.  These improvements enhance nearly every aspect of physics measurements at SK.

\begin{figure}[!ht]
\centering
\includegraphics[width=0.49\textwidth]{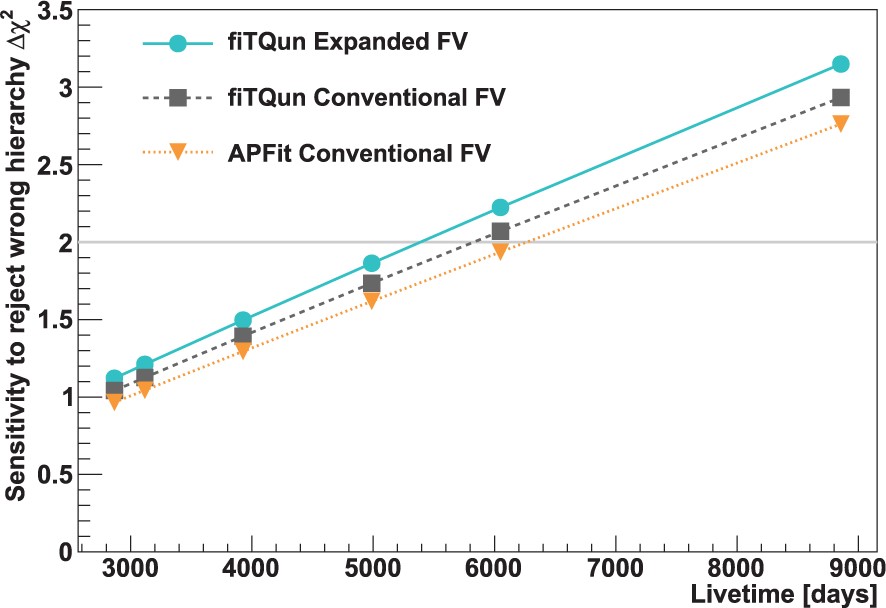}
\caption{SK expected sensitivity to the normal mass ordering as a function of livetime assuming $\sin^2\theta_{23}=0.5$. Gray and blue lines show the sensitivity of the atmospheric neutrino sample reconstructed with the latest fit algorithm using the conventional SK Fiducial Volume (FV) and expanded FV, respectively. Orange lines denote the sensitivity when events are reconstructed with an older fit algorithm using the conventional FV. Reproduced from \cite{Super-Kamiokande:2019gzr}.
}
\label{fig:SKMHsens}
\end{figure}

Recent SK atmospheric oscillation fit results are shown in Fig.~\ref{fig:nova_results}.  These results slightly prefer the normal mass ordering, and the oscillation parameter best fit values are \(\sin^2\theta_{23} = 0.588 ^{+0.031}_{-0.064} (0.575 ^{+0.036}_{-0.073})\),
\(|\Delta m ^2_{32,31}| = 2.50^{+0.13}_{-0.20} (2.50^{+0.08}_{-0.37}) \times10^{-3}\) eV\(^2\), and
\(\delta_{CP} = 4.18^{+1.41}_{-1.61} (4.18^{+1.52}_{-1.66})\) assuming the normal (inverted) ordering \cite{Super-Kamiokande:2017yvm}.  The precision of these measurements continues to improve as SK continues to collect data.  The expected SK sensitivity to reject the wrong mass ordering is given in Fig.~\ref{fig:SKMHsens}.

SK data also delivers data sets that are necessary for understanding water Cherenkov reconstruction algorithms, their efficiencies, and related systematic uncertainties, all of which are essential for precision T2K measurements, discussed in the following section. Atmospheric neutrino interactions are also used to validate neutrino interaction simulations on water nuclei. In addition, cosmic ray data is critical for tuning the detector simulation. The atmospheric neutrino single-\(\pi^0\) sample is the most important in-situ energy scale calibration in the T2K energy range, and is also a valuable sample for studying and improving software algorithms.

\subsubsection{Tokai to Kamioka (T2K)}
\label{sec:t2k}
The Tokai to Kamioka (T2K) experiment is a long-baseline neutrino oscillation experiment \cite{Abe:2011ks} which utilizes a high-intensity, highly pure muon or antimuon neutrino beam produced by 30~GeV protons from the Main Ring (MR) accelerator at J-PARC.  The neutrino flux and neutrino interactions are measured near the source by a suite of near detectors installed 280~m downstream of the neutrino production target.  The far detector is the SK detector, which sits 295~km away from the neutrino source and 2.5\(\degree\) off-axis from the neutrino beam center.  T2K hosts a broad physics program of precision neutrino oscillation measurements, neutrino cross section measurements, as well as searches for exotic physics.

\begin{figure}[H]
\centering
\includegraphics[width=0.45\textwidth]{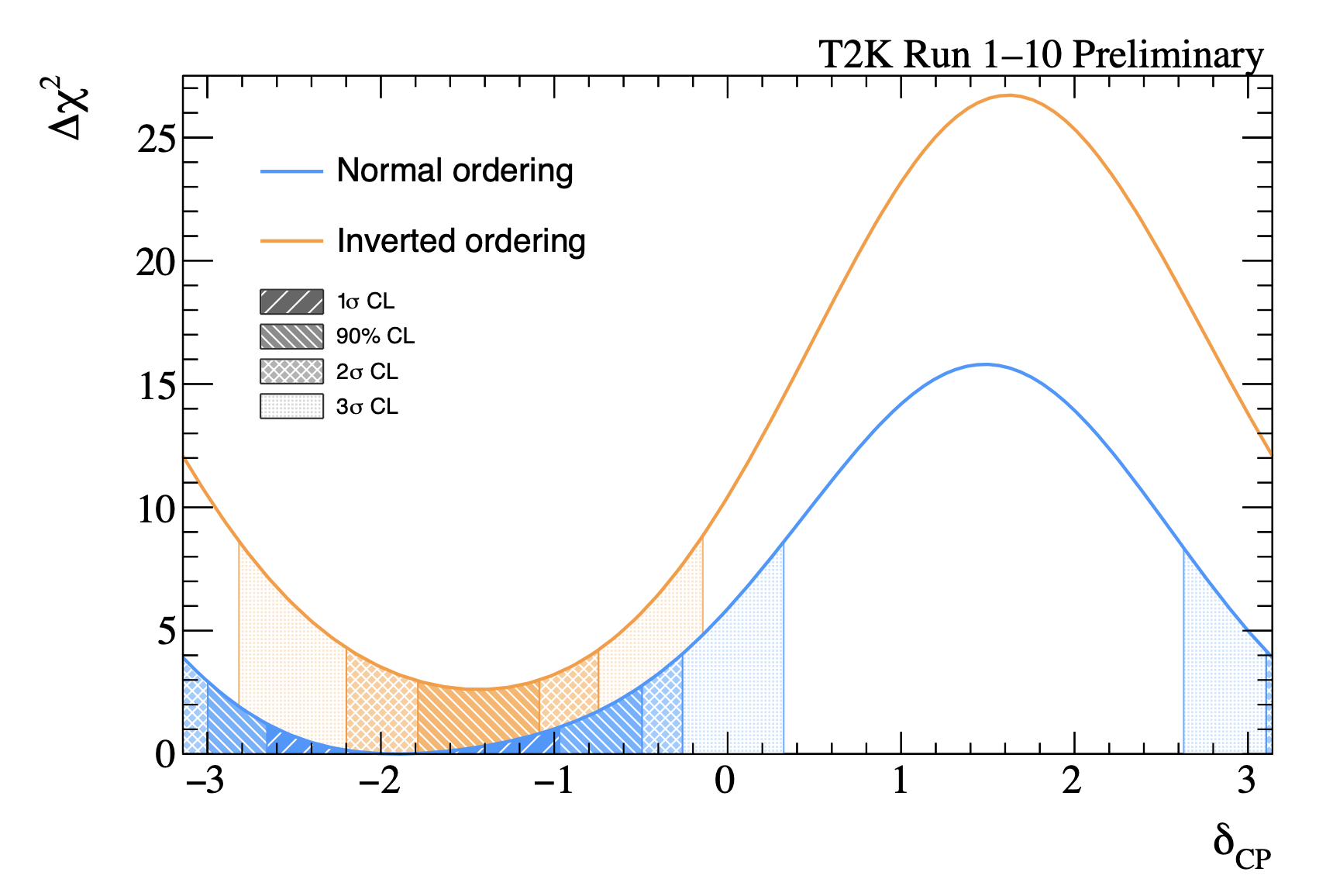}
~~~~~~
\includegraphics[width=0.35\textwidth]{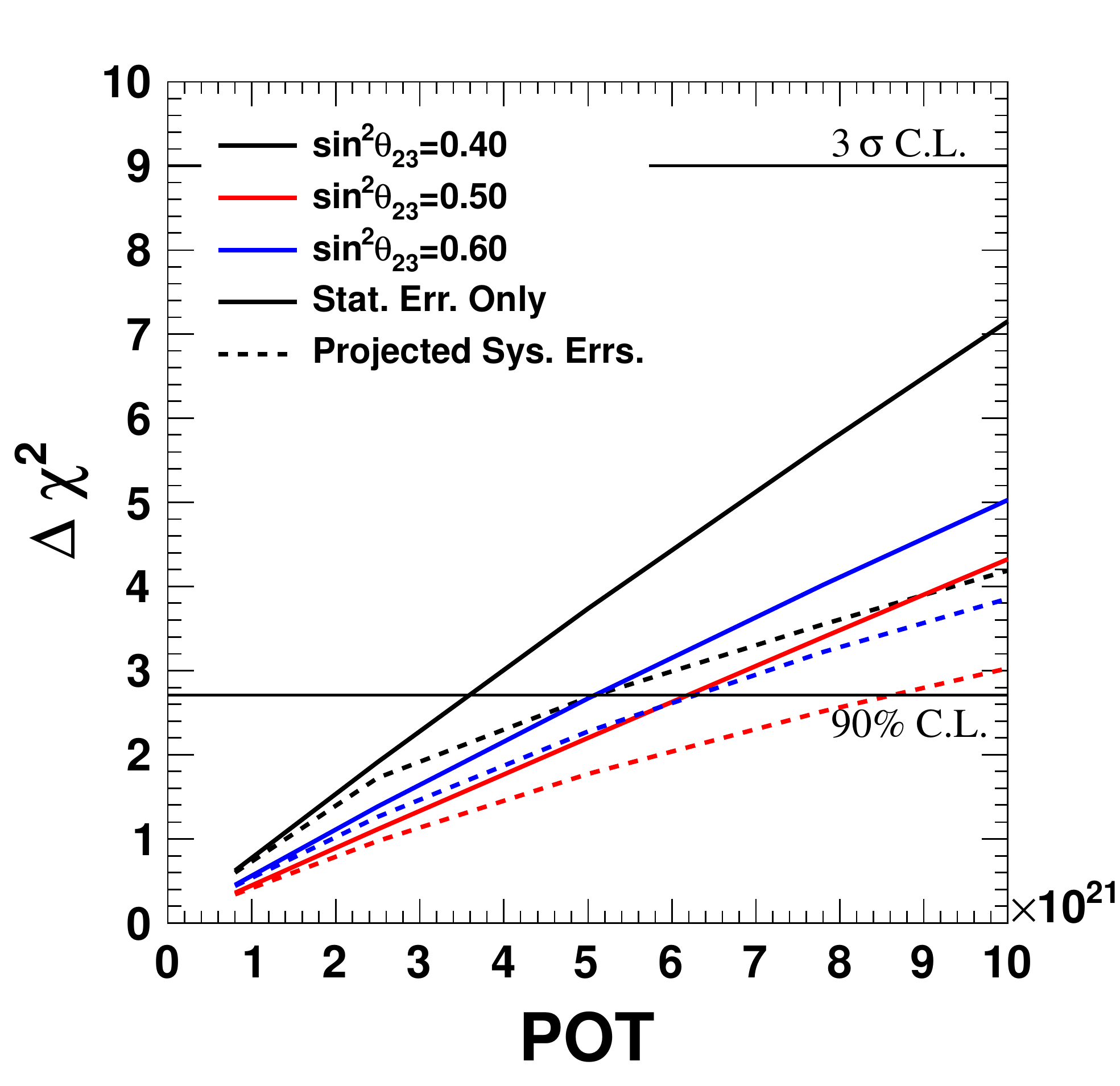}
\caption{T2K 2020 \(\delta_{CP}\) measurement (left) and expected sensitivity to resolving \(\sin\delta_{CP}\neq0\) plotted as a function of T2K POT assuming \(\delta_{CP}=-\pi/2\), normal MO, \(\Delta m^2_{32} = 2.4\times10^{-3}\)~eV\(^2\), and various true values of $\sin^2\theta_{23}$ (right). Reproduced from \cite{Berns:2021iss} and \cite{T2K:2014xyt}.
}
\label{fig:T2KdCPsens}
\end{figure}

The T2K experiment began commissioning in 2009 and physics data taking in 2010. % and is scheduled to run until 2027.  
In 2013, T2K discovered \(\nu_\mu\rightarrow\nu_e\) appearance \cite{T2K:2013ppw} and in 2019, T2K presented the most stringent constraints on CPV in the lepton sector \cite{T2K:2019bcf}, as shown in Fig.\ \ref{fig:T2KdCPsens}.  Recent T2K measurements of the three-flavor oscillation parameters are shown in Fig.~\ref{fig:nova_results} and can be summarized as \(\sin^2\theta_{23}=0.53_{-0.04}^{+0.03}\), \(\Delta m^2_{32}=(2.45\pm0.07)\times10^{-3}\)~eV\(^2\), and \(\delta = -1.89^{+0.70}_{-0.58} (-1.38^{+0.48}_{-0.54})\) with a 3\(\sigma\) confidence interval of [-3.41, -0.03] assuming the normal (inverted) mass ordering \cite{T2K:2021xwb,T2K:2019bcf}.

The J-PARC MR proton beam, which has stably operated at 515~kW, is used to produce the neutrino beam for T2K. A MR magnet power supply upgrade in 2021/2022 will decrease the time between proton beam spills from 2.48~s to 1.32~s, increasing the beam power to \(>\)700~kW. Additional RF upgrades and machine development should allow the beam power to reach \(>\)1~MW by 2025.  A series of three electromagnetic horns focus hadrons outgoing from the neutrino production target, allowing for either a neutrino-enhanced or an antineutrino-enhanced beam.  Upgrades to the horns and the horn power supplies are also taking place in 2021/2022.  These upgrades will allow the T2K experiment to increase the horn current from \(\pm\)250~kA to \(\pm\)320~kA, increasing the neutrino beam flux right-sign component (e.g.\ muon neutrinos in a neutrino-enhanced beam) by \(\sim\)10\% and decreasing the wrong-sign component (e.g.\ muon antineutrinos in a neutrino-enhanced beam) by \(\sim\)5\% at the flux peak.

As of 2020, T2K oscillation analyses incorporate the charged pion yields measured by the NA61/SHINE experiment using a T2K-replica target \cite{NA61SHINE:2016nlf} to constrain the beam flux prediction, and this analysis will be expanded to include more species and higher statistics in the near future \cite{NA61SHINE:2018rhe}. These improvements have reduced the T2K flux uncertainty from \(\sim\)10\% to \(\sim\)5\% near the flux peak prior to a near detector constraint. Future hadron production data at lower energies and high-statistics replica-target data will reduce these uncertainties further, and improvements in proton and tertiary beam monitoring and analysis are also being developed.  These efforts to improve flux uncertainties are useful not only to T2K, but rather are relevant to the full future global program of accelerator-based neutrino beams.

Multiple near detectors sample the neutrino beam at different locations 280~m from the neutrino production target. T2K’s near detector suite includes an on-axis detector colinear with the neutrino beam axis (INGRID), and two magnetized tracking off-axis detectors at different positions transverse to the beam (ND280 at 2.5\(\degree\) and WAGASCI+BabyMIND at 1.5\(\degree\)). The different positions in the beam result in three distinct energy spectra for use in physics analyses. A new, upgraded ND280 detector, with a fully active target (SuperFGD), surrounded by new horizontal TPCs,
is being installed in 2022 \cite{T2K:2019bbb}.
This detector will have significantly increased acceptance to particles emitted at high and backward angles, and to low energy protons. The improved capabilities of this detector will help to reduce systematic uncertainties due to the neutrino interaction model to an unprecedented level.

The SK far detector has also been undergoing upgrades since 2018, as discussed in Sec.~\ref{sec:sk}.

\begin{figure}[h]
\centering
\includegraphics[width=0.48\textwidth]{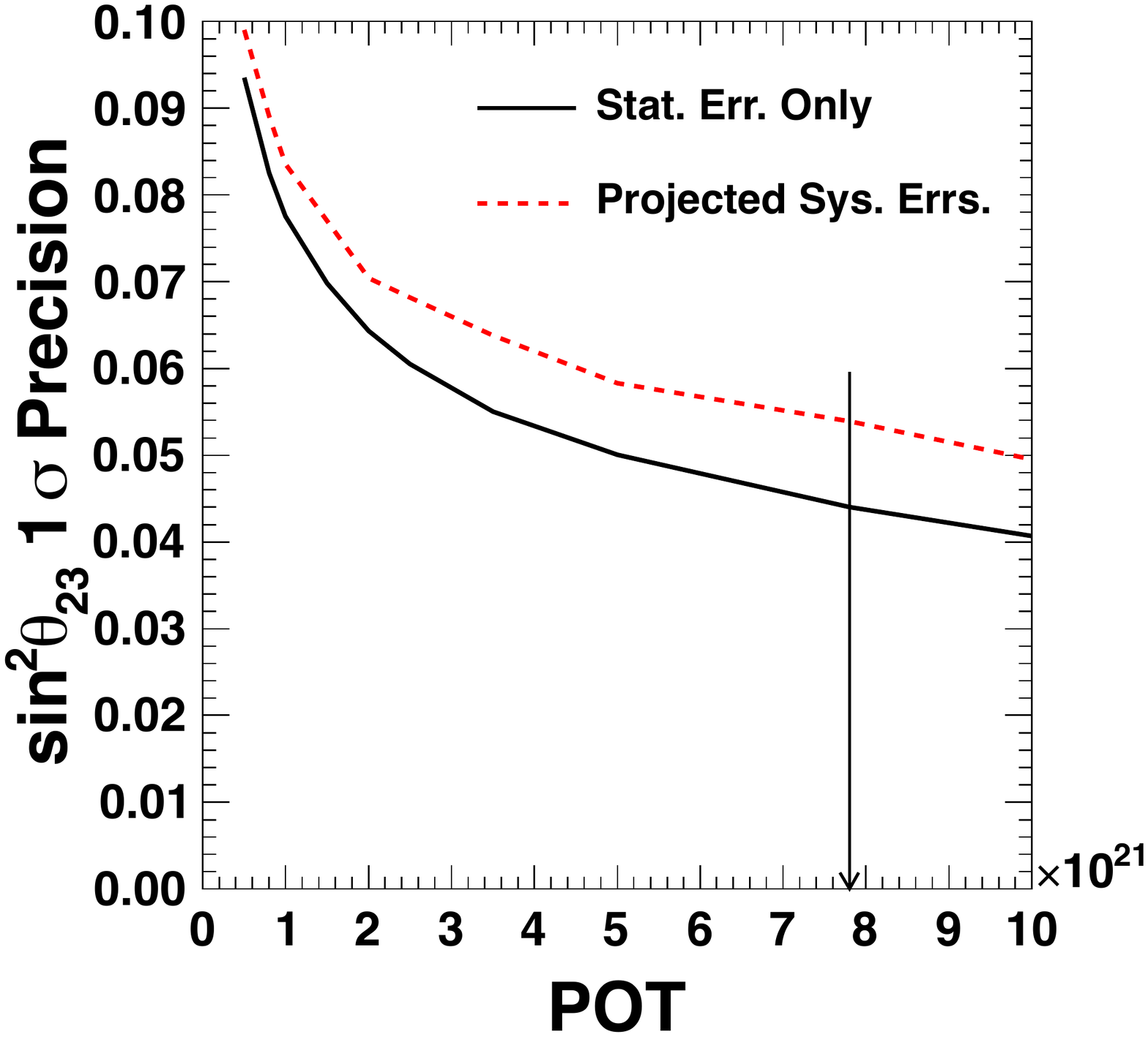}
\includegraphics[width=0.48\textwidth]{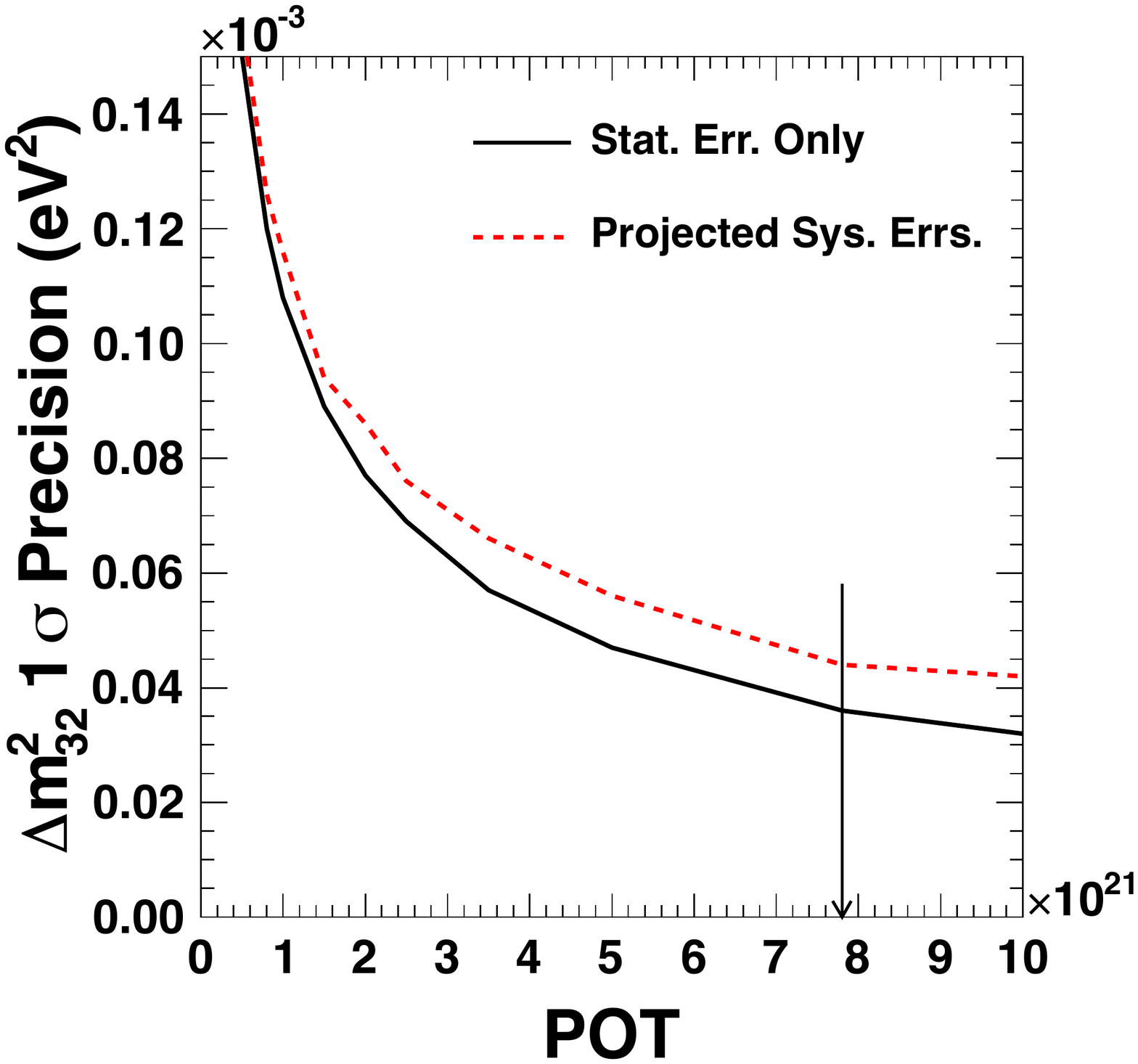}
\caption{T2K expected sensitivity to \(\sin^2\theta_{23}\) (left) and \(\Delta m^2_{32}\) (right) plotted as a function of T2K POT assuming \(\delta_{CP}=-\pi/2\), normal MO, \(\Delta m^2_{32} = 2.4\times10^{-3}\)~eV\(^2\), and $\sin^2\theta_{23} = 0.5$. Reproduced from \cite{T2K:2014xyt}.
}
\label{fig:T2Kth23dm23sens}
\end{figure}

T2K expects to take data with the upgraded beam and detectors until the start of Hyper-Kamiokande, aiming to collect data corresponding to \(10\times10^{21}\) protons-on-target (POT).  The physics program will continue to pursue $3~\sigma$ observation of CPV in neutrinos and enhanced sensitivity to the mass ordering and atmospheric parameters \cite{T2K:2014xyt,T2K:2016siu}.  The expected sensitivity of T2K to resolve CPV is given in Fig.~\ref{fig:T2KdCPsens}, and the expected sensitivity to the atmospheric mixing parameters is given in Fig.~\ref{fig:T2Kth23dm23sens}. Future T2K running will be used to make further improvements to the J-PARC neutrino beam, and future data will be used to develop analysis techniques and approaches for precision oscillation measurements, especially including the use of upgraded, high performance near detectors.  T2K's program is therefore essential to support the next  generation of long-baseline neutrino experiments.

\subsubsection{Hyper-Kamiokande (HK)}
As the next-generation underground water Cherenkov detector in Japan, Hyper-Kamiokande (HK)  \cite{Abe:2018uyc} builds on the highly successful Super-Kamiokande and T2K experiments.  
The HK experiment plans to utilize the upgraded J-PARC MR 1.3~MW proton beam, the upgraded 280~m near detector suite, a new intermediate detector to be installed \(\sim\)1~km away from the neutrino production target, and a new far detector 295~km away from the neutrino source.  The 260~kton Hyper-Kamiokande far detector has an 8.4 times larger fiducial volume than SK and is expected to acquire an unprecedented exposure of 3.8 Mton-year over a period of 20 years of operation.  The following also assumes collecting \(2.7\times10^{21}\) POT per year, running at a ratio of 1 to 3 neutrino-enhanced to antineutrino-enhanced beam.  Civil construction towards the HK tank is underway now.  Detector installation and commissioning is planned for 2025/2026, and datataking is scheduled to start in 2027.

The planned HK near detector suite will include INGRID, WAGASCI+BabyMIND, and the upgraded ND280 inherited from T2K\footnote{An MOU related to property transfer is under discussion now.}.  Additional upgrades to ND280 specific for HK are also under consideration.  A new Intermediate Water Cherenkov Detector (IWCD) is proposed to be newly constructed \(\sim\)1 km from the neutrino production target. This kiloton-scale water Cherenkov detector can be moved vertically to measure the neutrino beam intensity and energy spectra at different off-axis angles, yielding different true neutrino energy spectra. Precision neutrino cross section measurements, as well as measurements of the neutrino flux and intrinsic electron neutrino backgrounds will be performed using this suite of near detectors. 

The HK detector will be located 8~km south of the SK detector site in a cavern with a rock overburden of 650~m. Like SK for T2K, the HK detector will sit 2.5\(\degree\) off of the center of the J-PARC neutrino beam axis.  The detector tank will be 71~m in height and 68~m in diameter, with an expected fiducial volume of 188~kt.  Newly developed, high sensitivity 50~cm ``Box \& Line'' PMTs will provide 20\% photo-coverage for the HK inner detector, while additional photo-coverage using multi-PMT modules is also planned.  Like SK, a highly-instrumented outer detector will be used to veto incoming particles.

HK, like SK, will be sensitive to various physics processes, including nucleon decay, solar neutrinos, supernova burst neutrinos, relic supernova neutrinos, atmospheric neutrinos, and accelerator neutrinos.  Sensitivities to the last two are relevant to the subject of this report and are detailed here. 

\begin{figure}[H]
\includegraphics[width=0.45\textwidth]{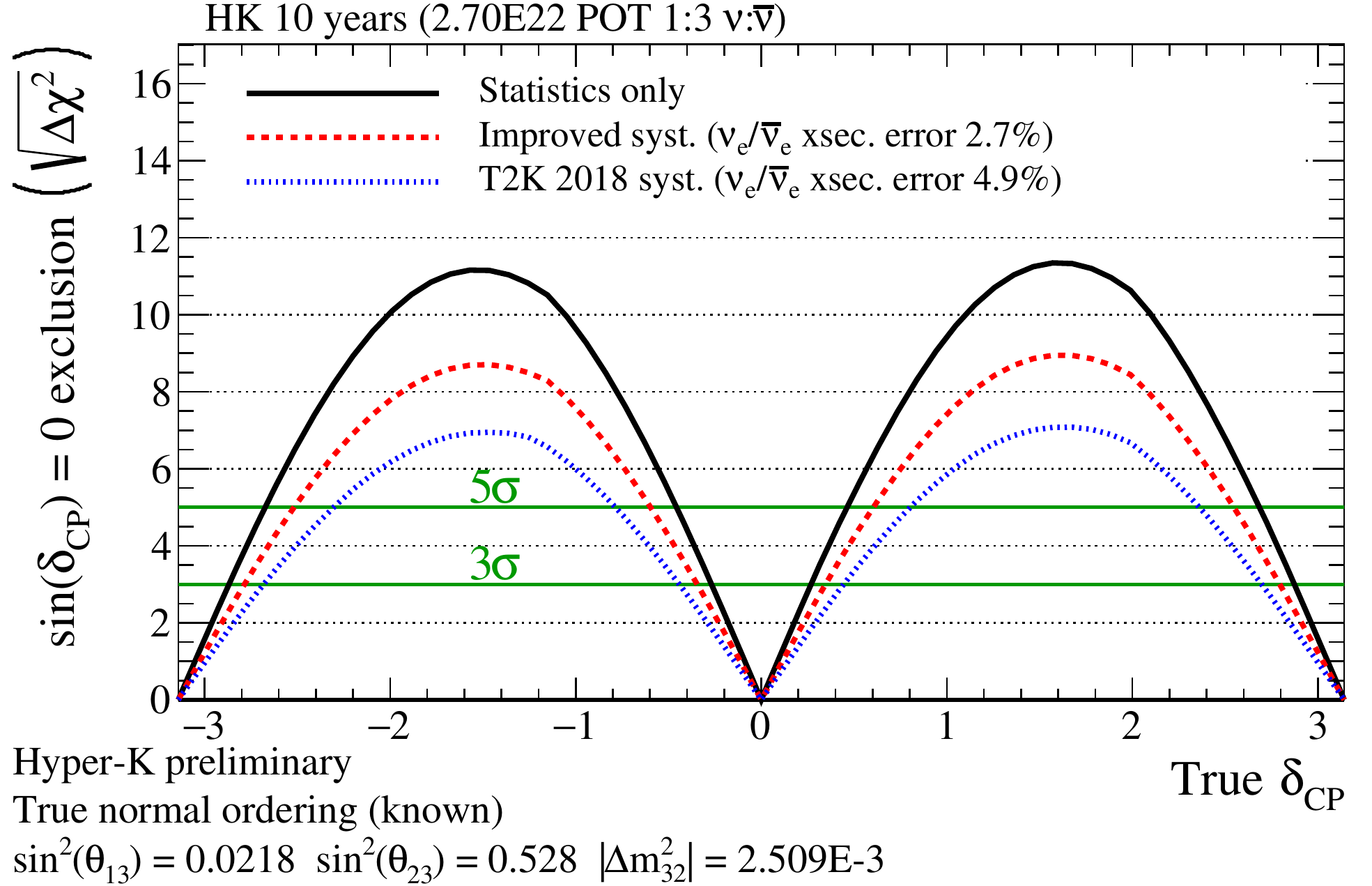}
~~~~~~~
\includegraphics[width=0.45\textwidth]{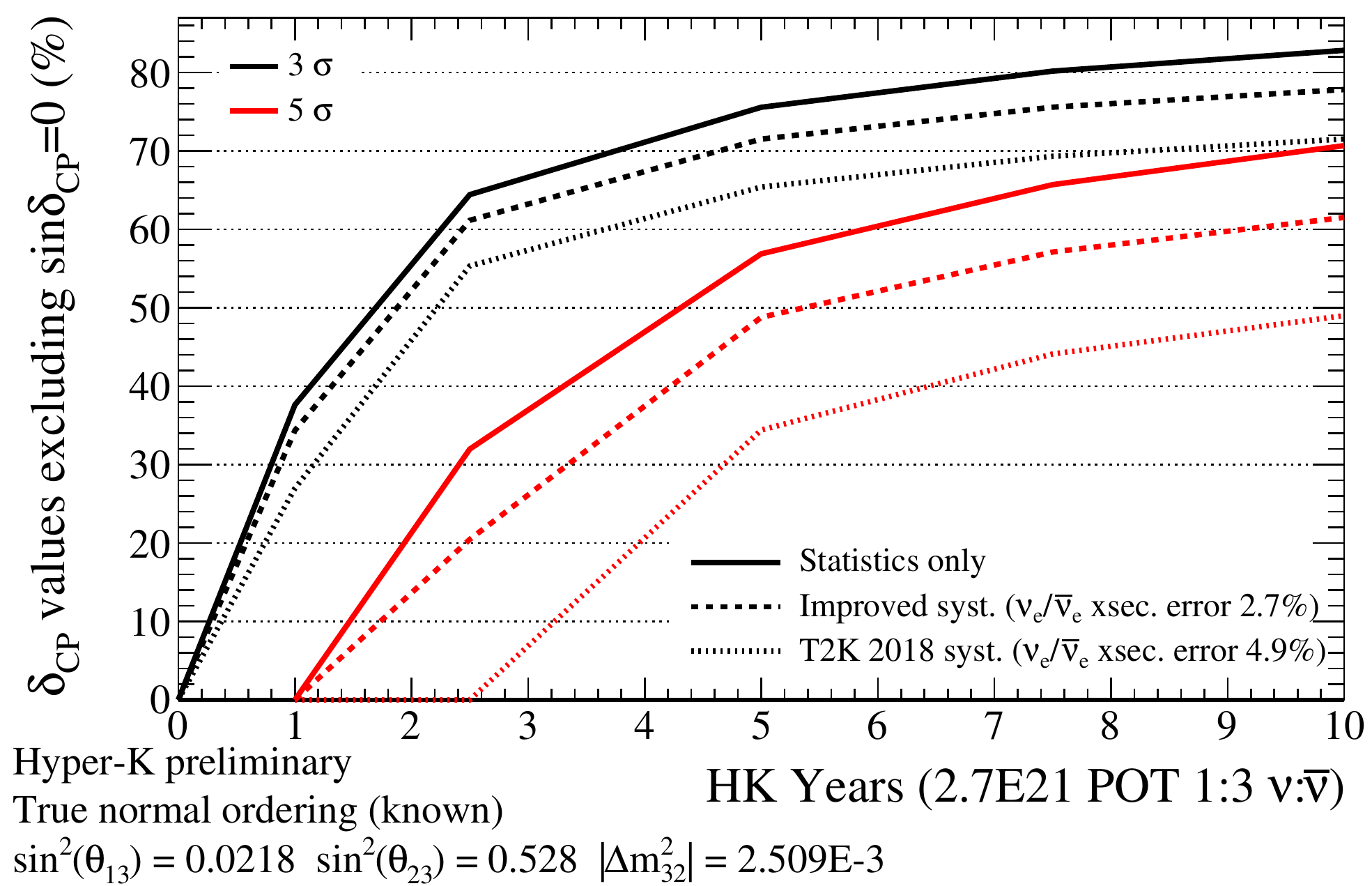} \\
\includegraphics[width=0.45\textwidth]{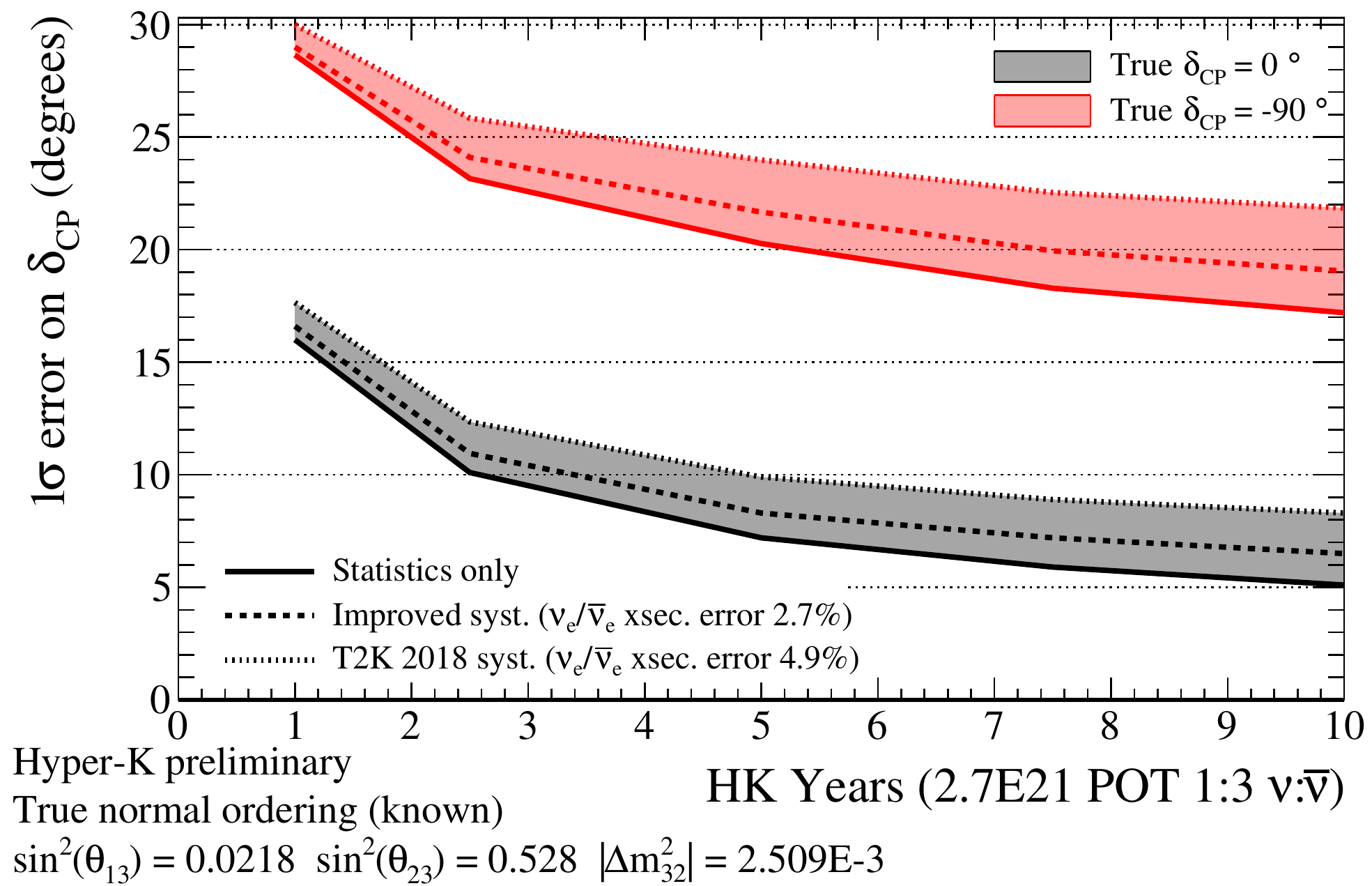}
~~~~~~~ 
\includegraphics[width=0.45\textwidth]{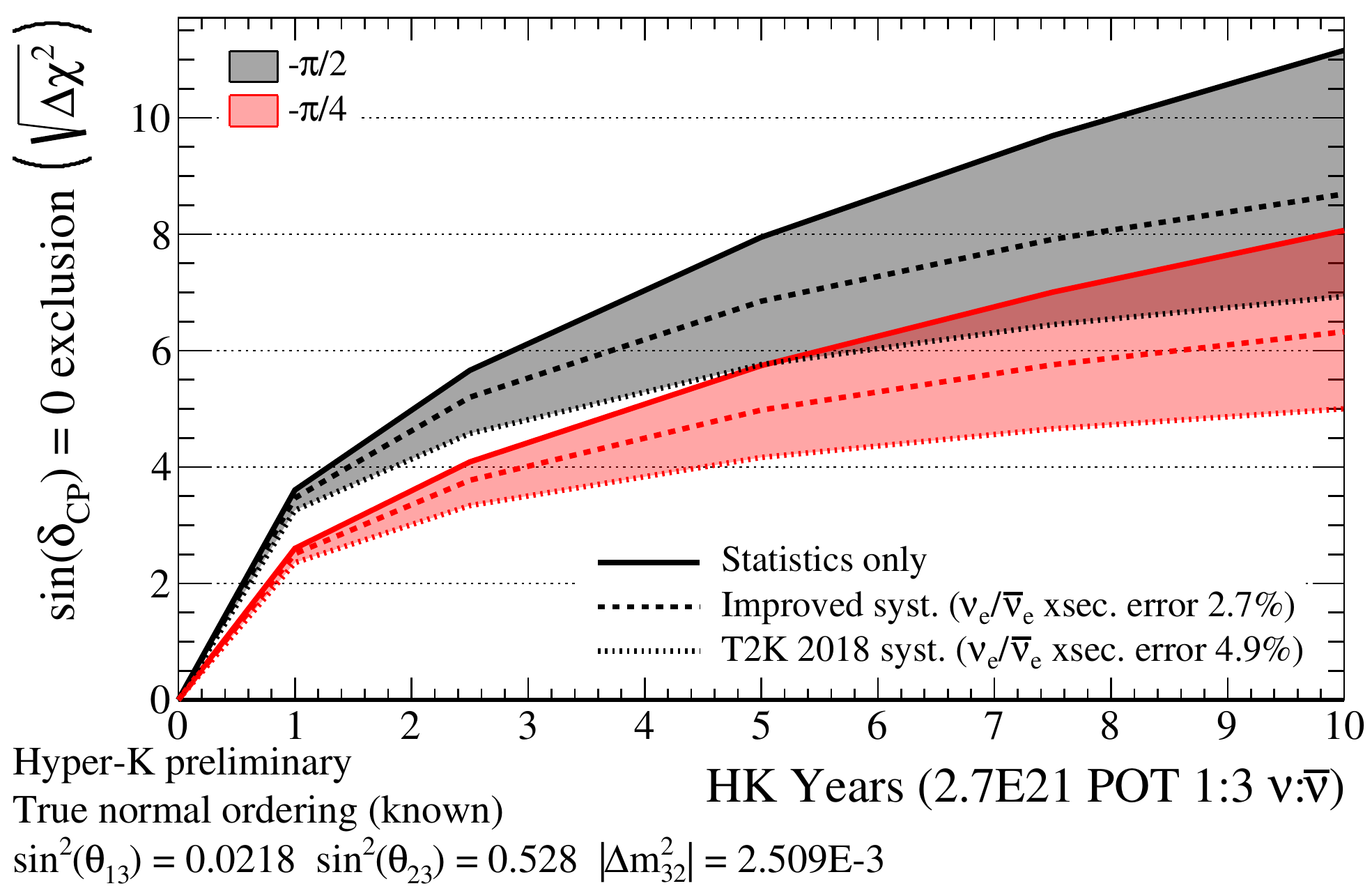}
\caption{HK expected sensitivity to exclude \(\sin\delta_{CP}=0\) plotted as a function of the true value of \(\delta_{CP}\) (top left) and the percent of true \(\delta_{CP}\) values for which \(\sin\delta_{CP}=0\) can be excluded plotted as a function of years of operation (top right). 
HK expected precision on \(\delta_{CP}\) plotted as a function of year (bottom left) and sensitivity to exclude \(\sin\delta_{CP}=0\) plotted as a function year in the case of true \(\delta_{CP}=-\pi/2\) and \(-\pi/4\) (bottom right).  Plots all compare different systematic error assumptions and assume that the MO is known.  From \cite{Scott:2020gng}.
}
\label{fig:HKsdcpneq0sens}
\end{figure}

Long-baseline neutrinos from the J-PARC accelerator in particular give sensitivity to the oscillation parameters \(\sin^2\theta_{23}\), \(\Delta m^2_{32}\), and \(\delta_{CP}\).  The HK sensitivity to excluding \(\sin\delta_{CP}=0\) assuming a known mass ordering for different possible true values \(\delta_{CP}\) is shown in Fig.~\ref{fig:HKsdcpneq0sens}.  As can be seen in this figure, the measurement of \(\delta_{CP}\) is particularly sensitive to the systematic error on the \(\nu_e/\bar{\nu}_e\) cross section ratio.  Precision measurements by the near detector suite to constrain this error are essential.  

\begin{figure}[H]
\includegraphics[width=0.45\textwidth]{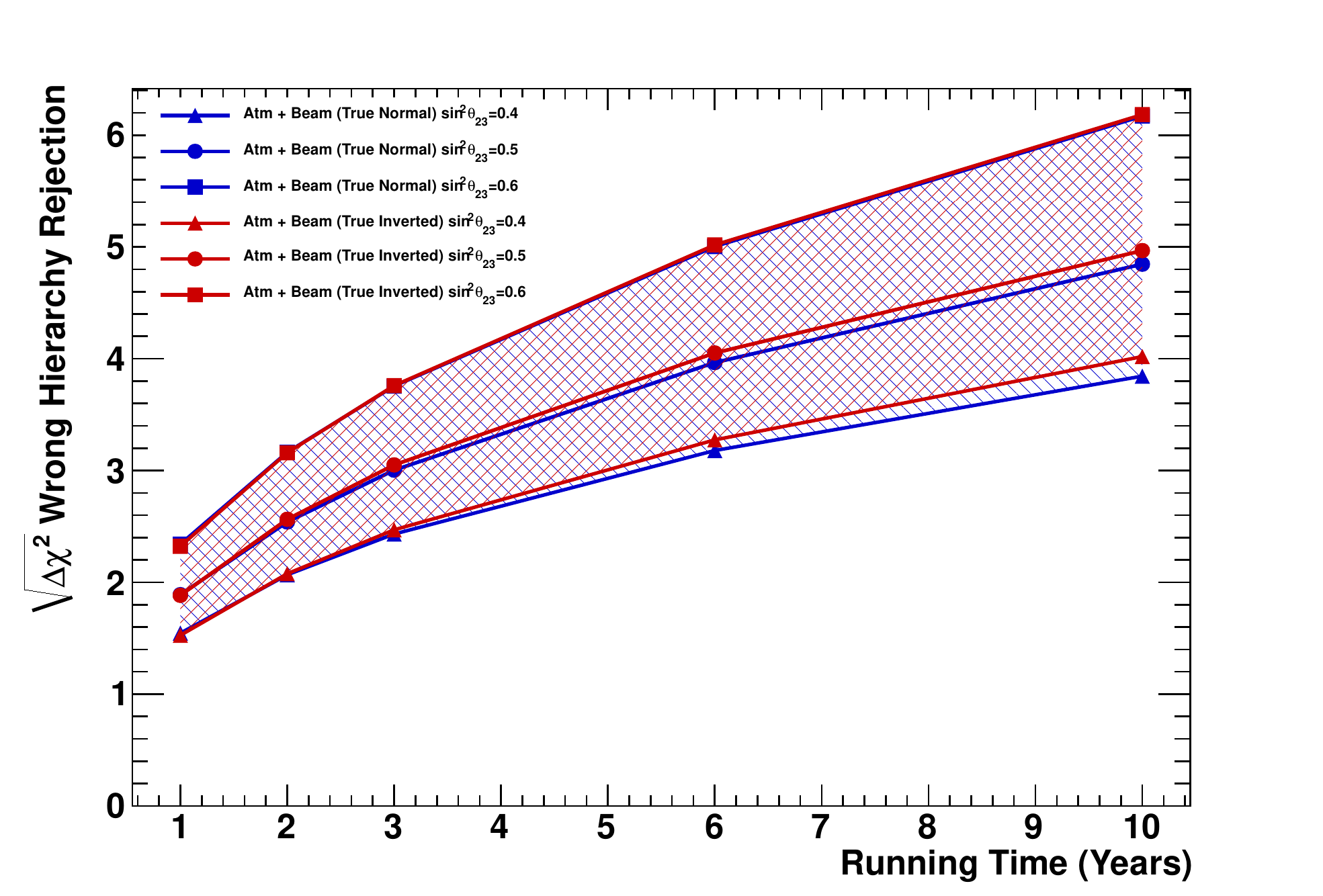}
\includegraphics[width=0.45\textwidth]{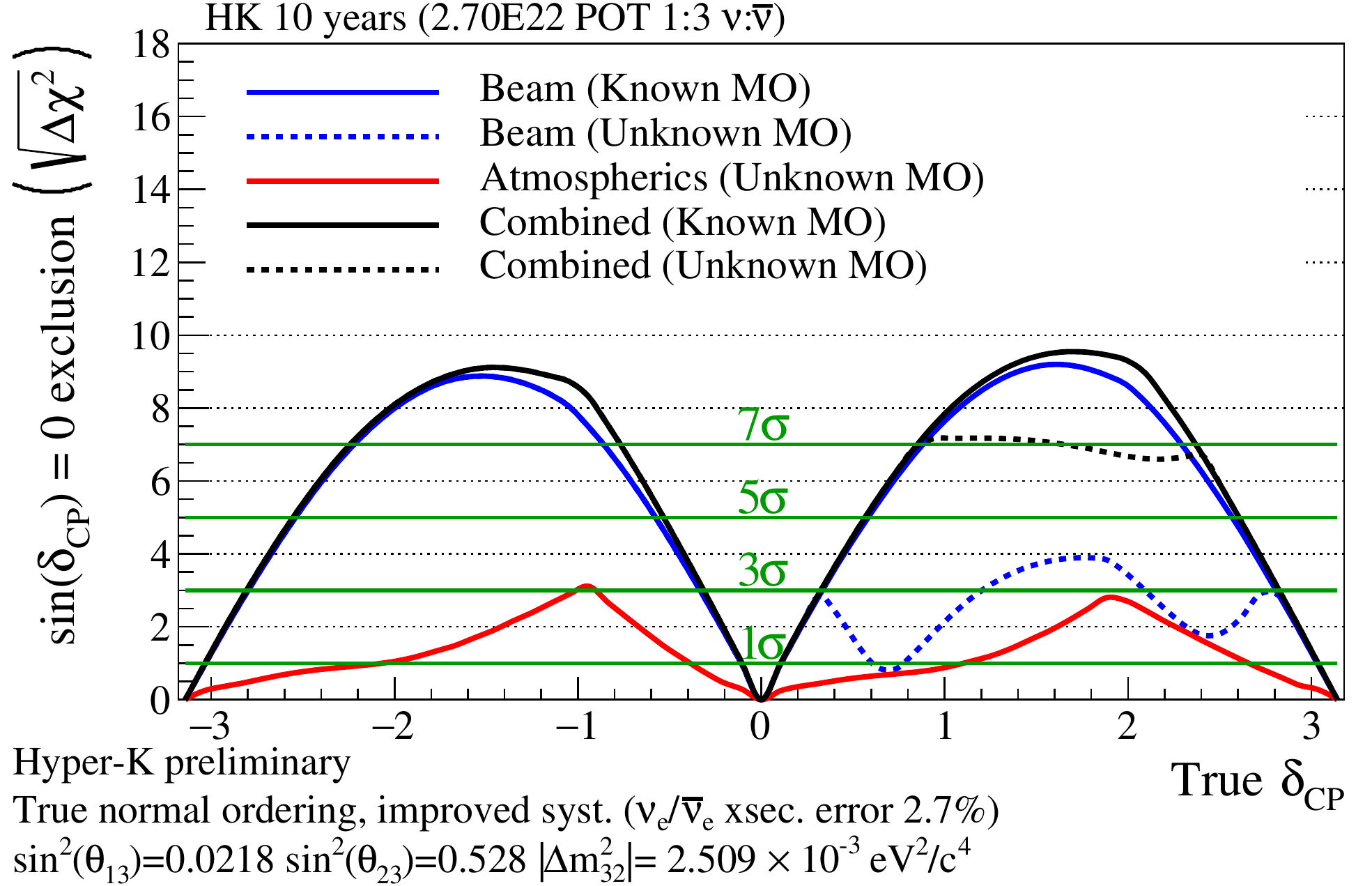}
\caption{
The HK sensitivity to the mass ordering plotted as a function of time for a combination of beam and atmospheric neutrinos (left) reproduced from \cite{Abe:2018uyc} and the HK sensitivity to exclude \(\sin\delta_{CP}=0\) plotted as a function of the true value of \(\delta_{CP}\) assuming the mass ordering is unknown (right).  A combined fit of HK beam and atmospheric neutrinos significantly enhances the HK sensitivity to \(\delta_{CP}\).}
\label{fig:HKatmplusbeam}
\end{figure}

Atmospheric neutrinos span a wide range of energies and path lengths, unlike the narrow-band accelerator neutrino beam, and matter effects for atmospheric neutrinos that travel through the earth give HK sensitivity to the neutrino mass ordering.  Simultaneously fitting HK beam and atmospheric neutrinos gives enhanced sensitivity to oscillation parameters, especially in the case that the mass ordering is unknown, as shown in Fig.~\ref{fig:HKatmplusbeam}.  

Hyper-Kamiokande also has sensitivity to measure the values of \(\Delta m^2_{32}\) and \(\sin^2\theta_{23}\).  HK can determine the \(\theta_{23}\) octant at 3\(\sigma\) significance if the true value of \(\sin^2\theta_{23}\) is between 0.47 and 0.55 \cite{Scott:2020gng}.

In order to reach this unprecedented precision in the oscillation parameter measurements, systematic errors must be controlled to unprecedented levels.  Highly precise detector calibration, both before detector installation and in-situ, is necessary.  Both HK internal measurements using near detectors, control samples, etc, and external measurements to constrain neutrino interactions and fluxes are also essential.  Precisely determining the \(\nu_e/\bar{\nu}_e\) cross section ratio will be key to a precision measurement of \(\delta_{CP}\), but other unexpected and/or incorrectly modeled systematics could also have a major impact on the HK precision.  Global effort towards understanding these potential ``unknown-unknowns'' will be important.

\subsection{South Pole: IceCube/DeepCore}

\subsubsection{Neutrino oscillation physics at IceCube}

The IceCube Neutrino Observatory~\cite{Aartsen:2016nxy} at the South Pole instruments a cubic km of clear glacial ice 1.5-2.5 km below the surface, using 5160 Digital Optical Modules (DOMs) to detect Cherenkov light resulting from neutrino interactions in the ice. The observatory detects vast numbers of GeV-TeV atmospheric neutrinos as well as a high energy (up to PeV) flux of astrophysical origin, with an overall neutrino detection rate of $\sim$10\,mHz. A more densely instrumented region known as DeepCore in the clear ice 2.1 km below the surface greatly improves the detector's low energy performance down between 5-100\,GeV~\cite{IceCube:2011ucd}.

\begin{figure}[H]
\centering
\includegraphics[width=0.49\textwidth]{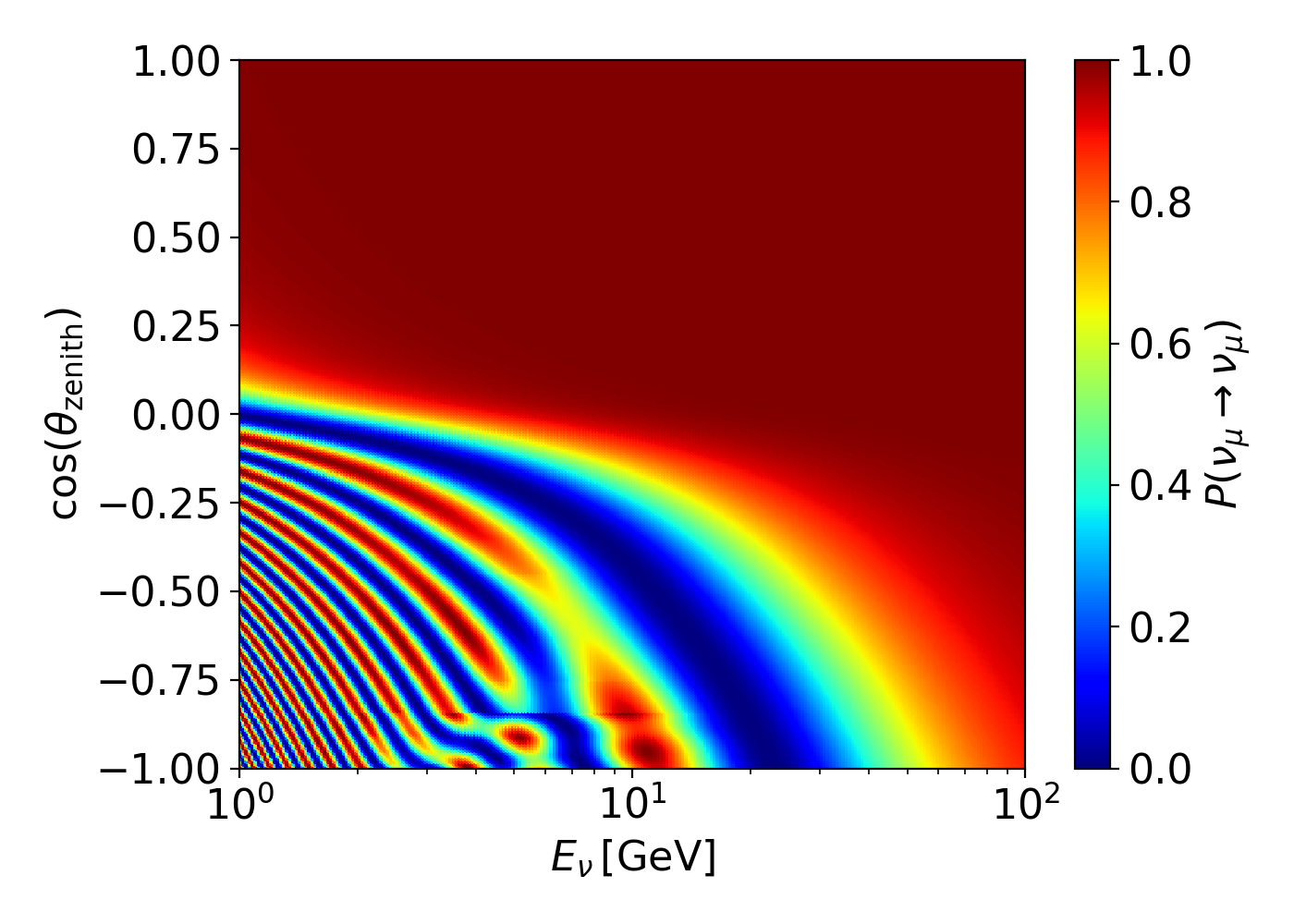}
\includegraphics[width=0.49\textwidth]{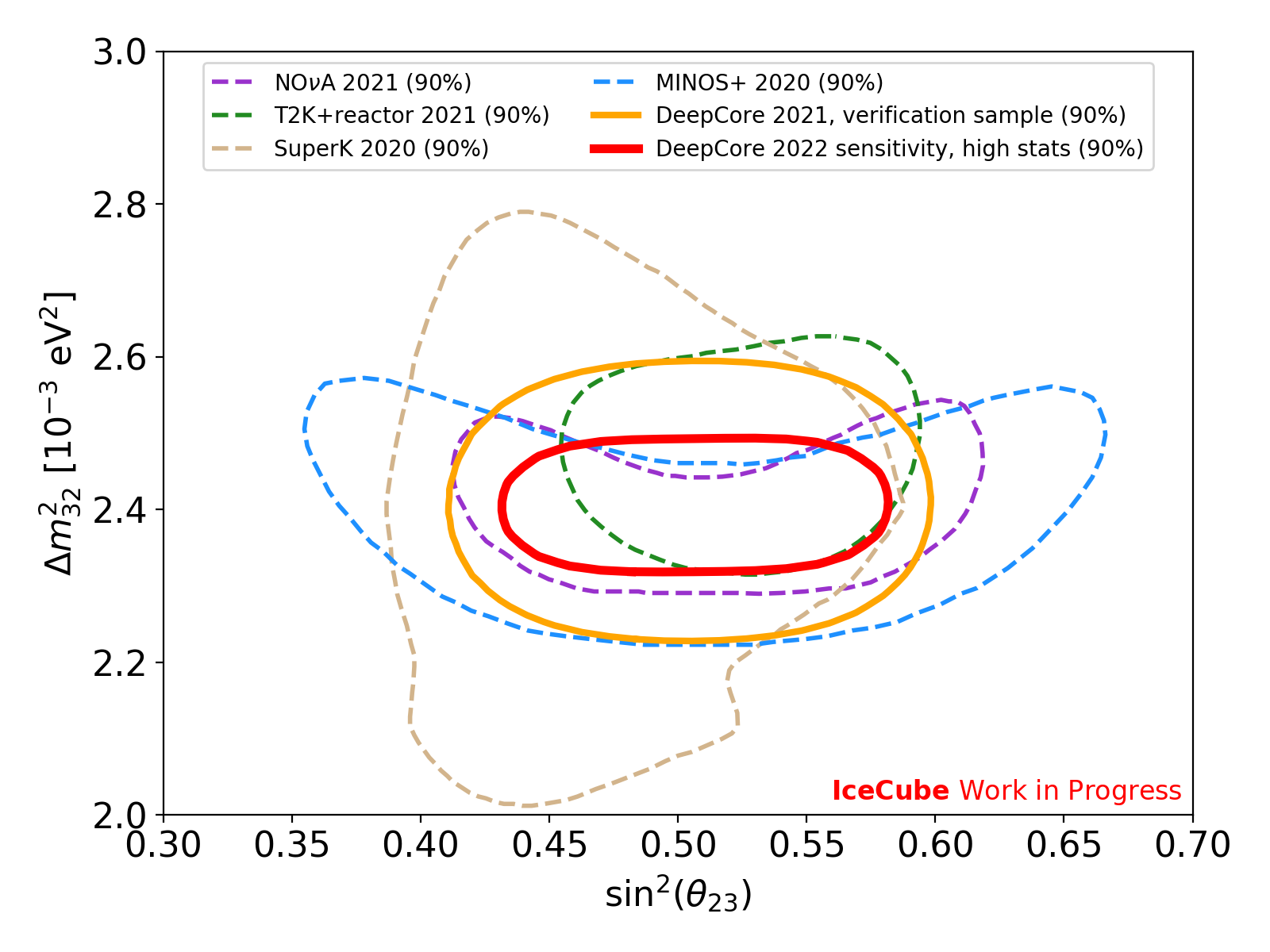}
\caption{\textbf{Left}: $\nu_\mu \rightarrow \nu_\mu$ survival probability in the energy range of interest for IceCube-DeepCore and the IceCube Upgrade, including effects of Earth matter. Both the disappearance of $\nu_\mu$ and the corresponding appearance of $\nu_\tau$ are observed by IceCube.
\textbf{Right}: The most recent IceCube oscillation measurement~\cite{oscnext_verif_sample} and sensitivity of an ongoing analysis, alongside results from other experiments~\cite{T2K:2021xwb,NOvA:2021nfi, superk_disappearance,MINOS:2020llm}.}
\label{fig:oscillogram}
\end{figure}

IceCube-DeepCore observes strong atmospheric neutrino oscillations in DeepCore's energy range for neutrinos crossing the Earth before detection, primarily in the $\nu_\mu \rightarrow \nu_\mu$ and $\nu_\mu \rightarrow \nu_\tau$ channels (Fig.~\ref{fig:oscillogram}). At higher energies ($\gtrsim$100\,GeV) no oscillations are expected in the standard three flavor paradigm as the baseline required exceeds the Earth's diameter. With a large range of energies, baselines, flavors and matter profiles encompassed by its dataset, IceCube is sensitive to broad range of standard oscillations -- $\nu_\mu$ disappearance, $\nu_\tau$ appearance, and the mass ordering.

The strengths of the IceCube-DeepCore oscillation program include:
\begin{itemize}
\item \textbf{High statistics:} The size of the DeepCore sub-detector (10 Mton) coupled with the copious natural atmospheric flux affords very high detection rates, yielding $>$200,000 neutrinos in the current all-flavor 9-year data sample; orders of magnitude more events than current accelerator experiments.
\item \textbf{$\nu_\tau$ detection:} Charged-current $\nu_\tau$ interactions are difficult to access at many other neutrino experiments due to the kinematic suppression resulting from the large $\tau$ lepton mass. DeepCore is uniquely able to detect large numbers of $\nu_\tau$ in the $\nu_\mu \rightarrow \nu_\tau$ appearance channel at $\mathcal{O}$(10-20)\,GeV, with $>$9,000 $\nu_{\tau,\rm{CC}}$ events expected in the current 9-year dataset.
While Deepcore's particle identification for $\nu_\tau$ does not reach the levels of e.g.~DONuT \cite{DONuT:2007bsg} or OPERA \cite{Agafonova:2018auq}, Deepcore's existing data set already far exceeds the $\sim$300 $\nu_\tau$ detected to date by all other neutrino experiments combined~\cite{DONuT:2007bsg,Agafonova:2018auq,Li:2017dbe,Abraham:2022jse}.
\item \textbf{Complementarity:} DeepCore probes comparable $L/E$ oscillations to long baseline accelerator experiments,  but at an order of magnitude higher energy (in the deep inelastic scattering regime) and with very different systematic uncertainties, and is thus highly complementary to other global measurements. Moreover, the flavor composition of the astrophysical neutrino flux above 100\,TeV also constrains neutrino mixing~\cite{Ahlers:2018yom} (see Sec.~\ref{sec:other}), providing an additional level of complementary.
\end{itemize}

\subsubsection{Neutrino oscillation measurements}

IceCube measures the atmospheric mass splitting, $\Delta m^2_{32}$, and mixing angle, $\theta_{23}$, via the simultaneous measurement of $\nu_\mu$ disappearance and corresponding $\nu_\tau$ appearance, with a background of largely unoscillated $\nu_e$. The dataset is binned in three dimensions: reconstructed neutrino energy, cosine of the reconstructed zenith angle (a proxy for baseline), and particle ID (PID, a proxy for flavor), with oscillations producing 3D deformations in these dimensions that cannot be easily mimicked by systematic uncertainties, providing robust measurements. Neutrino properties are reconstructed using a maximum likelihood fit utilizing tabulated expectations of the detector response to light sources at a given location. PID utilizes boosted decision tree (BDT) ensembles to predict whether an event resulted from a $\nu_{\mu,\rm{CC}}$ interaction, largely based on the presence of \textit{track-like} light emission from the secondary $\mu$ produced~\cite{DeHolton:2021fyh}. All other neutrino interaction types exhibit more spherical, \textit{cascade-like} emission profiles.

In addition to upgoing events, the data samples used for these measurements include neutrinos from above the detector which travel only $\mathcal{O}$(10-100 km) and thus provide unoscillated neutrino events that can constrain systematic uncertainties related to the neutrino flux cross-sections and detector effects, in essence somewhat like a near detector in long baseline accelerator experiments. Overwhelming backgrounds of atmospheric muons and random noise coincidences at trigger level are suppressed by seven orders of magnitude via multiple layers of event selection, ultimately resulting in $\sim$1 mHz neutrino rates in current GeV-scale oscillation analyses, with vanishing noise and $<$1\% atmospheric muon contamination.

Fig.~\ref{fig:oscillogram} shows the most recent IceCube oscillation measurement using 8 years of data, which uses a sub-sample (known as the \textit{verification sample}) of $\sim$23,000 predominantly track-like events with minimally scattered light, making it extremely robust against uncertainties in modeling the optical properties of the ice~\cite{oscnext_verif_sample}. Fig.~\ref{fig:oscillogram} also shows the expected sensitivity of a 9-year high-statistics IceCube measurement that is currently underway, which features $>$200,000 neutrinos and improved reconstruction and background rejection methods. The verification sample result is already approaching the precision of the latest long baseline accelerator measurements (NO$\nu$A and T2K), whilst the high-statistics analysis is expected to achieve precision commensurate with current world-leading results.

IceCube also detects $\nu_\tau$ appearance in atmospheric neutrinos providing an additional test of the oscillation parameters.

\subsubsection{Next-generation oscillation physics with the IceCube Upgrade}

\begin{figure}
\centering
\includegraphics[trim={6cm 0cm 6cm 2cm},clip,width=0.3\linewidth]{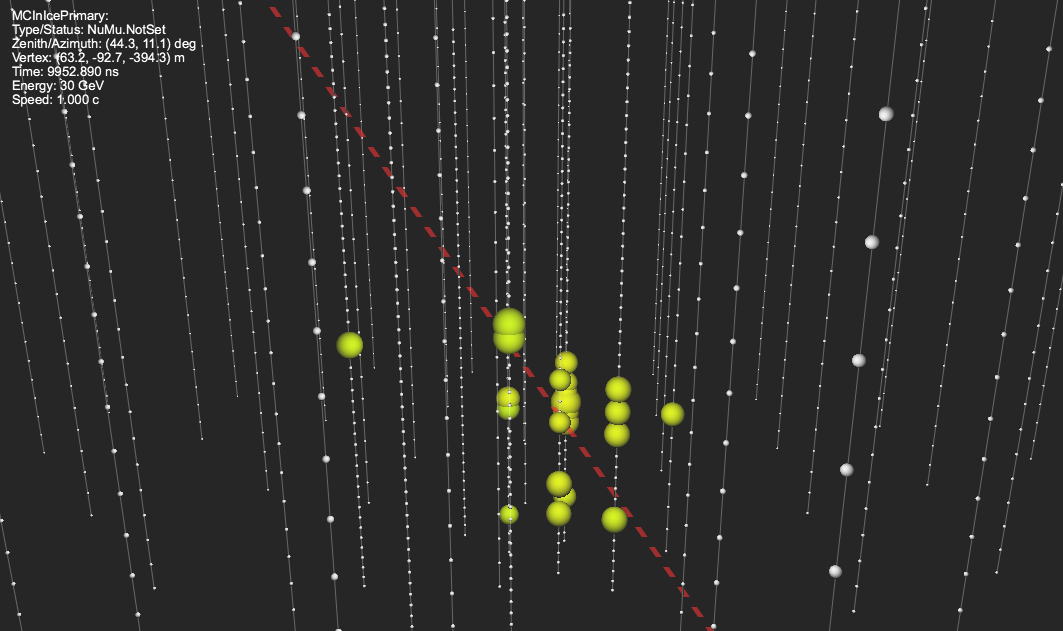}
~~~~~~~
\includegraphics[trim={6cm 0cm 6cm 2cm},clip,width=0.3\linewidth]{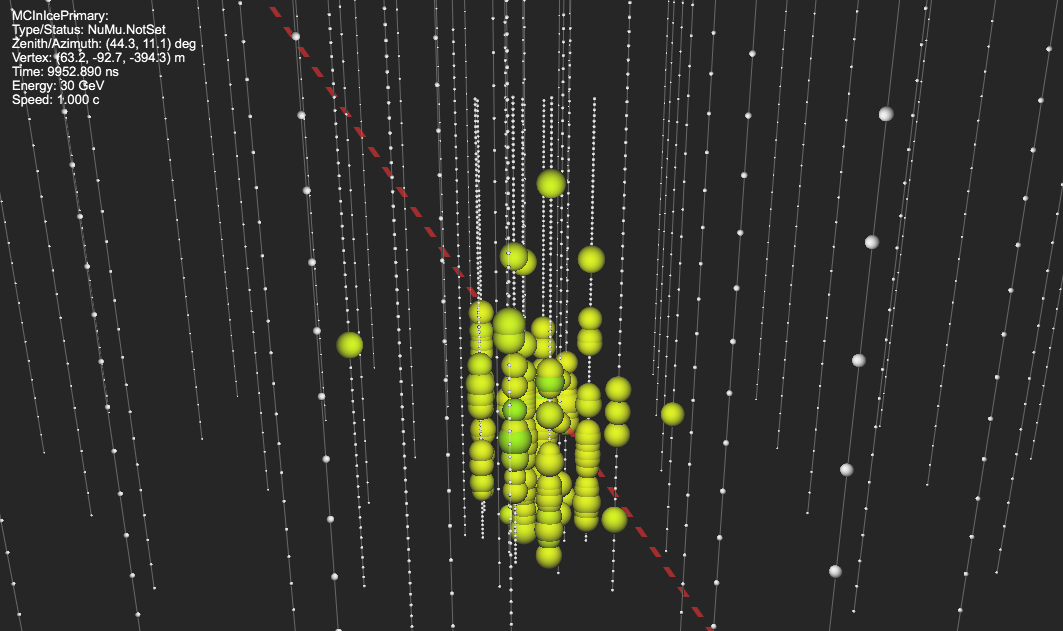}
\caption{Event displays of a 30\,GeV $\nu_{\mu,\rm{CC}}$ event in DeepCore (left) and the Upgrade (right). White (yellow) spheres represent optical modules (photon hits) respectively, with the neutrino direction in red. }
\label{fig:event_display}
\end{figure}

In 2025-26, a dense new sub-array of 7 strings featuring $\sim$700 multi-PMT optical modules, spaced 3\,m apart vertically, will be deployed in IceCube, vastly improving the photocathode density in a 2 Mton fiducial volume within the already dense 10 Mton DeepCore volume. Known as the \textit{IceCube Upgrade}~\cite{Ishihara:2019aao}, this next-generation detector has the following advantages:

\begin{figure}
\centering
\includegraphics[trim={0cm 0.2cm 0 0},clip,width=0.45\textwidth]{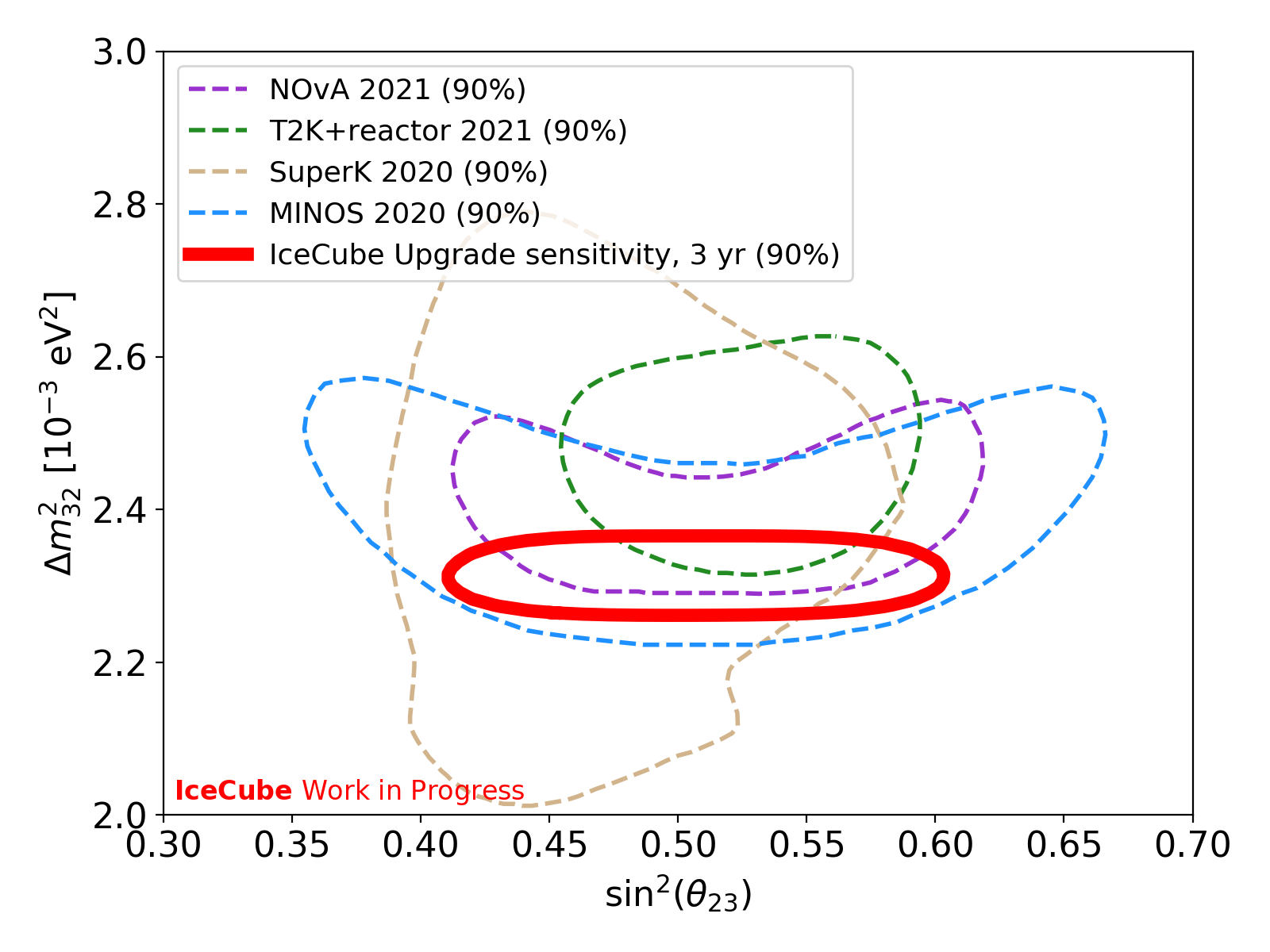}
\caption{Initial (conservative) estimate of oscillation sensitivity with the IceCube Upgrade~\cite{Ishihara:2019aao}, alongside other experiments~\cite{T2K:2021xwb,NOvA:2021nfi, superk_disappearance,MINOS:2020llm}. The 2018 IceCube fit value is assumed~\cite{IceCube:2017lak}.}
\label{fig:upgrade_disappearance_sensitivity}
\end{figure}

\begin{itemize}
\item \textbf{Lower energy threshold:} The denser instrumentation will allow the Upgrade to measure atmospheric neutrinos down to $\sim$1\,GeV, and more generally increase the $<$10\,GeV neutrino rate by an order of magnitude compared to DeepCore~\cite{Ishihara:2019aao}. This allows the Upgrade to probe higher order oscillation bands, and significantly increases the fraction of detected neutrinos that have oscillated.\\
\item \textbf{Reconstruction:} Denser instrumentation will also result in major improvements in neutrino energy and direction resolution, by a factor 2-4 at the $\mathcal{O}$(10-20)\,GeV energies relevant for current DeepCore oscillation measurements (see Fig.~\ref{fig:event_display} for an example of the increased light detection)~\cite{Ishihara:2019aao, Stuttard:2020zsj}. The neutrino rates at these energies will also increase by a factor $\sim$2-4, depending on fiducial volume.\\
\item \textbf{Calibration:} A plethora of new and densely deployed calibration devices will dramatically improve detector and ice property calibration, also allowing re-calibration of more than a decade of existing IceCube data.
\end{itemize}

This precision neutrino physics detector will facilitate the significant reduction in systematic uncertainties required to fully exploit IceCube's huge statistical power. Initial studies of the oscillation physics potential of the Upgrade are highly conservative, using only the 2 Mton Upgrade fiducial volume (e.g.~excluding the remaining 8 Mton of DeepCore), simplified reconstruction, PID and background rejection methods (not yet exploiting the segmented new optical sensors or recent advances in DeepCore analyses), and neglecting both improvements in detector calibration and the existing 10+ years of DeepCore data. Nonetheless, even these conservative estimates indicate that the Upgrade can achieve $\Delta m^2_{23}$ precision exceeding current long baseline accelerator results~\cite{T2K:2021xwb,NOvA:2021nfi} (see Fig.~\ref{fig:upgrade_disappearance_sensitivity}) and $\lesssim$6\% $\nu_\tau$ normalization precision by the end of this decade~\cite{Ishihara:2019aao}.

\subsubsection{Neutrino mass ordering}
A measurement of the matter effect is required to determine the mass orderings.
For atmospheric neutrino experiments such as IceCube however, matter effects in the Earth's core ($\cos\theta<-0.8$ in Fig.~\ref{fig:oscillogram}) create additional potential that provides some information. 

While neutrino telescopes like IceCube cannot distinguish between $\nu$ and $\bar{\nu}$ on an per-event level, two effects -- an energy-dependent asymmetry in atmospheric flux of $\nu$ and $\bar{\nu}$ of up to 25\%~\cite{Honda:2015fha} and a $\nu$-nucleon cross-section that is about twice as large for $\nu$ as $\bar{\nu}$~\cite{Formaggio:2012cpf} -- combine into a net signature in the $\sim 2-15\,\mathrm{GeV}$ energy range for core-crossing events that IceCube is sensitive to.

An analysis of 3 years of IceCube-DeepCore data yields a weak sensitivity to the mass ordering of $0.45 - 0.65\sigma$ in the preferred range for the mixing angle $ 0.45 < \sin^2(\theta_{23}) < 0.55$, with an insignificant preference for the normal ordering~\cite{IceCube:2019dyb}. In contrast, the vastly improved low energy performance of the IceCube Upgrade dramatically increases the sensitivity in the normal ordering, with an initial study predicting that $3.8\sigma (1.8\sigma)$ can be reached in case of true normal (inverted) mass ordering after 6 years of operation.

Additionally, studies are underway to evaluate the Upgrade's $\nu_\mu / \bar{\nu}_\mu$ separation capabilities using a combination of reconstructed inelasticity in charged-current interactions and Michel electron tagging, which could offer significant gains in mass ordering sensitivity.

\subsection{KM3NeT}
KM3NeT is an international research project building a series of neutrino telescopes in the depths of the Mediterranean Sea. The main goals of KM3NeT are twofold: measuring the neutrino mass ordering; and identifying high-energy neutrino sources in the Universe.

The KM3NeT detectors consist of 3D arrays of detection units (DUs) each comprising 18 digital optical modules (DOM) made of 31~photo-multiplier tubes (PMTs). Various neutrino energy ranges can be accessed with this technology by simply adjusting the spacing between lines and between DOMs. Two detectors are under construction: one offshore Toulon (France) and one offshore Capo Passero (Italy). The former is called ORCA (Oscillation Research with Cosmics in the Abyss) and features a dense DOM geometry optimised to study neutrino oscillations in the GeV energy range with atmospheric neutrinos. The latter is called ARCA (Astronomy Research with Cosmics in the Abyss) and features a sparser DOM geometry adapted to the study of high-energy neutrino sources in the Universe.

Beyond the two main objectives mentioned above, the detectors offer a wide range of scientific opportunities not only in fundamental physics but also in astrophysics, earth and sea sciences. The full description of the physics program can be found in the Letter of Intent~\cite{KM3Net:2016zxf}.

\subsubsection{Scientific Status and Prospects}
\label{sec:sci}

\subsubsubsection{Neutrino oscillation with the first KM3NeT/ORCA data}
The modular design of KM3NeT allows physics analyses to be performed during detector construction. An oscillation measurement using the first data collected with 6 DUs  during 355 days ($\sim$0.3 Mton-year) was presented in 2021 \cite{KM3NeT:2021hkj}.

\begin{figure}[!ht]
\centering
\includegraphics[width=0.49\textwidth]{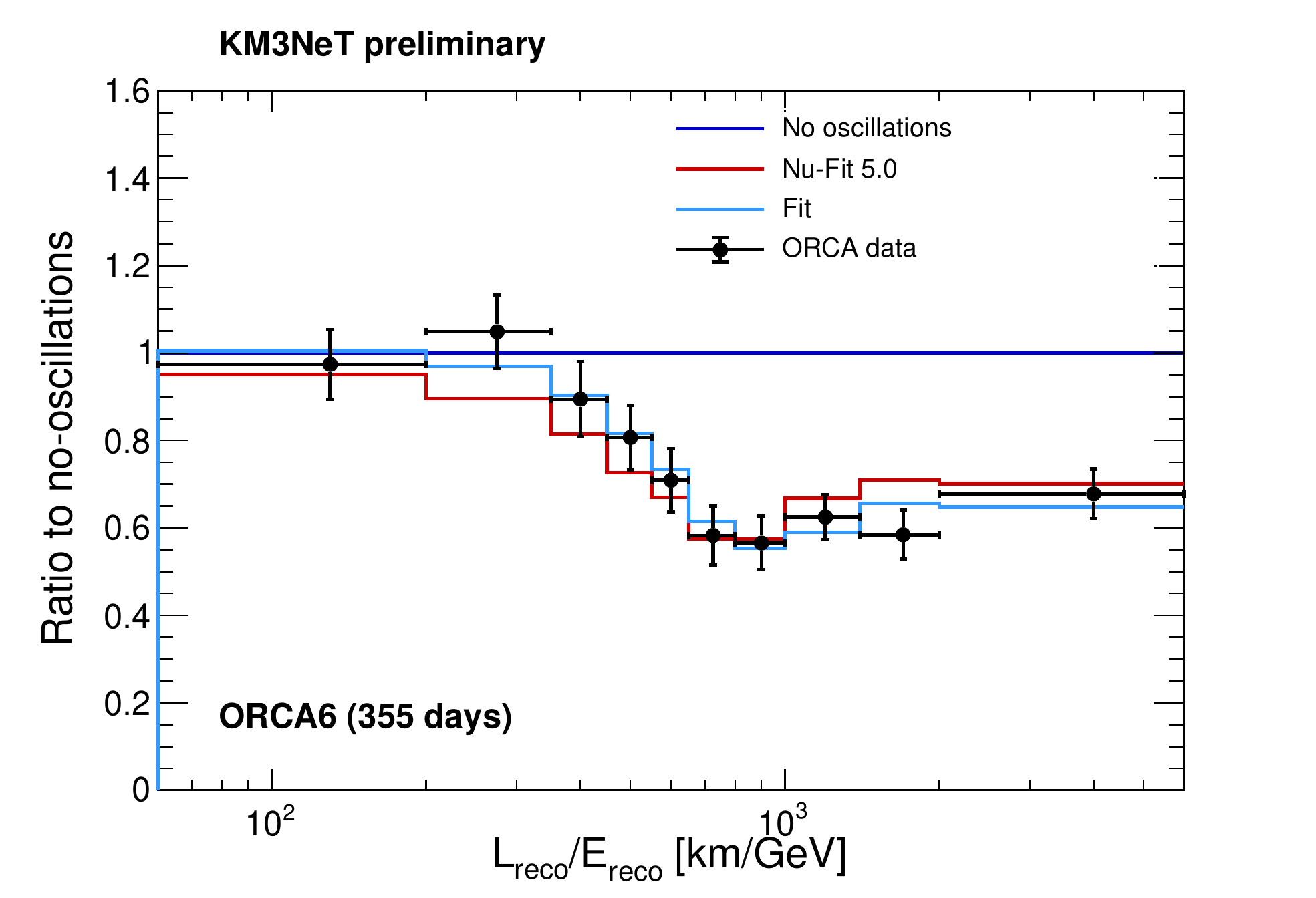}
\includegraphics[width=0.49\textwidth]{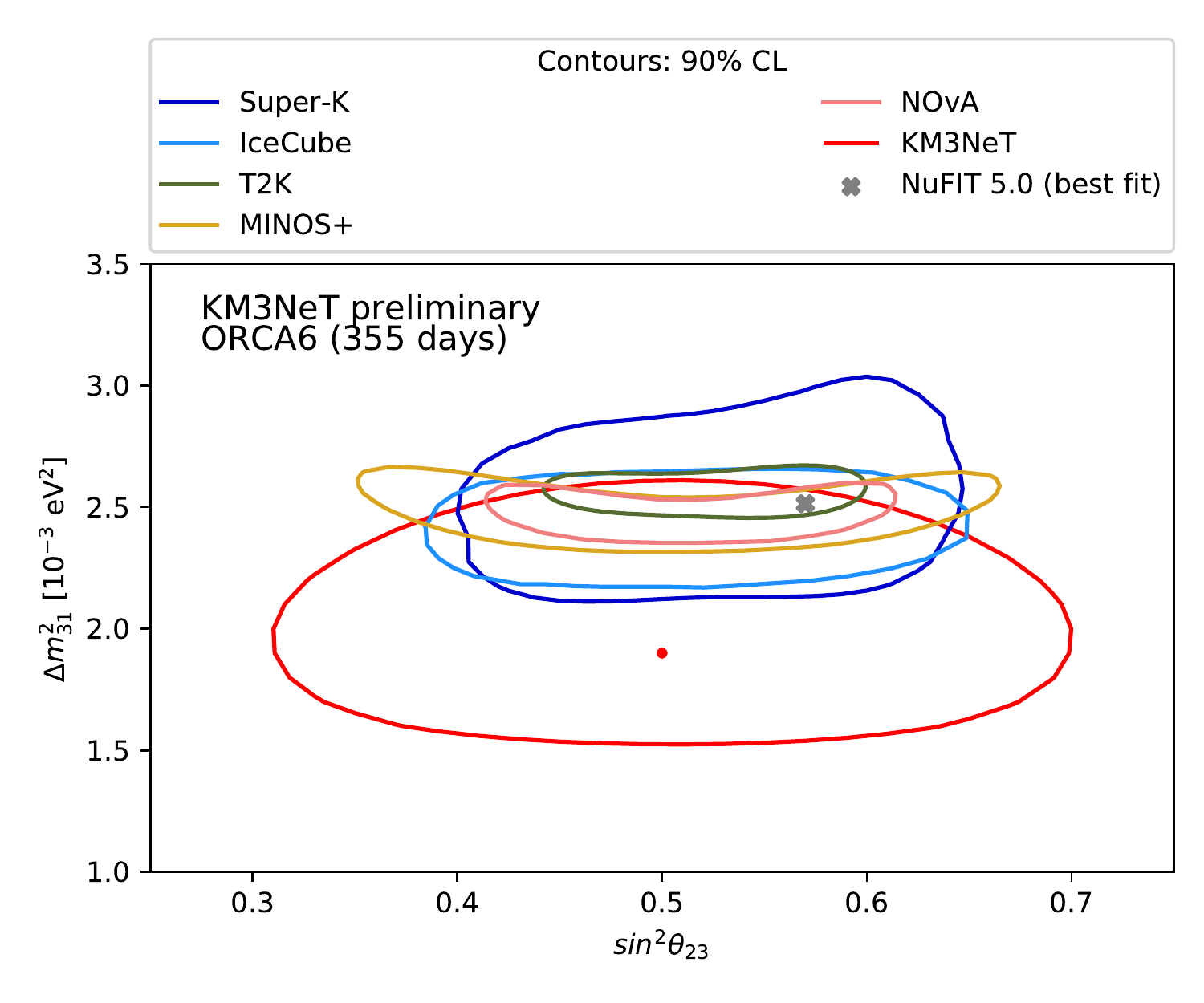}
\label{fig:orca6-cont}
\caption{\textbf{Left}: L/E distribution of the expected number of events relative to the no-oscillation hypothesis for the data collected with the KM3NeT/ORCA first six DUs (black) and the fit distribution (light-blue). The expected number of events assuming the no-oscillation hypothesis (purple) and the global best fit oscillation parameters~\cite{Esteban:2018azc} (red) are also shown for comparison.
\textbf{Right}: Contour at 90\% CL of the oscillation parameters obtained with the data collected with the KM3NeT/ORCA first six DUs (red). Contours of other experiments ~\cite{IceCube:2017lak,Super-Kamiokande:2017yvm,T2K:2019bcf,aurisano_adam_2018_1286760,NOvA:2019cyt} have been added for comparison purposes as well as the global best fit value~\cite{Esteban:2018azc}.}
\label{fig:orca oscs}
\end{figure}

Fig.~\ref{fig:orca oscs} shows the distribution of events as a function of the baseline (L) to energy (E) ratio, relative to the expectation without oscillations. A clear depletion of events is visible as L/E increases, indicating a characteristic oscillation pattern. The data excludes the no-oscillation hypothesis with a confidence level of 5.9$\sigma$. The atmospheric neutrino oscillation parameters are measured to be $\Delta{m^2_{31}} = 1.95^{+0.24}_{-0.22}\times10^{-3}\ \mbox{eV}^2$ and $\sin^2\theta_{23} = 0.50 \pm 0.10$. The allowed 90\% CL region is reported in Fig.~\ref{fig:orca oscs} and is compatible with other experiments.

\subsubsubsection{Sensitivity studies to neutrino oscillation with the complete detector}
Upon completion (115 DUs), KM3NeT/ORCA will have an instrumented mass of 7 Mton. In the following, we present results in terms of Mton-years, with typical benchmarks 7 and 21 Mton-year (i.e.~1 and 3 years of operation of the full detector). These results are based on a recent study~\cite{KM3NeT:2021ozk} reporting the sensitivity to neutrino oscillation measurements updated with respect to the Letter of Intent~\cite{KM3Net:2016zxf} to include improvements in detector modeling, trigger, reconstruction and particle identification methods.

\begin{figure}[!ht]
\includegraphics[width=0.49\textwidth]{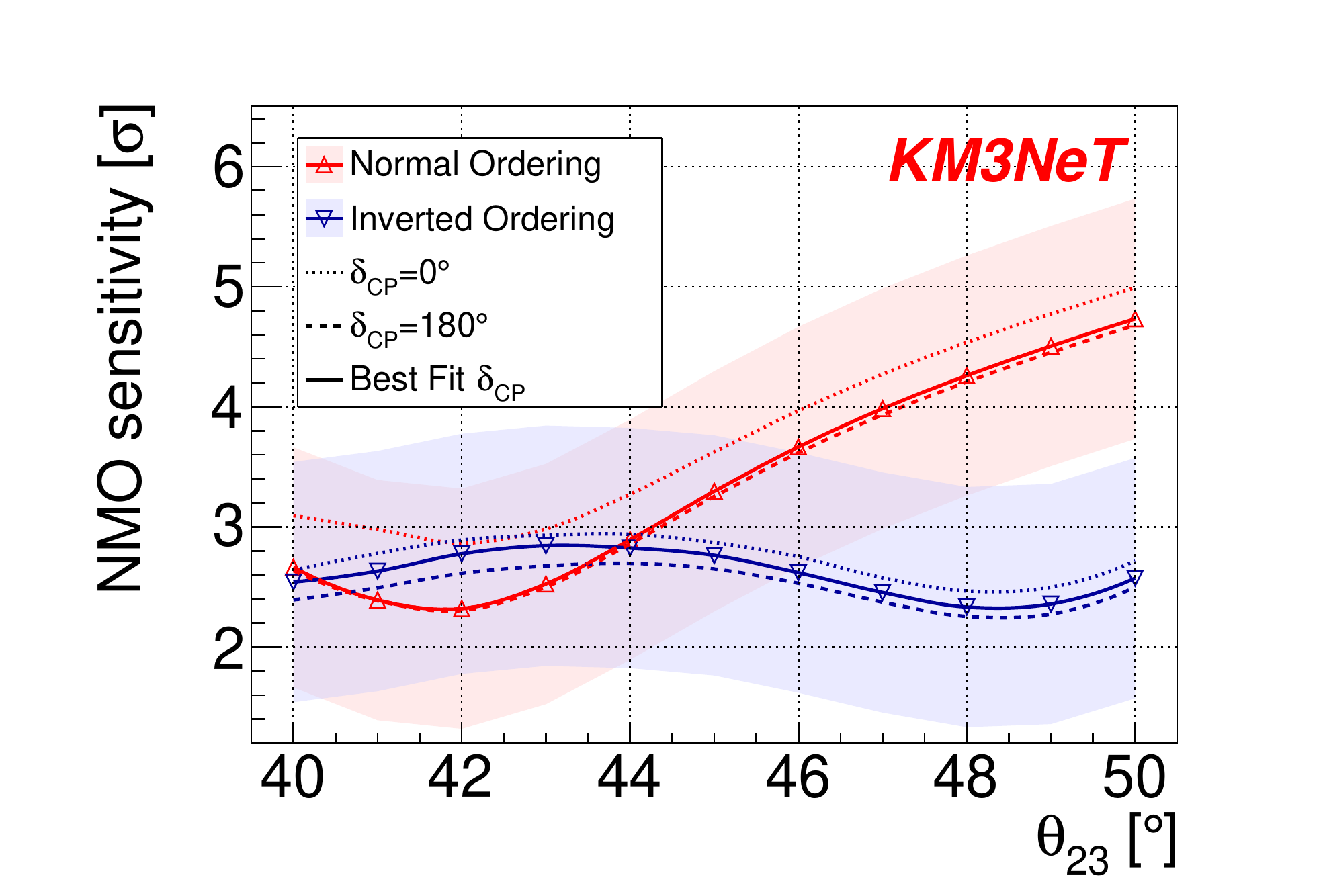}
\includegraphics[width=0.49\textwidth]{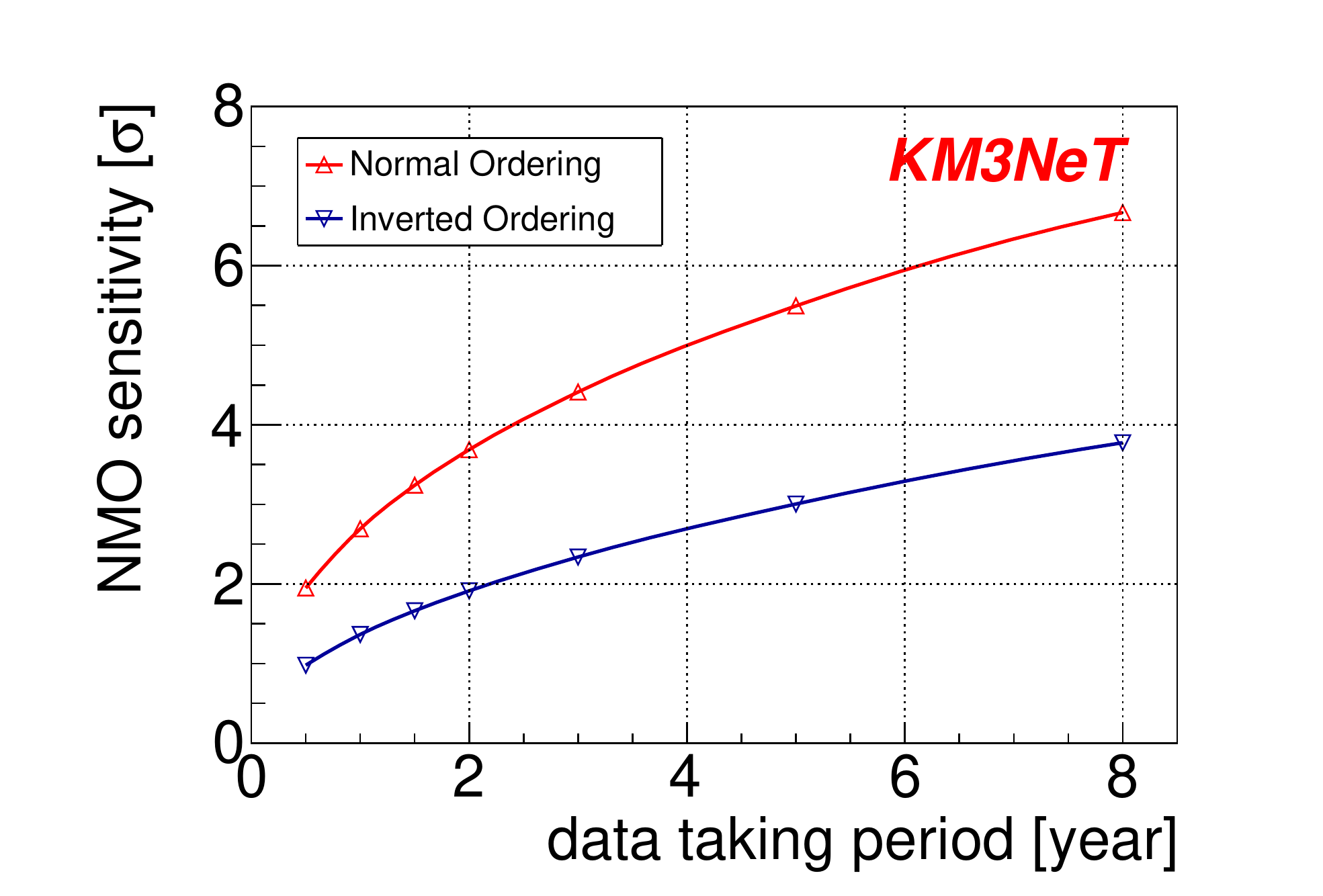}
\caption{\textbf{Left}: Sensitivity to the mass ordering for an exposure of 21~Mton-years as a function of the true $\theta_{23}$ value, for both normal (red) and inverted (blue) orderings under three assumptions for the $\delta$ value: the world best fit points for NO and IO reported in \cite{Esteban:2018azc} (solid line), \ang{0} (dotted line) or \ang{180} (dashed line).  The coloured shaded areas represent the range of significance values that would occur with 68\% probability according to the Asimov approach~\cite{Cowan:2010js}.
\textbf{Right}: Sensitivity to the mass ordering as a function of data taking time with a 7~Mton detector for both normal (red) and inverted (blue) orderings and assuming the oscillation parameters reported in \cite{Esteban:2018azc}.}
\label{fig:SensitivityNMH}
\end{figure}

\paragraph{Mass ordering}
The sensitivity to the atmospheric mass ordering with an exposure of 21~Mton-years is reported as a function of $\theta_{23}$ for both orderings in Fig.~\ref{fig:SensitivityNMH}. Assuming the best estimates for $\theta_{23}$ from \cite{Esteban:2018azc}, the mass ordering sensitivity is 4.4$\sigma$ if the true ordering is normal and 2.3$\sigma$ if it is inverted.
Fig.~\ref{fig:SensitivityNMH} shows the sensitivity for both orderings as a function of data taking time with a complete detector. The mass ordering can be determined at 3$\sigma$ level with 9~Mton-years exposure if the true ordering is normal, and with 35~Mton-years exposure if it is inverted.

\begin{figure}[!ht]
\includegraphics[width=0.49\textwidth]{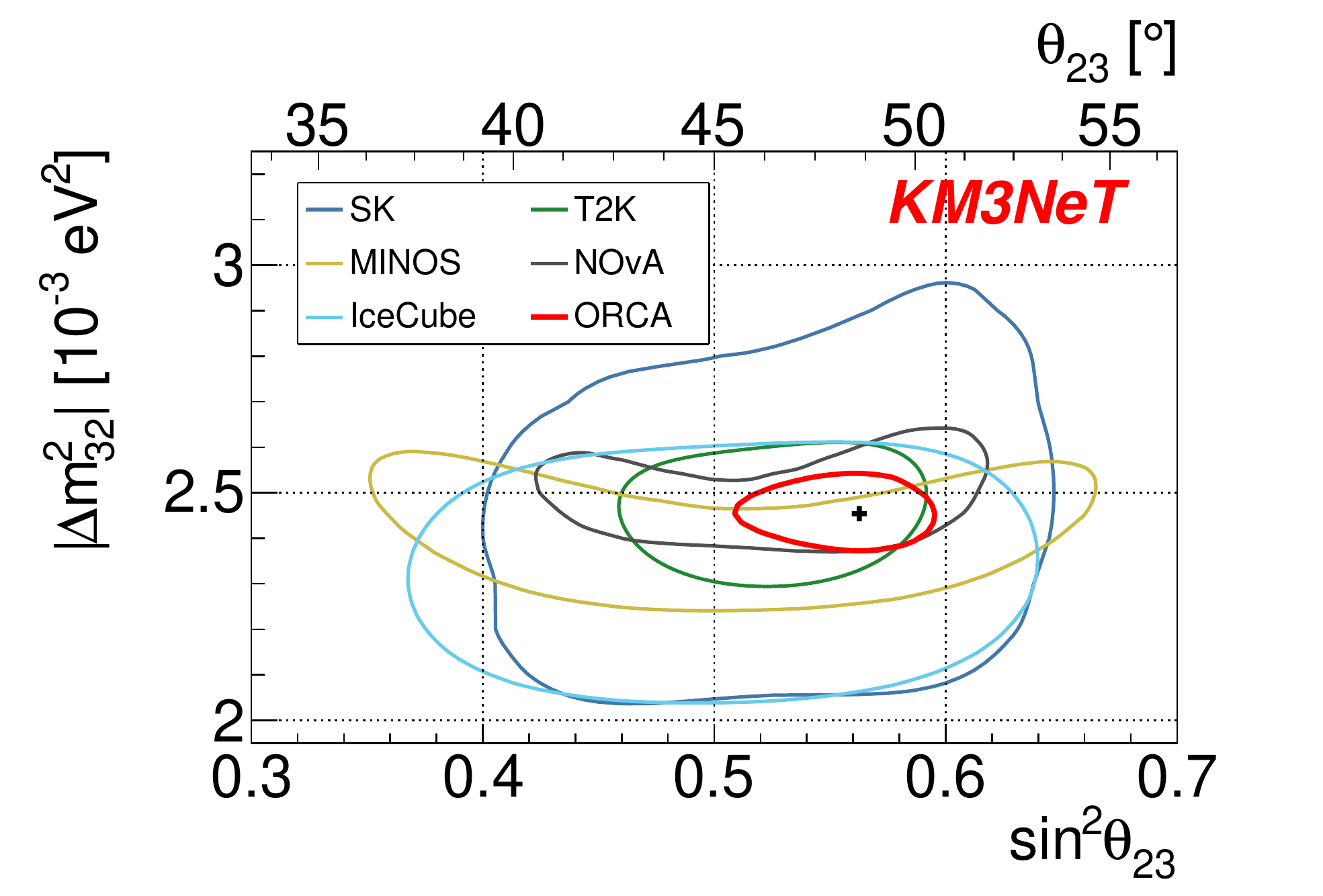}
\includegraphics[width=0.49\textwidth]{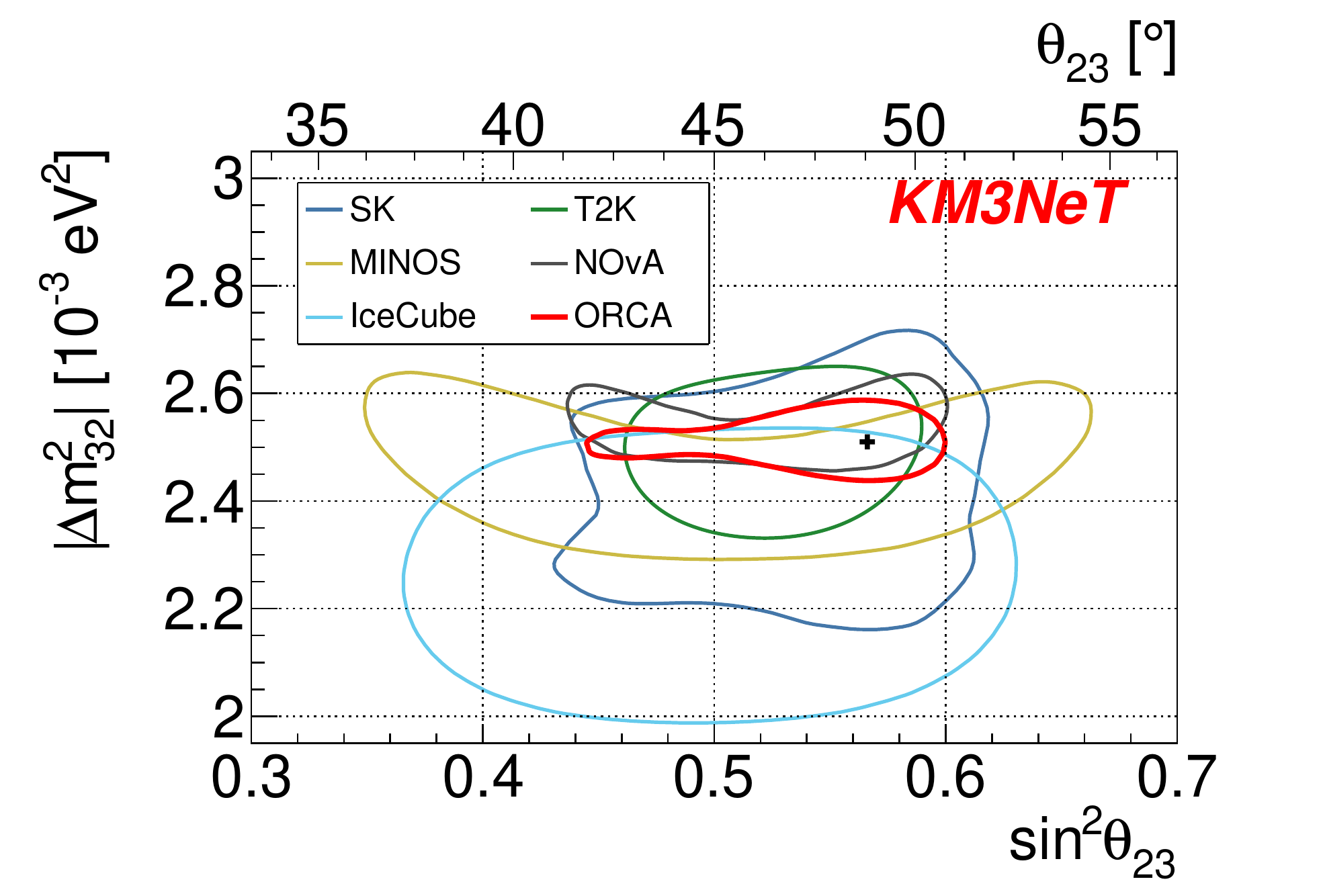}
\caption{Expected precision of $\Delta m^2_{32}$ and $\theta_{23}$ for both NO (left) and IO (right) with 3 years of KM3NeT/ORCA data taking with the complete detector (21~Mton-years) at 90\% confidence level (red) overlaid with existing results from other experiments~\cite{IceCube:2017lak,Super-Kamiokande:2017yvm,T2K:2019bcf,aurisano_adam_2018_1286760,NOvA:2019cyt} and the oscillation parameters reported in \cite{Esteban:2018azc} (black cross).}
\label{fig:contour}
\end{figure}

\paragraph{Oscillation Parameter Measurements}
We also present the expected sensitivity of KM3NeT/ORCA to the atmospheric oscillation parameters. Fig.~\ref{fig:contour} shows the expected contours in the $\Delta m^2_{32}-\sin^2\theta_{23}$ plane and illustrates the precision that KM3NeT/ORCA will be able to provide with 21~Mton-years exposure compared to the current knowledge.

\paragraph{$\nu_\tau$ appearance}
KM3NeT/ORCA will also be able to detect $\nu_\tau$ appearance at $>5\sigma$ within one year of running which will also contribute to the oscillation results.

\subsubsection{Technical Status and Schedule}
\label{sec:pros}

Since November 2021, ten DUs are operational at the KM3NeT/ORCA site. Before that, a configuration with 6 DUs was operated for 22 months continuously, between January 2020 and November 2021, corresponding to a total exposure of approximately 0.6~Mton-years. As of February 2022, a total exposure of approximately 0.8~Mton-years has been collected.

The experience gained in assembling and deploying the first ten KM3NeT/ORCA DUs indicates the full ORCA detector (115 DUs) can be completed by the end of 2025, as long as no limitations are imposed due to availability of funding. Since data taking proceeds during construction, under this schedule a total of 14.5 Mton-years would be accumulated by the time the detector is complete, assuming a constant DU production rate and a data-taking efficiency of 95\%, in line with the performance achieved with the first 6 DUs.

At the best-fit oscillation parameters reported in \cite{Esteban:2018azc}, a 3$\sigma$ measurement would then be possible by 2026 in the NO and by 2030 in the IO. In the NO scenario, a 5$\sigma$ determination could be achieved by 2028 if $\theta_{23}>\ang{48}$. These projections assume the detector instrumented mass grows linearly with the number of operational DUs.

\subsection{Expected Sensitivity Milestones}
% blah Mass ordering: JUNO+NOvA+T2K: 2008.11280

\begin{figure}
    \centering
    \includegraphics[width=0.85\textwidth]{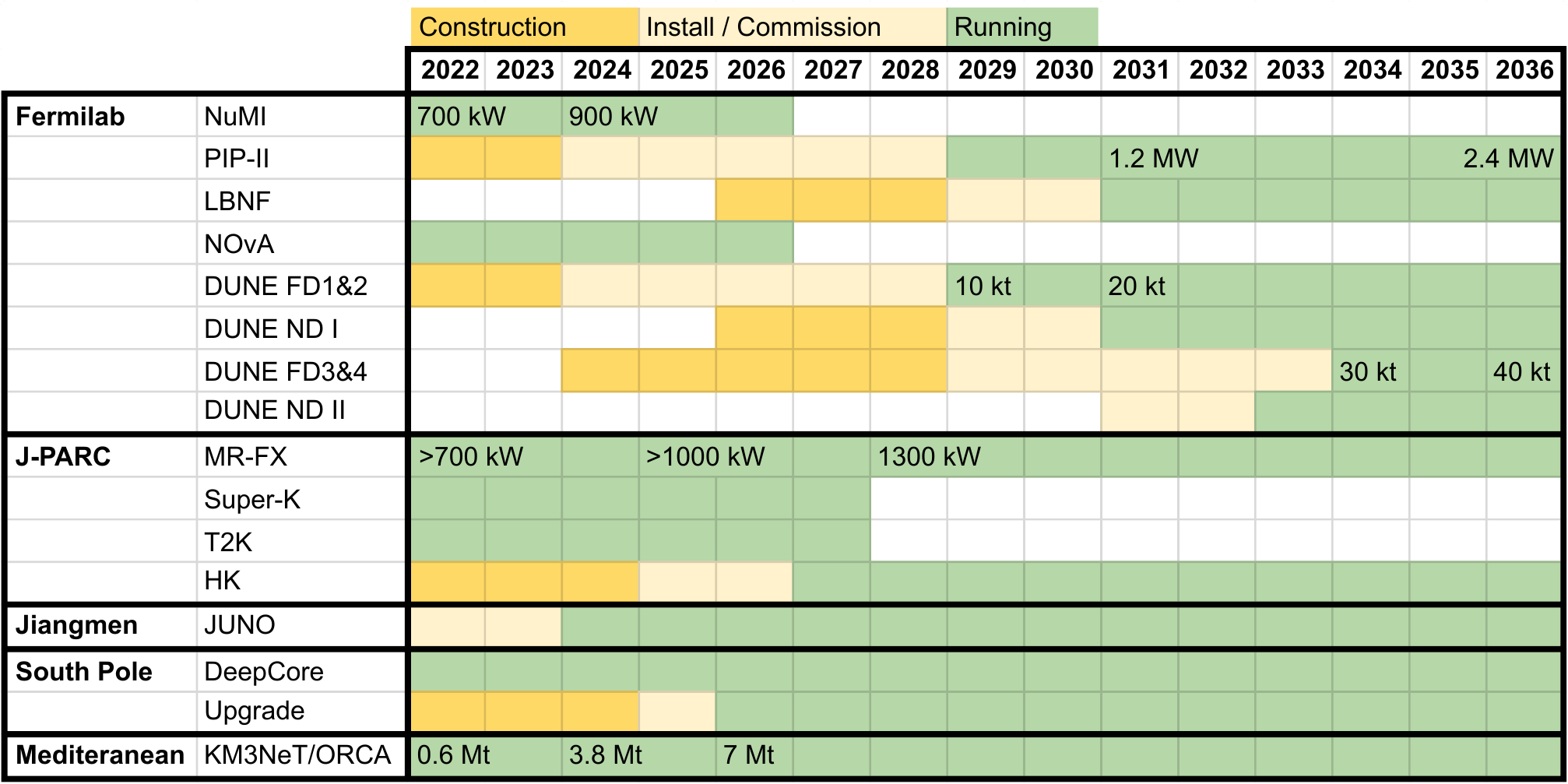}
    \caption{Assumptions about the construction timeline for future neutrino projects.}
    \label{fig:timeline}
\end{figure}

Fig.~\ref{fig:timeline} shows the current best estimate of the construction and data taking timelines of experiments in the current program.

We estimate the year at which experiments will reach major milestones ($3\sigma$ and $5\sigma$ or the best achievable) for the three known unknowns in neutrino oscillations in table \ref{tab:milestones}.
Note that the results show optimistic and conservative results which take into account both staging (detector construction, accelerator construction, and reactor power construction) as well as the oscillation parameters, some of which can significantly enhance the sensitivities.

\begin{table}
\centering
\caption{The year individual next-generation experiments are currently expected to reach key milestones for the three known unknowns in neutrino oscillations.
Optimistic vs.~conservative includes both staging effects as well as effects from the true oscillation parameters.
Combined fits will be significantly more powerful than the naive sum, especially in the case of the atmospheric mass ordering.
Experiments estimate their sensitivities somewhat differently.
Additional time for analysis is not included.}
\begin{tabular}{c|c|c|c|c}
 & & Atmospheric Mass Ordering & $\theta_{23}$ Octant & $\sin\delta\neq0$ for 50\% of $\delta$ \\\hline
\multirow{2}{*}{JUNO} & Optimistic & 2030: $3\sigma$ & - & -\\
 & Conservative & 2030: $2.5\sigma$ & - & -\\\hline
\multirow{2}{*}{DUNE} & Optimistic & 2030: $5\sigma$ & 2036: $3\sigma$, 2040: $5\sigma$ & 2035: $3\sigma$, 2039: $5\sigma$\\
 & Conservative & 2032: $5\sigma$ & 2040: $2\sigma$ & 2037: $3\sigma$\\\hline
\multirow{2}{*}{HK} & Optimistic & 2033: $5\sigma$ & 2033: $5\sigma$ & 2029: $3\sigma$, 2032: $5\sigma$\\
 & Conservative & 2032: $3\sigma$ & 2034: $3\sigma$ & 2029: $3\sigma$, 2037: $5\sigma$\\\hline
\multirow{2}{*}{IceCube} & Optimistic & 2030: $3\sigma$, 2033: $4\sigma$ & - & -\\
 & Conservative & 2033: $2\sigma$ & - & -\\\hline
\multirow{2}{*}{KM3NeT} & Optimistic & 2026: $3\sigma$, 2029 $5\sigma$ & - & -\\
 & Conservative & 2030: $3\sigma$, 2032: $4\sigma$ & - & -
\end{tabular}
\label{tab:milestones}
\end{table}

The choice of assumed true oscillation parameters can affect the sensitivity quite a bit.
The KM3NeT \cite{KM3NeT:2021ozk} and IceCube \cite{IceCube-Gen2:2019fet} mass ordering sensitivities both assume the best fit oscillation parameters from \cite{Esteban:2018azc}, notably the upper octant while we note that one of the three global fits, \cite{Capozzi:2021fjo}, prefers the lower octant which would significantly reduce their sensitivity.
The HK sensitivity to the mass ordering \cite{Hyper-Kamiokande:2018ofw} includes atmospheric and accelerator neutrinos, is largely independent of the true mass ordering, and is better in the upper octant than the lower.
For the octant fiducial values of $\sin^2\theta_{23}=0.45,0.55$ are taken; the impact of $\delta$ is small and median values are used.
For CPV improved systematics of $\nu_e$ cross section uncertainties are assumed, along with the upper octant and the normal mass ordering both of which are assumed to be known.
A start date of 2027 is assumed for HK.
The DUNE sensitivity to $\delta$ \cite{DUNE:2021mtg} assumes that the accelerator starts at 1.2 MW and only the currently funded baseline program, while the optimistic numbers show the improvement if the full scope is achieved after six years.
The DUNE sensitivity to the mass ordering also assumes 1.2 MW initially, that the true ordering is inverted, and is valid for any value of $\delta$.

Finally, we want to draw special attention to the fact that combined fits, either by the experiments or by theorists, can perform significantly better than the naive comparison.
Comparing $\nu_e\to\nu_e$ measurements with $\nu_\mu\to\nu_\mu$ measurements provides a powerful way to probe the atmospheric mass ordering since the dominant $\Delta m^2$ measured in each differs in different directions depending on the mass ordering \cite{Nunokawa:2005nx}.
The details of how, when, and which experiments are to be combined is subtle and will depend on exactly when each experiment has reported data in the coming years.
Further, it should be noted that as precision in the knowledge of oscillation parameters improves, searches for additional physics may likely require measurements over a wide energy range and at the longest baselines; this is, e.g.~the approach of the DUNE experiment.

\section{Three Flavor Oscillation Supporting Program}
\subsection{Experiments}
Auxiliary measurements, particularly hadron production measurements to constrain neutrino fluxes and flux systematic errors, as well as precision measurements of cross sections to constrain interaction systematic errors, are necessary for precision three-flavor oscillation measurements.  Reports from NF06 (Neutrino Interaction Cross Sections) \cite{Balantekin:2022jrq} and NF09 (Artificial Neutrino Sources) \cite{Fields:2022pxk} detail some of these essential measurements.

The CERN Neutrino Platform is another important supporting program towards precision three-flavor oscillation measurements.  From the CERN website: ``The CERN Neutrino Platform is CERN’s undertaking to foster and contribute to fundamental research in neutrino physics at particle accelerators worldwide, as recommended by the 2013 European Strategy for Particle Physics. It includes the provision of a facility at CERN to allow the global community of neutrino experts to develop and prototype the next generation of neutrino detectors. The CERN Neutrino Platform is CERN’s main contribution to a globally coordinated programme of neutrino research.''

\subsection{Oscillation Theory}
\label{sec:osc theory support}
Understanding the physics behind three flavor oscillation has taken a convoluted path with missteps on the way.
The role of resonances in solar neutrinos is now understood \cite{Mikheev:1986gs} as is the probability to jump from one eigenstate to another \cite{Parke:1986jy,Bethe:1986ej,Rosen:1986jy}.
The three-flavor oscillation probability in sufficiently uniform matter density, relevant in the Earth for $\nu_e$ appearance experiments, was initially thought to be fairly intractable \cite{Barger:1980tf} but has since been increasingly well understood \cite{Zaglauer:1988gz,Kimura:2002wd,Barenboim:2019pfp,Denton:2019ovn}.
Nonetheless, in order to perform statistically robust estimates of confidence intervals \cite{Feldman:1997qc} a very large number of pseudo-experiments must be thrown and calculating the oscillation probabilities for each set of the oscillation parameters is often the limiting factor.
To this end, some initial studies of the computational speed of different expressions has been carried out in the context of long-baseline accelerator neutrino experiments \cite{Barenboim:2019pfp}.
Preliminary implementations suggest a speed up of 3-10 is realistic, suggesting hat there may be more progress possible.
Atmospheric and nighttime solar neutrinos are even more computationally expensive due to both the range of zenith angles (baselines) as well as the shell structure of the Earth.
As HK, DUNE, IceCube, and KM3NeT look to considerably increase the precision of these channels, additional improvements in computational methods would also assist these calculations.

\subsection{Joint Fits}
Joint fits of data from different experiments give enhanced sensitvities to oscillation parameters by breaking degeneracies.  A clear example can be found in Fig.~\ref{fig:IC JUNO}, where a combination of JUNO and IceCube data under a joint oscillation parameter hypothesis, beyond a simple statistical sum, gives a clear enhancement to the mass ordering sensitivity.

When experiments use similar neutrino sources, energy scales, detectors, etc., it is extremely important to correctly take into account correlations between systematic errors.  Especially, incorrect/inconsistent treatment of the neutrino interaction model has a significant impact on resulting joint fits.

Producing joint fits in a timely manner is important for quickly realizing measurements of various oscillation parameters.  Even in the case of highly different energy scales, clarity in how data inputs are used and how fits are performed is important.

Several currently running experiments are working on dedicated joint fits including consistent treatment of inputs and correct treatment of correlations between systematics.  In other cases, joint fits by phenomenologists are proposed.

\subsubsection{T2K + SK}
An MOU between the T2K and SK collaborations has been established and a joint working group has been formed to produce joint fits of T2K long-baseline and SK atmospheric neutrino oscillation data.  These joint fits use a unified set of inputs and properly take into account correlations in systematics between the two experiments.  In particular, significant correlations between experiments are expected from the neutrino interaction model, and these correlations  are especially important given that the two experiments use the same detector.  

With a joint fit, sensitivity to oscillation parameters, in particular the mass ordering, \(\theta_{23}\), and \(\delta_{CP}\), is enhanced beyond the naive sum of \(\chi^2\) values published by the two experiments separately.  First joint fit results are expected to be released publicly in 2022.

\subsubsection{T2K + NOvA}
An MOU between the T2K and NOvA collaborations has been established and a joint working group has been formed to produce joint fits of T2K and NOvA oscillation data.  Combinations of experimental data from different baselines and neutrino energies resolves degeneracies in oscillation parameters in a complementary way, which provides improved sensitivity to the physics and/or physics reach outside the parameter space accessible by each experiment. In particular, a joint fit of T2K and NOvA is sensitive to the (unknown) mass ordering, \(\theta_{23}\), \(\theta_{13}\), and \(\delta_{CP}\).  As significant correlations between experiments are expected from the neutrino interaction model, direct collaboration between the experiments is necessary.

\subsubsection{JUNO + NOvA + T2K}
A joint fit of JUNO + NOvA + T2K data could provide a first \(\geq5\sigma\) measurement of the mass ordering, as proposed by phenomenologists in \cite{Cabrera:2020own}.

\subsubsection{DUNE + HK}
There is interest in exploiting the complementarity of the HK and DUNE experiments. This complementarity is similar to the one of the T2K and the NOvA experiments: different baselines, beam energies (narrow-band vs.~wide-band), detector technology, detection mechanism and detector size.  A joint fit of the data from both experiments including common inputs and correct correlations would be beneficial following several years of data-taking by both experiments.  Setting up analysis frameworks with this in mind now would greatly benefit the future generation of analyzers.
It is imperative that the lessons learned in the NOvA+T2K working group extends to each experiment's successor.

\subsubsection{JUNO + ORCA}
\begin{figure}[!ht]
\centering
\includegraphics[width=0.49\textwidth]{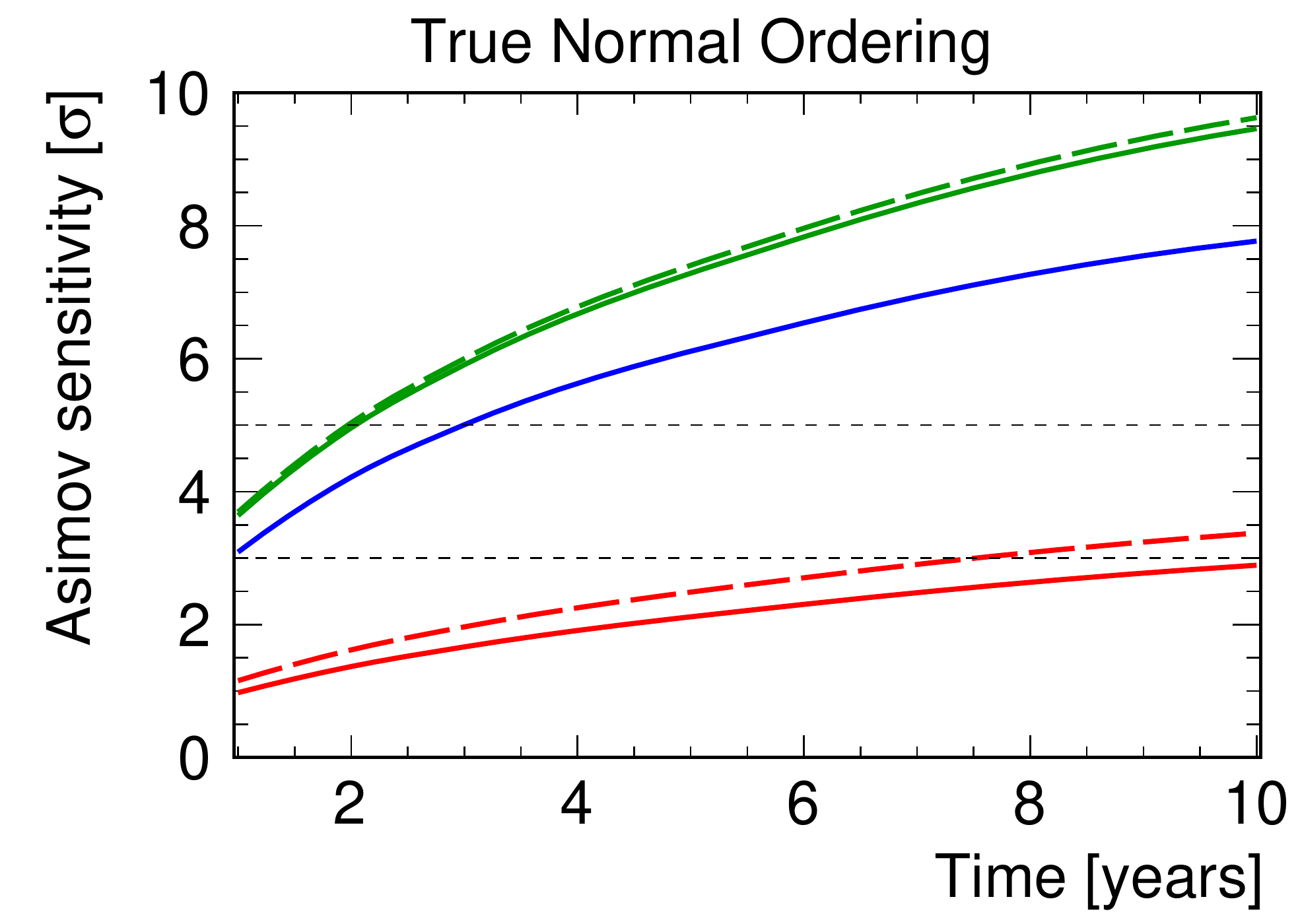}
\includegraphics[width=0.49\textwidth]{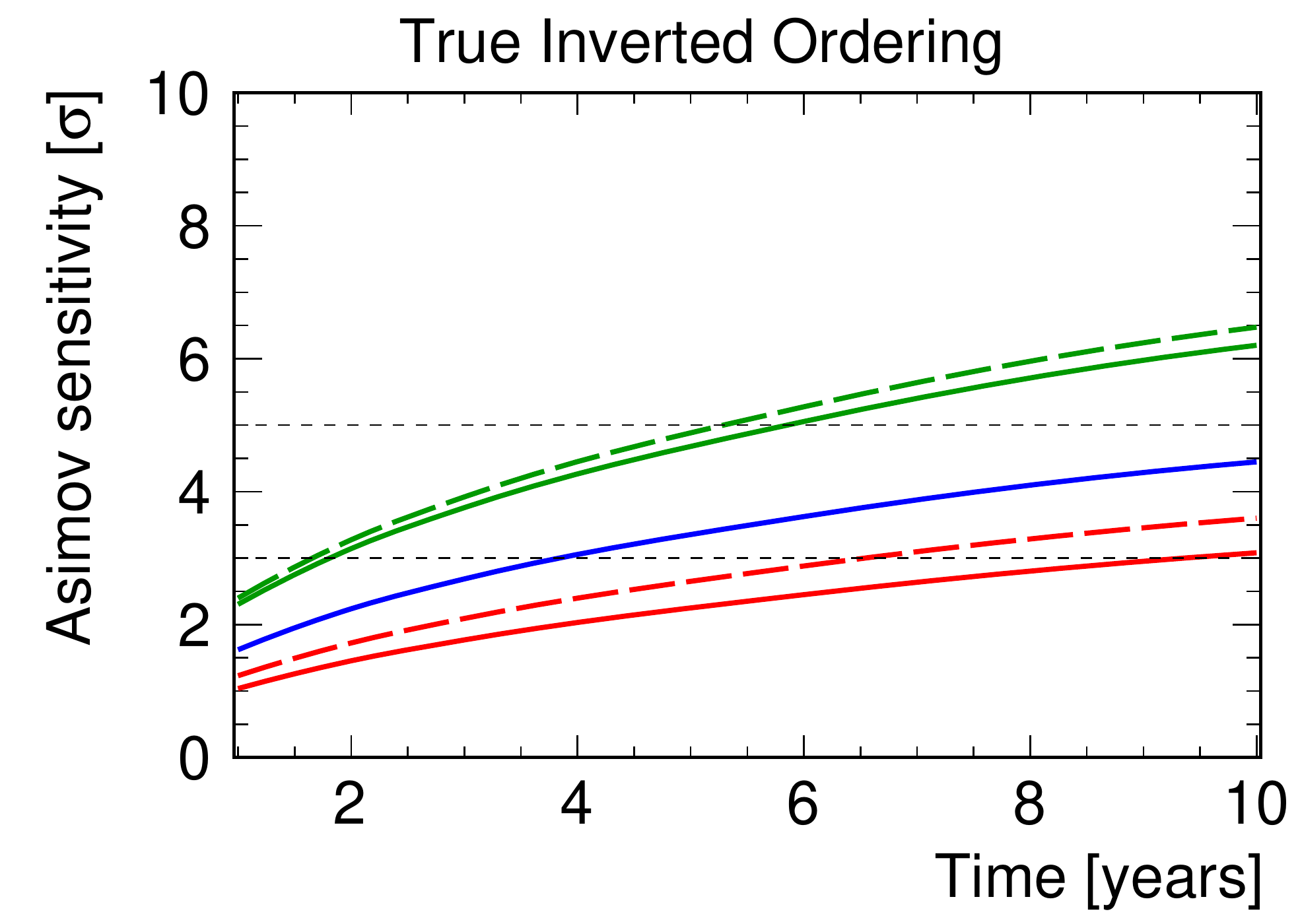}
\label{fig:JUNO-8_10cores_ORCA115_IO_JP}
\caption{NMO sensitivity as a function of time for only JUNO (red), only KM3NeT/ORCA (blue),
and the combination of JUNO and KM3NeT/ORCA (green), considering 2 (solid) or 4 (dashed)
Taishan NPP reactors, corresponding respectively to 8 or 10~reactor cores at 53~km from JUNO for NO (left) and IO (right).}
\label{fig:orcajuno}
\end{figure}

The sensitivity of a combined analysis of the JUNO and KM3NeT/ORCA experiments to determine the neutrino mass ordering was performed~\cite{KM3NeT:2021rkn}. This combination is particularly interesting as it significantly boosts the potential of either detector, beyond simply adding their neutrino mass ordering sensitivities, by removing a degeneracy in the determination of $\Delta m_{31}^2$ between the two experiments when assuming the wrong ordering. A $5\sigma$ determination of the neutrino mass ordering is expected after 6~years of joint data taking for any value of the oscillation parameters. As shown in Fig.~\ref{fig:orcajuno}, this sensitivity would be achieved after only 2 years of joint data taking assuming the current global best-fit values for those parameters for normal ordering.

\subsubsection{IceCube/DeepCore + JUNO}

\begin{figure}[!ht]
\centering
\includegraphics[trim={0cm 0 0 0.2cm},clip,width=0.39\textwidth]{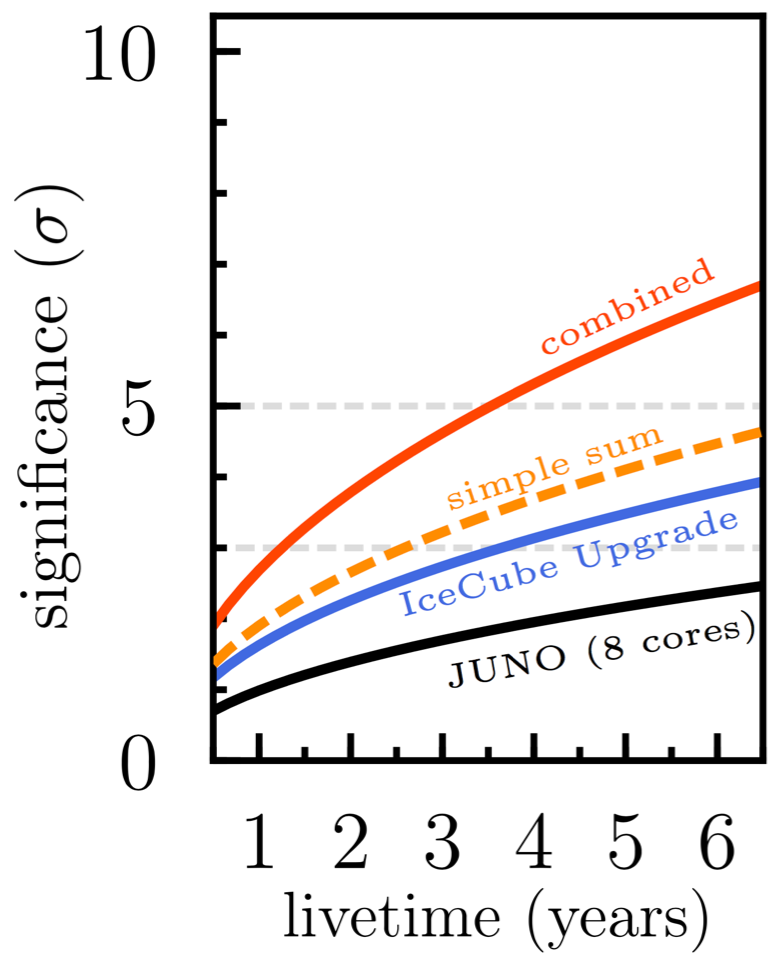}
\caption{The sensitivity to the mass ordering for the IceCube Upgrade and JUNO, both individually and in a combined analysis~\cite{IceCube-Gen2:2019fet}.}
\label{fig:IC JUNO}
\end{figure}

Similarly to ORCA + JUNO, IceCube/DeepCore data can also be advantageously combined with JUNO.
This can be seen in the  mass ordering analyses with both the IceCube Upgrade and with future JUNO data~\cite{Blennow:2013vta}, which is sensitive to the mass ordering via sub-leading survival probability terms arising from the difference between $\Delta m^2_{31}$ and $\Delta m^2_{32}$ for $\bar{\nu}_e$ \cite{Petcov:2001sy}.
When analyzing IceCube Upgrade data in combination with JUNO data under a joint oscillation parameter hypothesis, the sensitivity to the mass ordering improves significantly beyond the pure statistical combination of independent measurements~\cite{IceCube-Gen2:2019fet}, primarily due to the correlation of $\Delta m^2_{31}$ with the mass ordering hypothesis in both experiments. Such a combined analysis has an expected $6.5\sigma (5.1\sigma)$ mass ordering sensitivity for the normal (inverted) ordering after 6 years of data taking of each fully constructed and powered experiment, giving the possibility of a conclusive determination of the mass ordering in the early 2030s shown in Fig.~\ref{fig:IC JUNO}.
Similar results would be expected for the combination of any measurements of $\Delta m^2_{31}$ from $\nu_e$ disappearance and $\nu_\mu$ disappearance.

\FloatBarrier
\section{Possible Upgrades to Planned Experiments}
\label{sec:upgrades}

\subsection{DUNE}
\subsubsection{Realizing the P5 DUNE}

\begin{figure}[!htb]
\begin{centering}
\includegraphics[width=7in]{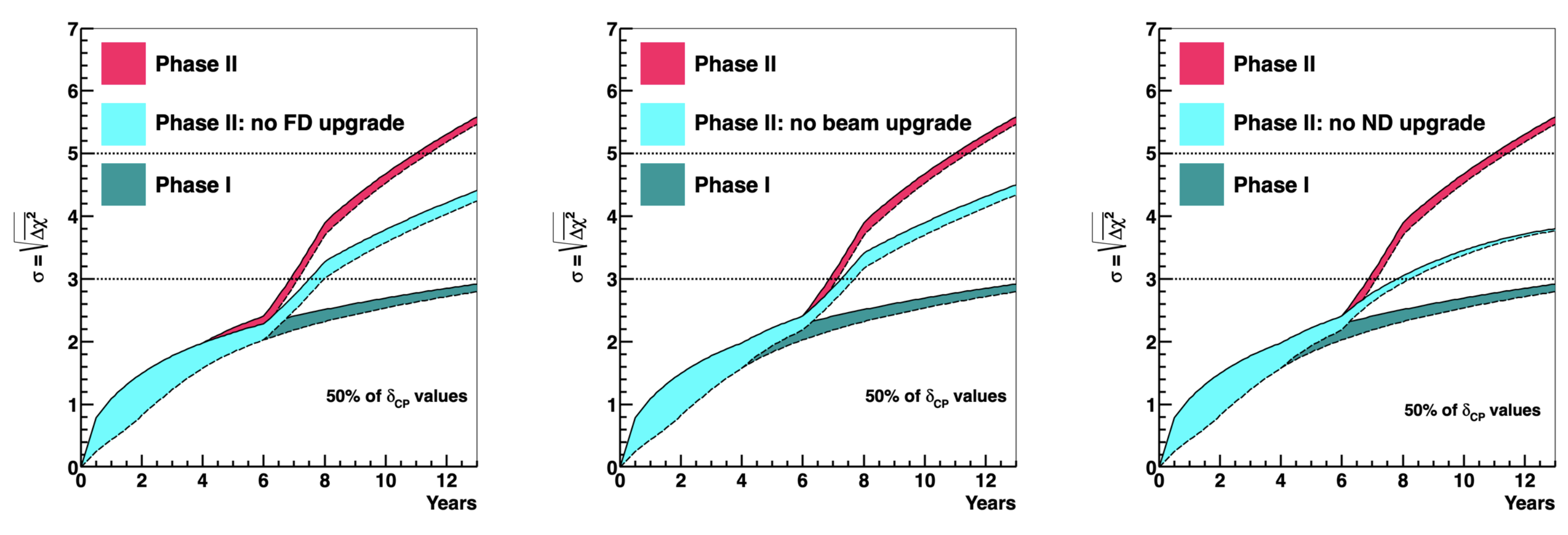}
\caption{
The significance in units of $\sigma$ with which the DUNE experiment
could observe CP violation for at least 50\% of the possible
$\delta_{\rm CP}$ values as a function of running time.
In each of the three panels the lower green curve shows the
evolution of the DUNE Phase-I experiment and the upper red
curve shows the evolution for the DUNE Phase-II experiment.
The intermediate cyan curves forego one of the three 
pieces of the Phase-II program; 
modules 3 and 4 at the far detector on the left, 
the 2.4 MW beam power upgrade in the center,
and the near detector upgrades on the right.
\label{fig:dune-phase2-reach}
}
\end{centering}
\end{figure}

DUNE Phase I consists of two far detector modules, 
totaling 35 kton, a beam 
capable of 1.2~MW of proton delivery, 
and a suite of near detectors adequate for the initial physics goals
of resolving the mass ordering and beginning the search for
CP violation with 3$\sigma$ sensitivity. To fully realize the
neutrino program imagined by the 2014 P5 panel, DUNE Phase II will
upgrade these three pieces of the experiment~\cite{DUNE:2022aul}. 
Two additional
modules will be added to the far detector, the beam intensity will 
be upgraded to 2.4~MW, and the near detector will be augmented
to gain additional precision in neutrino cross-section measurements
with limit the neutrino oscillation measurements. As illustrated
in Fig.~\ref{fig:dune-phase2-reach}, each of these three
improvements are required to achieve 5$\sigma$ sensitivity to 
CP violation and to realize the full oscillation 
measurement program.

\subsubsection{Near Detector}
The Phase-I DUNE near detector 
will have a liquid argon TPC system with a muon spectrometer, ND-LAr + TMS. 
These sub-detectors will  move off axis as part of PRISM to constrain flux and 
cross-section uncertainties. These components are sufficient
for the DUNE Phase-I goals to resolve the neutrino mass hierarchy
and to reach 3$\sigma$ sensitivity to CP violation. However, to reach
5$\sigma$ sensitivity across a wide range of $\delta$ values, improved 
knowledge of neutrino-Ar interactions
will be required. In DUNE Phase-II, TMS is planned to be replaced
with a magnetized high-pressure gaseous argon TPC. This 
detector, ``ND-GAr''~\cite{DUNE:2022yni}, 
will enable a precise measurement of the momenta
of muons produces in the ND-LAr and will additionally allow for high 
resolution reconstruction of the vertex activity produced by 
neutrino interactions on the low-density gaseous argon target.
The improvement in neutrino cross-section knowledge 
will enable the search for CP violation to be extended to 5$\sigma$.

\subsubsection{Third and Fourth Modules}
During DUNE Phase-I two drifts sized to house
the four DUNE liquid argon modules will be excavated and 
two modules will be installed; one per drift which allows for
outfitting, installation, and operations to proceed in
parallel. DUNE requires two additional detector modules
to realize 5$\sigma$ sensitivity to CP violation, 
to search for physics beyond the standard model in neutrino
oscillations, and to fully realize the non-accelerator 
physics measurement program of solar neutrinos, 
proton decay, and atmospheric neutrinos.
Modules three and four will need to be at least as capable 
of executing the DUNE accelerator physics program as 
modules one and two but need not be identical to these initial modules
and could implement improved technologies
or modifications to the detector design targeted at specific topics
in the DUNE physics program.
Possibilities include several ideas for novel 
TPC readout technologies~\cite{Lowe:2020wiq,Para:2022gju,Dwyer:2018phu,Kubota:2022mtx},
introduction of Xe for neutrinoless double beta decay~\cite{Mastbaum:2022rhw}, and
deployment of liquid scintillator~\cite{Theia:2022uyh}.
The DUNE collaboration has hosted one workshop focused on 
these ideas in 2019~\cite{MOOD} and another is planned in November 2022.

\subsubsection{Accelerator Upgrades}

The Main Injector at Fermilab recently (May 2022) set a new record
power of 0.9~MW of protons delivered to the NOvA experiment at 120~GeV. 
The PIP-II upgrades~\cite{Lebedev:2017vnu} to the Fermilab proton source
are now underway and when completed later this decade will 
enable delivery of
1.2~MW for the DUNE experiment. Beam delivery beyond 1.2~MW is limited
by the Booster ring which is entering its 50$^{\rm th}$ year of service.
Fermilab is investigating two options~\cite{Eldred:2022vxi} 
for future investments to realize the 2.4~MW beam power 
envisioned by the 
2014 P5 for DUNE. Both options increase the PIP-II linac energy to 2~GeV
and call for the construction of a new rapid-cycling synchrotron.
The ``Initial Configuration Document 2'', ICD-2~\cite{Nagaitsev:2021xzy},
option is likely the most cost effective route to 2.4~MW power for
the neutrino program.
The ``Booster Science Replacement'', BSR~\cite{Ainsworth:2021ahm},
could deliver of 2.4~MW for DUNE while also increasing the 
power available to experiments 
at 8~GeV to 750~kW over the 170~kW possible with ICD-2.

\subsection{Hyper-K}

\subsubsection{HK Near Detector}
Upgrades to the T2K near detector ND280 are underway now, as discussed in Sec.~\ref{sec:t2k}.  Measurements made by the upgraded detector will be used to understand the detector performance in the coming few years.  Further detector upgrades or configuration changes towards dedicated precision measurements for the HK experiment are now under consideration, and near-future ``ND280 Upgrade'' measurements will inform design decisions for a future HK ``ND280++'' project.

\subsubsection{HK-Gd}
As discussed in Sec.~\ref{sec:sk}, the SK detector has recently been upgraded to SK-Gd, giving SK enhanced access to various new physics processes.  Similarly, Gd doping of the HK detector is under consideration as a possible upgrade of the HK detector.  

The main motivator towards HK-Gd is a significant enhancement in the sensitivity to Diffuse Supernova Neutrino Backgrounds (DSNB), since HK-Gd will be able to significantly reduce spallation backgrounds at \(E < 20\)~MeV, measure the DSNB spectrum at \(E > 10\)~MeV with 100s of events, and study the supernova rate down to \(Z\sim1\), see e.g.~\cite{Suliga:2022ica} for a recent DSNB review.  HK-Gd could also offer direct benefits to the HK atmospheric and long-baseline measurements by allowing for enhanced neutrino-antineutrino separation and improving the detector energy reconstruction via neutron counting.  These improvements can directly impact the HK precision of three-flavor oscillations.

\subsubsection{Detector in Korea}
A second HK tank located at the Mt. Bisul site in Korea, \(\sim\)1100~km away and 1.3\(\degree\) off-axis from the J-PARC neutrino beam source, is proposed to serve as a far detector for T2HKK (Tokai to Hyper-K Korea). This second tank would enhance HK sensitivities by doubling the detector volume, but would also give access to the second oscillation maximum for long-baseline neutrinos, allowing for significantly enhanced sensitivity to oscillation parameters beyond a simple increase in statistics \cite{Hyper-Kamiokande:2016srs}.  Differences in tank geometry, off-axis angle, and overburden between the detectors in Japan and Korea would further break degeneracies and enhance sensitivities when fitting data from both detectors.  

Design of the Korea Neutrino Observatory (KNO), which will serve as the T2HKK far detector tank, is now underway. %with plans to begin construction soon.

\subsubsection{J-PARC Accelerator Upgrades}
Currently planned upgrades to the J-PARC MR are aiming to achieve 1.3~MW at this point, but further future upgrades beyond that could achieve higher beam powers.  For example, installation of a new 8 GeV booster ring at J-PARC for injection into the MR could allow the MR beam power to be increased beyond 3.2~MW.

Several main components of the J-PARC neutrino beamline infrastructure (secondary beamline shielding, decay volume, and hadron absorber) are no longer human accessible due to irradiation, and will never be accessed again.  Those components were all designed to be able to accept a 3\(\sim\)4~MW proton beam from the J-PARC MR.

Construction of a new Continuous Wave (CW), 10~MW beam has also been considered at the KEK site in Tsukuba, Ibaraki prefecture, Japan.  This new accelerator would use a series of 4 CW linear accelerators installed in the tunnel currently in use for KEKB.  One major challenge for this design would be development of a CW horn to focus secondary particles produced by a CW beam.  Another challenge is that neutrino beamline and near detector infrastructure do not exist at the KEK site now, meaning that construction of a new neutrino beamline and near detector suite would be necessary.

\FloatBarrier
\section{Other Probes of the Oscillation Parameters}
\label{sec:other}
Beyond the traditional probes there are various other ways to learn about the oscillation parameters in astrophysical and nuclear environments.

\subsection{Galactic Supernova}
When the next supernova goes off in our galaxy, a wealth of information will be immediately acquired about astrophysics, astroparticle physics, and particle physics.
Since the mass ordering affects the flavor transitions in the supernova, a measurement of the neutrino signal will provide crucial evidence about the mass ordering; either as the first measurement or as a confirmation of terrestrial results.
For a typical supernova distance of $\sim10$ kpc one can expect that one of the current or upcoming experiments will have between $2-6\sigma$ sensitivity to the mass ordering subject to possibly additional systematic uncertainties \cite{DeSalas:2018rby}.
Note that the dependence of the diffuse supernova neutrino background on the mass ordering is likely too small to measure even with many years of HK \cite{Moller:2018kpn}.

\subsection{Astrophysical Neutrinos}
IceCube has detected high energy astrophysical neutrinos \cite{IceCube:2013low} from beyond our galaxy \cite{Denton:2017csz,IceCube:2017trr}.
These neutrinos are expected to be some combination of $\nu_e$ and $\nu_\mu$ at the source and by the time they reach the Earth they have decohered with probabilities $P_{\alpha\beta}=\sum_{i=1}^3|U_{\alpha i}|^2|U_{\beta i}|^2$, thus measurements of the flavor ratio detected at the Earth provides information about the mixing matrix.
Existing constraints are not sensitive \cite{IceCube:2015gsk,Stachurska:2019srh} but it is conceivable that with IceCube's upgrade they could provide some information on the mixing parameters \cite{IceCube-Gen2:2020qha}.

\subsection{Absolute Mass-Scale Based Probes}
There are various other means of probing the six oscillation parameters.
In particular, measurements of the absolute mass scale can provide information about the mass ordering in some cases.
\begin{itemize}
\item \textbf{Kinematic end-point} experiments such as KATRIN, ECHo, HOLMES, and Project-8 \cite{KATRIN:2019yun,Gastaldo:2017edk,Alpert:2014lfa,Project8:2017nal} are sensitive to $\sqrt{\sum_i|U_{ei}|^2m_i^2}$ which is $\gtrsim10$ meV in the NO and $\gtrsim50$ meV in the IO, although these experiments may not have sensitivity to the mass ordering before oscillation experiments do.
\item \textbf{Cosmological} measurements of the cosmic microwave background temperature and polarization information, baryon acoustic oscillations, and local distance ladder measurements lead to an estimate that $\sum_im_i<90$ meV at 90\% CL which mildly disfavors the inverted ordering over the normal ordering \cite{DiValentino:2021hoh} since $\sum_im_i\gtrsim60$ meV in the NO and $\gtrsim110$ meV in the IO; although these results depend on one's choice of prior of the absolute neutrino mass scale \cite{DeSalas:2018rby,Gariazzo:2022ahe}.
Significant improvements are expected to reach the $\sigma(\sum m_\nu)\sim0.04$ eV level with upcoming data from DESI and VRO, see the CF7 report \cite{Adhikari:2022sve}, which should be sufficient to test the results of local oscillation data in the early universe at high significance, depending on the true values.
\item If lepton number is violated via an effective operator related to neutrino mass, then we expect \textbf{neutrino-less double beta decay} to occur proportional to $m_{\beta\beta}=|\sum_iU_{ei}^2m_i|$ which could be as low as zero in the NO but is expected to be $\gtrsim1$ meV in the IO, thus a detection below 1 meV would imply the mass ordering is normal.
The latest data from KamLAND-Zen disfavors some fraction of the inverted hierarchy for favorable nuclear matrix element calculations which are fairly uncertain \cite{KamLAND-Zen:2022tow}.
\item Finally, a measurement of the \textbf{cosmic neutrino background} is sensitive, in principle, to a combination of the absolute mass scale, whether neutrinos are Majorana or Dirac, and the mass ordering \cite{Long:2014zva,Roulet:2018fyh}.
\end{itemize}
Among these non-oscillation measurements, only the cosmological sum of the neutrino masses is likely to be sensitive to the atmospheric mass ordering within the next decade.

\FloatBarrier
\section{Possible Future Experiments}

\label{sec:far future}

\subsection{DUNE to THEIA}

THEIA is a proposal to build a large mass
detector ($\gg 10$~kilotons) using 
water-based liquid scintillator.
Although it is not strictly an upgrade
to the DUNE experiment, if it were built
at SURF as a fourth DUNE module, 
THEIA could add to the long baseline
mission of DUNE.
Scintillator has enhanced 
detection efficiency at MeV energies and 
THEIA would primarily be focused on the 
measurement of solar neutrinos and 
supernova neutrinos.
However, using a combination of
scintillation light and Cherenkov light,
event reconstruction of GeV neutrino
interactions is possible.
A 17-kt THEIA module would have
$>5\sigma$ sensitivity to the neutrino
mass ordering after seven years of running
at SURF, and could observe CP violation
at $>2\sigma$ for 50\% of
$\delta_{\rm CP}$ values after 
seven years of running~\cite{Theia:2022uyh}.

\subsection{ESSnuSB}
The European Spallation Source neutrino Super-Beam (ESSnuSB) is a proposed long-baseline accelerator neutrino experiment in Sweden designed for measuring CP violation using the existing ESS facility \cite{ESSnuSB:2021azq,Alekou:2022mav,Abele:2022iml}.
Two proposed baselines of 360 km and 540 km are being considered for a neutrino beam peaked at $\sim0.3$ GeV targeting the second oscillation maximum.
The experiment imagines a later water Cherenkov far detector similar to HyperK or the MEMPHYS project \cite{MEMPHYS:2012bzz} with an assumed detector mass of 538 kt, roughly the size of two HyperK tanks, and with 40\% PMT coverage.
The accelerator facility, under construction for other uses, could be upgraded to 5 MW with $2.7\times10^{23}$ POT/year with a proton energy of 2.5 GeV.
With a preliminary systematics analysis, ESSnuSB can achieve 5 $\sigma$ discovery of CPV for 75\% of $\delta$ values in about eight years and can make a high precision measurement of $\delta$ as well.

\subsection{ICAL at INO}
The ICAL detector at the India-based Neutrino Observatory (INO) is a proposed magnetized iron calorimeter to measure atmospheric neutrinos \cite{ICAL:2015stm}.
The detector would be 50 kt and instrument with resistive plate chambers.
The detector would also be inside a 1.5 T field allowing for charge identification of muons produced in atmospheric $\nu_\mu$ CC interactions.
This would allow for a measurement of the mass ordering at 3 $\sigma$ with 10 years go data.
An 85 t mini-ICAL was commissioned in 2018 including a muon veto.
In addition to the mass ordering, ICAL at INO is also well suited for various new physics studies such as Lorentz invariance violation \cite{Sahoo:2021dit}.

\subsection{Neutrinos from muons}
A neutrino factory is a class of proposed experiments using neutrinos from muon storage rings, see e.g.~\cite{Bogacz:2022xsj}.
While the experimental challenges are considerable, they build on an effort within the energy frontier to study electroweak physics with a muon collider \cite{DeBlas:2022wxr,Aime:2022flm} and could be a natural intermediate step on the way.
The advantages of such a machine are extremely high neutrino intensities with an extremely well characterized beam coming from muon decay which contains equal numbers of muon neutrinos and electron neutrinos providing a unique flavor combination in the flux allowing for measurements of up to six out of the nine oscillation channels.
Such a proposal would likely lead to the highest precision measurement of the three-flavor oscillation parameters as well numerous additional oscillation and BSM searches, although the details depend strongly on many design choices which have to be made.
See also nuSTORM which uses as similar approach and would assist the accelerator oscillation program by measuring neutrino cross sections and possibly provide a first step towards a neutrino factory \cite{nuSTORM:2022div}.

In China, the MuOn-decay MEdium-baseline NeuTrino beam facility (MOMENT) is another proposed high intensity neutrino beam \cite{Vihonen:2022thz,Cao:2014bea}.
A 15 MW 1.5 GeV proton beam would lead to a beam of neutrinos from muon decays of 200-300 MeV targeting a baseline of 150 km.
The far detector could be Gd-doped water Cherenkov or other designs.
Such an experiment is expected to be sensitive to $\delta$ at the level of 8$^\circ$-18$^\circ$ in 10 years \cite{Tang:2019wsv}.

% \subsection{Beta beams}
% {\it Is this required? No LOIs or whitepaper on this topics to our 
% knowledge and it was not mentioned in previous P5.}

\subsection{Protvino to Orca: P2O}
Studies were made on the possibilities to use KM3NeT/ORCA as far detector for a long baseline neutrino experiment with a neutrino beam made at the U70 accelerator complex in Protvino (Russia).
It has been shown that P2O offers sensitivity to $\delta$ comparable to or beyond that of DUNE or HK \cite{Akindinov:2019flp,hep-ph_Perrin-Terrin_2021,Perrin-Terrin:2021jtl}, reaching about $\sigma(\delta)\sim6^\circ-8^\circ$ for all values of $\delta$ including tagging of the initial state given 10 years of 450 kW beam.

At modest beam power, it appears that the technology being developed for the HL-LHC would be sufficient to instrument the beam line with silicon trackers. Those trackers would allow to reconstruct the kinematics of each and all neutrinos produced by the $\pi^+ \to \mu^+ \nu_\mu$ decays with energy resolutions better than 1\%. Using time and angular coincidence each neutrino observed in KM3NeT/ORCA could be individually matched with the one reconstructed kinematically. Such an association would remove most of the systematic uncertainties and would allow to reach energy resolutions inaccessible with conventional neutrino detectors.

\FloatBarrier
\section{Conclusions}
\label{sec:conclusions}
Since the discovery of neutrino oscillations almost 25 years ago, there has been a massive global effort to measure the six oscillation parameters.
This effort has been extremely successful.
Since then, three of the parameters have become fairly well measured: $\theta_{13}$, $\theta_{12}$, and $\Delta m^2_{21}$, we have partial information about two parameters: $\theta_{23}$ and $\Delta m^2_{31}$, and we are just starting to be sensitive to the final parameter: $\delta$.
As the next generation experiments aim to measure the final parameters, we expect the three-flavor oscillation scenario to either fall into stark relief, or holes will emerge pointing us towards new physics.

With the support of the last P5 report, the US is becoming a world leader in measuring and understanding these elusive particles.
In particular, the report set the stage to build the Deep Underground Neutrino Experiment (DUNE) which will be the most sophisticated neutrino oscillation experiment ever constructed.
This will build on the US's success of long-baseline accelerator experiments MINOS and NOvA and the LArTPC detectors of MicroBooNE and the rest of the short-baseline neutrino program.
DUNE will provide competitive measurements of all six oscillation parameters and will likely be world leading in several of them.

In Japan, building of the success of T2K and SuperK, HyperK will be built which will also have sensitivity to the majority of the oscillation parameters.
In China, building on the success of Daya Bay, JUNO will have excellent sensitivity to the solar oscillation parameters and some sensitivity to the atmospheric mass ordering.
At the south pole and in the Mediterranean IceCube and KM3NeT respectively will have very good sensitivity to the atmospheric oscillation parameters.
In addition, oscillation parameters can be somewhat probed in astrophysical environments in very orthogonal ways.
Finally, there exist several ideas for neutrino oscillation experiments beyond even those under construction now to further refine and improve our understanding neutrino physics.

In addition to the impressive strengths of individual experiments, the combined fit of several or all of them will be extraordinarily powerful.
Such combined fits and joint fits as well as the initial experimental analyses are crucial to support to ensure a timely presentation of results.

If the proposed experiments are fully constructed in a timely fashion, the atmospheric mass ordering will be robustly tested in multiple different oscillation scenarios.
We will also have excellent sensitivity to determining the octant of $\theta_{23}$ values of $\theta_{23}$ fairly close to maximal.
Finally, we will have the capability to make two truly independent measurements of $\delta$ and, depending on the true parameters, discover CP violation at high significance $>5$ $\sigma$.

In order to fully realize the goals of understanding the neutrino sector, the community needs to continue the development and construction of existing experiments.
This includes high power accelerator facilities, large state-of-the-art far detectors, and comprehensive near detectors for the long-baseline experiments.
To augment these challenging measurements, smaller experiments to improve our understanding of detector technologies and neutrino interactions will also be necessary.
These experiments will also need the personnel to carry out the analyses in a timely fashion to ensure that the data collected by the machines is properly analyzed.
Finally, as this program is measuring unknown parameters of nature which are extremely difficult to carry out and sensitive to involved systematics including neutrino cross sections, two or more distinct measurements of the same quantities are required to have a firm understanding of the oscillation parameters.

%\input{sections/9-acknowledgements}

% this is added just after end of document

\renewcommand{\refname}{References}

%\printglossary

\bibliographystyle{utphys}

\bibliography{common/tdr-citedb}

\appendix
\newpage
\section{LOIs tagged in NF01, ``Neutrino Oscillations''}

\begin{table}[h]
\newcounter{aa_row}
\setcounter{aa_row}{0}
\small
\begin{tabular}{r|p{0.90\textwidth}}
% =========================================================
\multicolumn{2}{l}{Neutrino experiments and facilities \hrulefill} \\
% =========================================================
\href{https://www.snowmass21.org/docs/files/summaries/NF/SNOWMASS21-NF4_NF1-RF4_RF0-CF7_CF1_SUPERK-050.pdf}{
\addtocounter{aa_row}{1} \arabic{aa_row}} &
Ongoing Science Program of Super-Kamiokande
\\
\href{https://www.snowmass21.org/docs/files/summaries/NF/SNOWMASS21-NF1_NF2_Daya_Bay-086.pdf}{
\addtocounter{aa_row}{1} \arabic{aa_row}} &
Legacy of the Daya Bay Reactor Neutrino Experiment
\\
\href{https://www.snowmass21.org/docs/files/summaries/NF/SNOWMASS21-NF1_NF4_SNOplus-185.pdf}{
\addtocounter{aa_row}{1} \arabic{aa_row}} &
Reactor and Geo Neutrinos at SNO+
\\
\href{https://www.snowmass21.org/docs/files/summaries/NF/SNOWMASS21-NF1_NF3__NF06_T2KCollab-130.pdf}{
\addtocounter{aa_row}{1} \arabic{aa_row}} &
T2K Experiment: future plans and capabilities
\\
\href{https://www.snowmass21.org/docs/files/summaries/NF/SNOWMASS21-NF1_NF3_Patricia_Vahle-145.pdf}{
\addtocounter{aa_row}{1} \arabic{aa_row}} &
The NOvA Physics Program through 2025
\\
\href{https://www.snowmass21.org/docs/files/summaries/NF/SNOWMASS21-NF1_NF3_Jeremy_Wolcott-088.pdf}{
\addtocounter{aa_row}{1} \arabic{aa_row}} &
Expected Final Sensitivity of the NOvA Experiment to 3-Flavor Neutrino Oscillations
\\
\href{https://www.snowmass21.org/docs/files/summaries/NF/SNOWMASS21-NF1_NF0_Ryan_Patterson-093.pdf}{
\addtocounter{aa_row}{1} \arabic{aa_row}} &
Development of a joint oscillation analysis by the NOvA and T2K collaborations 
\\
\href{https://www.snowmass21.org/docs/files/summaries/NF/SNOWMASS21-NF1_NF4_Pedro_Ochoa-034.pdf}{
\addtocounter{aa_row}{1} \arabic{aa_row}} &
The JUNO Experiment
\\
\href{https://www.snowmass21.org/docs/files/summaries/NF/SNOWMASS21-NF1_NF0_DUNE-052.pdf}{
\addtocounter{aa_row}{1} \arabic{aa_row}} &
Long-Baseline Physics in DUNE
\\
\href{https://www.snowmass21.org/docs/files/summaries/NF/SNOWMASS21-NF1_NF4-RF4_RF5_Aurisano-154.pdf}{
\addtocounter{aa_row}{1} \arabic{aa_row} } &
Atmospheric $\nu_\tau$ Appearance in the Deep Underground Neutrino Experiment
\\
\href{https://www.snowmass21.org/docs/files/summaries/NF/SNOWMASS21-NF1_NF0_Tom_Stuttard-058.pdf}{
\addtocounter{aa_row}{1} \arabic{aa_row}} &
Neutrino oscillations with IceCube-DeepCore and the IceCube Upgrade
\\
\href{https://www.snowmass21.org/docs/files/summaries/NF/SNOWMASS21-NF7_NF1-IF2_IF9_Adam_Bernstein-099.pdf}{
\addtocounter{aa_row}{1} \arabic{aa_row} } &
 Neutrino Detection and Ranging
\\
\href{https://www.snowmass21.org/docs/files/summaries/NF/SNOWMASS21-NF1_NF0-205.pdf}{
\addtocounter{aa_row}{1} \arabic{aa_row} } &
Neutrino physics with muon-decay medium-baseline neutrino beam facility (MOMENT)
\\
%
% ===================================================
\multicolumn{2}{l}{Simulation and reconstruction \hrulefill} \\
% ====================================================
\href{https://www.snowmass21.org/docs/files/summaries/CompF/SNOWMASS21-CompF2_CompF1-NF1_NF5-CF1_CF2-IF8_IF2_Monzani-085.pdf}{
\addtocounter{aa_row}{1} \arabic{aa_row}} &
Fast Simulations for Noble Liquid Experiments
\\
\href{https://www.snowmass21.org/docs/files/summaries/CompF/SNOWMASS21-CompF3_CompF2-EF0_EF0-NF1_NF6_Kagan-129.pdf}{
\addtocounter{aa_row}{1} \arabic{aa_row}} &
Differentiable Simulators for HEP
\\
% ==
\href{https://www.snowmass21.org/docs/files/summaries/CompF/SNOWMASS21-CompF3_CompF2-NF1_NF5-CF1_CF2-IF8_IF3_Monzani-084.pdf}{
\addtocounter{aa_row}{1} \arabic{aa_row} } &
The Future of Machine Learning in Rare Event Searches
\\
\href{https://www.snowmass21.org/docs/files/summaries/NF/SNOWMASS21-NF1_NF3-CompF3_CompF0_Aurisano-152.pdf}{
\addtocounter{aa_row}{1} \arabic{aa_row} } &
$\nu_\tau$ Reconstruction in the Deep Underground Neutrino Experiment
\\
\href{https://www.snowmass21.org/docs/files/summaries/NF/SNOWMASS21-NF1_NF6-CompF3_CompF4_HarryBool-191.pdf}{
\addtocounter{aa_row}{1} \arabic{aa_row} } &
Scalable, End-to-End Optimizable Data Reconstruction and Physics Inference Techniques for 
Large-scale Particle Imaging Neutrino Detectors
\\
% ============================================
\multicolumn{2}{l}{Theory and phenomenology \hrulefill} \\
% =============================================
\href{https://www.snowmass21.org/docs/files/summaries/NF/SNOWMASS21-NF1_NF3-TF0_TF0_Peter_Denton-010.pdf}{
\addtocounter{aa_row}{1} \arabic{aa_row} } &
Direct Probes of the Matter Effect in Neutrino Oscillations
\\
\href{https://www.snowmass21.org/docs/files/summaries/NF/SNOWMASS21-NF1_NF3_POONAM_MEHTA-027.pdf}{
\addtocounter{aa_row}{1} \arabic{aa_row} } &
Role of higher order maxima of oscillation probabilities at long baseline neutrino experiments
\\
\href{https://www.snowmass21.org/docs/files/summaries/NF/SNOWMASS21-NF1_NF3_Poonam_Mehta-204.pdf}{
\addtocounter{aa_row}{1} \arabic{aa_row} } &
Role of high energy beam tunes in optimizing the sensitivity to current unknowns at DUNE
\\
\href{https://www.snowmass21.org/docs/files/summaries/NF/SNOWMASS21-NF1_NF5-TF11_TF0_Julia_Gehrlein-025.pdf}{
\addtocounter{aa_row}{1} \arabic{aa_row} } &
Leptonic Sum Rules
\\
\href{https://www.snowmass21.org/docs/files/summaries/NF/SNOWMASS21-NF1_NF5-TF11_TF0_Kevin_J_Kelly-126.pdf}{
\addtocounter{aa_row}{1} \arabic{aa_row} } &
Tau Neutrino Physics
\\
\href{https://www.snowmass21.org/docs/files/summaries/NF/SNOWMASS21-NF1_NF8_Ivan_Martinez_Soler-176.pdf}{
\addtocounter{aa_row}{1} \arabic{aa_row} } &
Physics with Sub-GeV Atmospheric Neutrinos
\\
\href{https://www.snowmass21.org/docs/files/summaries/NF/SNOWMASS21-NF2_NF1_Joint_Oscillation_Analyses_at_Reactors-115.pdf}{
\addtocounter{aa_row}{1} \arabic{aa_row} } &
Joint Experimental Oscillation Analyses in Search of Sterile Neutrinos
\\
\href{https://www.snowmass21.org/docs/files/summaries/NF/SNOWMASS21-NF2_NF1_Rosner-045.pdf}{
\addtocounter{aa_row}{1} \arabic{aa_row} } &
Three Sterile Neutrinos in E6
\\
\href{https://www.snowmass21.org/docs/files/summaries/NF/SNOWMASS21-NF3_NF1-CF2_CF0-TF11_TF0_Pedro_Machado-203.pdf}{
\addtocounter{aa_row}{1} \arabic{aa_row} } &
Ultralight dark matter and neutrinos
\\
\href{https://www.snowmass21.org/docs/files/summaries/NF/SNOWMASS21-NF3_NF1-CF7_CF0-TF11_TF8_Peter_Denton-023.pdf}{
\addtocounter{aa_row}{1} \arabic{aa_row} } &
Neutrino Non-Standard Interactions
\\
\href{https://www.snowmass21.org/docs/files/summaries/NF/SNOWMASS21-NF3_NF1-EF9_EF0-RF4_RF6-CF1_CF3-TF11_TF9-AF5_AF0-195.pdf}{
\addtocounter{aa_row}{1} \arabic{aa_row} } &
Neutrino Minimal Standard Model — a unified theory of microscopic and cosmic scales
\\
% ==============================================================
\multicolumn{2}{l}{Supporting program (theory and experiment) \hrulefill} \\
% ==============================================================
\href{https://www.snowmass21.org/docs/files/summaries/NF/SNOWMASS21-NF6_NF1-TF11_TF0-CompF2_CompF0_Katori-094.pdf}{
\addtocounter{aa_row}{1} \arabic{aa_row} } &
Neutrino-induced Shallow- and Deep-Inelastic Scattering
\\
\href{https://www.snowmass21.org/docs/files/summaries/NF/SNOWMASS21-NF6_NF1-TF11_TF0_Kendall_Mahn-147.pdf}{
\addtocounter{aa_row}{1} \arabic{aa_row} } &
Electron scattering and neutrino programs
\\
\href{https://www.snowmass21.org/docs/files/summaries/NF/SNOWMASS21-NF6_NF1-TF5_TF11-CompF2_CompF0_Aaron_Meyer-111.pdf}{
\addtocounter{aa_row}{1} \arabic{aa_row} } &
Nucleon Form Factors for Neutrino Physics
\\
\href{https://www.snowmass21.org/docs/files/summaries/NF/SNOWMASS21-NF6_NF1_Mayly_Sanchez-139.pdf}{
\addtocounter{aa_row}{1} \arabic{aa_row} } &
Physics Opportunities in ANNIE
\\
\end{tabular}
\end{table}

\newpage
\section{Snowmass whitepapers related to NF01}

\begin{table}[h]
\newcounter{bb_row}
\setcounter{bb_row}{0}
\small
\begin{tabular}{r|p{0.90\textwidth}}
\href{https://arxiv.org/abs/2203.06100}{
\addtocounter{bb_row}{1} \arabic{bb_row} } &
Snowmass Neutrino Frontier: DUNE Physics Summary
\\
\href{ https://arxiv.org/abs/2203.06281 }{
\addtocounter{bb_row}{1} \arabic{bb_row} } &
A Gaseous Argon-Based Near Detector to Enhance the Physics Capabilities of DUNE 
\\
\href{ https://arxiv.org/abs/2203.06853 }{
\addtocounter{bb_row}{1} \arabic{bb_row} } &
Electron Scattering and Neutrino Physics 
\\
\href{ https://arxiv.org/abs/2203.07214}{
\addtocounter{bb_row}{1} \arabic{bb_row} } &
High Energy Physics Opportunities Using Reactor Antineutrinos
\\
\href{ https://arxiv.org/abs/2203.08276}{
\addtocounter{bb_row}{1} \arabic{bb_row} } &
Design Considerations for Fermilab Multi-MW Proton Facility in the DUNE/LBNF era
\\
\href{ https://arxiv.org/abs/2111.06932}{
\addtocounter{bb_row}{1} \arabic{bb_row} } &
A Cost-Effective Upgrade Path for the Fermilab Accelerator Complex
\\
\href{ https://arxiv.org/abs/2106.02133 }{
\addtocounter{bb_row}{1} \arabic{bb_row} } &
An Upgrade Path for the Fermilab Accelerator Complex
\\
\href{ https://arxiv.org/abs/2203.08094 }{
\addtocounter{bb_row}{1} \arabic{bb_row} } &
The Physics Case for a Neutrino Factory
\\
\href{ https://arxiv.org/abs/2202.12839 }{
\addtocounter{bb_row}{1} \arabic{bb_row} } &
Theia: Summary of physics program
\\
\href{ https://arxiv.org/abs/2203.03925 }{
\addtocounter{bb_row}{1} \arabic{bb_row} } &
Physics Opportunities for the Fermilab Booster Replacement 
\\
\href{ https://arxiv.org/abs/2203.13979 }{
\addtocounter{bb_row}{1} \arabic{bb_row} } &
Japan's Strategy for Future Projects in High Energy Physics
\\
\href{ https://arxiv.org/abs/2203.05591 }{
\addtocounter{bb_row}{1} \arabic{bb_row} } &
Tau Neutrinos in the Next Decade: from GeV to EeV
\\
%
%
% \href{ [link] }{
% \addtocounter{bb_row}{1} \arabic{bb_row} } &
% [title] 
% \\
%
\end{tabular}
\end{table}

\end{document}